\documentclass[11pt]{article}
\usepackage{epsfig}
\usepackage{amsfonts}
\usepackage{amsmath}
\usepackage{amssymb}
\usepackage{dsfont}
\usepackage{hyperref}
\usepackage{color}
\usepackage{cite}
\usepackage{pdflscape}
\usepackage{pifont}

    \renewcommand{\arraystretch}{1.5}

\newcommand{\al}{\alpha}
\newcommand{\bt}{\beta}
\newcommand{\g}{\gamma}

\newcommand{\simu}{\sigma^{\mu\nu}}

\newcommand{\MQCD}{\Lambda_\chi}

\newcommand{\Or}{\mathcal O}

\newcommand{\vL}{\ensuremath{\mathcal{L}}}
\newcommand{\vp}{\varphi}
\newcommand{\sq}{^{2}}

\newcommand{\dHg}{d_{\rm Hg}}
\newcommand{\dXe}{d_{\rm Xe}}
\newcommand{\dRa}{d_{\rm Ra}}

\newcommand{\dslash}[1]{#1 \llap{/\kern-0.5pt}}
\newcommand{\Dslash}[1]{#1 \llap{/\kern+1.5pt}}
\newcommand{\DDslash}[1]{#1 \llap{/\kern+2.3pt}}
\newcommand{\dslashh}[1]{#1 \llap{/\kern+1pt}}

\newcommand{\bea}{\begin{eqnarray}}
\newcommand{\eea}{\end{eqnarray}}
\newcommand{\bma}{\begin{pmatrix}}
\newcommand{\ema}{\end{pmatrix}}
\newcommand{\nn}{\nonumber}

\newcommand{\Ex}[1]{\cdot 10^{#1}}

\makeatletter
\renewcommand*\env@matrix[1][*\c@MaxMatrixCols c]{%
  \hskip -\arraycolsep
  \let\@ifnextchar\new@ifnextchar
  \array{#1}}
\makeatother

\begin{document}
\begin{titlepage}

\begin{flushright}
LA-UR-16-23433\\
NIKHEF 2016-017
\end{flushright}

\vspace{2.0cm}

\begin{center}
{\LARGE  \bf 
Constraining the top-Higgs sector\\[0.3cm] of the Standard Model Effective Field Theory
}
\vspace{2.4cm}

{\large \bf  V. Cirigliano$^a$, W. Dekens$^{a,b}$, J. de Vries$^c$, and E. Mereghetti$^a$ } 
\vspace{0.5cm}

{\large 
$^a$ 
{\it Theoretical Division, Los Alamos National Laboratory,
Los Alamos, NM 87545, USA}}

\vspace{0.25cm}
{\large 
$^b$ 
{\it 
New Mexico Consortium, Los Alamos Research Park, Los Alamos, NM 87544, USA
}}

\vspace{0.25cm}
{\large 
$^c$ 
{\it 
Nikhef, Theory Group, Science Park 105, 1098 XG, Amsterdam, The Netherlands
}}

\end{center}

\vspace{1.5cm}

\begin{abstract}

Working in the framework of the Standard Model Effective Field Theory, we  study chirality-flipping couplings of the top quark to Higgs and gauge bosons. We discuss in detail the renormalization group evolution to lower energies and investigate direct and indirect contributions to high- and low-energy CP-conserving and CP-violating observables.   Our analysis includes  constraints from collider  observables, precision electroweak tests, flavor physics, and electric dipole moments.  We find that indirect probes are competitive or dominant for both CP-even and CP-odd observables, even after accounting for uncertainties associated with hadronic and nuclear matrix elements, illustrating the importance of including operator mixing in constraining the Standard Model Effective Field Theory.   We also  study  scenarios where multiple anomalous top couplings are generated at the high scale, showing that while the bounds on individual couplings relax, strong correlations among couplings survive.    Finally, we find that enforcing minimal flavor violation does not significantly affect the bounds  on the top couplings.

\end{abstract}

\vfill
\end{titlepage}

\section{Introduction}

More than twenty years after its discovery at the Tevatron~\cite{Abe:1995hr,Abachi:1995iq}, 
the top quark is nowadays copiously produced at the Large Hadron Collider (LHC) 
and  is at the forefront of searches for physics beyond the Standard Model.  
In fact,  in several scenarios,   including   partial compositeness~\cite{Kaplan:1991dc}, 
warped extra dimensions~\cite{Agashe:2006wa},  
supersymmetric  models with  light stops  relevant for electroweak baryogenesis~\cite{Carena:2008vj}, 
enhanced  deviations from the SM  are expected  in the top sector. 
The top quark is the fermion with the strongest coupling to the Higgs. The dominant Higgs production
mechanism in the SM, the Higgs width, and several important decay channels, are therefore sensitive probes of top-Higgs couplings.
At the same time, the anomalous top-Higgs couplings affect via quantum corrections processes that do not  necessarily involve a top quark and/or a Higgs boson.  Such ``indirect probes''   give valuable complementary information and,   
as we demonstrate here, can  constrain non-standard top-Higgs couplings more strongly than direct searches.

In this  article  we discuss  direct and indirect probes of  chirality-flipping 
top  couplings to gauge bosons and the Higgs, including  both CP-conserving  (CPC) and CP-violating (CPV) interactions. This set of interactions gives rise to an interesting list of phenomena ranging from signals in low-energy precision tests to deviations from SM predictions in proton-proton collisions at the LHC. Additional interactions that do not change the chirality of the top quark also appear at the dimension-six level. 
We will not consider these terms in this work.
A lot has been written already about top-gluon~\cite{Atwood:1992vj,Choudhury:2012np,Baumgart:2012ay,Biswal:2012dr,Bernreuther:2013aga,Hioki:2013hva,Aguilar-Saavedra:2014iga,Bramante:2014gda,Englert:2014oea,Rindani:2015vya,Gaitan:2015aia,Bernreuther:2015yna},
top-photon~\cite{CorderoCid:2007uc,Fael:2013ira,Bouzas:2012av,Bouzas:2013jha,Rontsch:2015una},  
top-$W$~\cite{Grzadkowski:2008mf,Drobnak:2010ej,GonzalezSprinberg:2011kx,Drobnak:2011aa,Rindani:2011pk,Rindani:2011gt,Cao:2015doa,Hioki:2015env,Aguilar:2015vsa,Schulze:2016qas,Birman:2016jhg},
top Yukawa~\cite{BhupalDev:2007ftb,Brod:2013cka,Dolan:2014upa,Demartin:2014fia,Kobakhidze:2014gqa,Khatibi:2014bsa,Demartin:2015uha,Chen:2015rha,Buckley:2015vsa,Mileo:2016mxg} couplings, 
and  global analyses~\cite{Kamenik:2011dk,Zhang:2012cd,deBlas:2015aea,Buckley:2015nca,Buckley:2015lku,Bylund:2016phk}. 
Throughout this work we often use results and insights from these papers. 
The main features of our work are: 
\begin{itemize}
\item We perform a systematic analysis of the renormalization-group equations of the top-Higgs operators including QCD and electroweak corrections. This is crucial in obtaining the strongest constraints on the set of top-Higgs couplings as in many instances indirect observables are more sensitive than direct observables.
\item We investigate in detail the  impact of measurements of electric dipole moments (EDMs) ~\cite{CorderoCid:2007uc,Kamenik:2011dk,Brod:2013cka,Gorbahn:2014sha,Chien:2015xha}. 
A major finding of our analysis is that  even after taking into account the  hadronic and nuclear  uncertainties~\cite{Chien:2015xha},  
EDMs dominate the bounds on all  the CPV  top couplings. 
In particular, as we reported in Ref.~\cite{Cirigliano:2016njn},  
bounds on the  top EDM  (weak EDM)  are improved by  three orders of magnitude over the previous literature. The new constraints on the CPV top couplings lie well below prospected sensitivities of collider searches.
\item  We present a  comprehensive analysis of direct and indirect constraints, the latter arising from both high- and low-energy,
including   up-to-date  indirect constraints from Higgs production and decay at the LHC. We show that constraints from Higgs production and decay signal strengths are competitive with respect to observables involving top final states.
\item  We derive bounds on anomalous couplings under three assumptions: first, we only allow one operator at a  time to 
be generated at the high scale  (including both CPC and CPV couplings). 
Second, we perform a global analysis where all chirality-flipping top-Higgs couplings are generated at the high scale and we investigate how the constraints on the individual couplings are softened due to cancellations between different operators. Finally, we apply the framework of Minimal Flavor Violation in which the top-Higgs couplings are related to couplings involving lighter quarks.
\end{itemize}

We work in the framework of  the linear SM Effective 
Field Theory (SM-EFT)~\cite{Buchmuller:1985jz,Grzadkowski:2010es,Jenkins:2013zja,Jenkins:2013wua, Alonso:2013hga}.
We assume that there is a gap between the scale of new physics $\Lambda$ and 
the electroweak scale $v=246$~GeV and keep only the  leading terms in $(v/\Lambda)^2$, 
corresponding to  dimension-six operators. 
We assume that at the high-scale  $\Lambda$ the largest non-standard  effects appear in the top sector, 
and hence set to zero all other couplings.
We then evolve the  non-standard top couplings to lower scales through renormalization group 
flow and heavy SM particle thresholds.  The evolution induces  operators 
that impact a number of high-energy and low-energy phenomena, 
thus leading to  constraints  on non-standard top-Higgs couplings  at the scale $\Lambda$. 

Direct probes involve top quark production 
(single top,  $t \bar t$,  and $t \bar t h$) and  decay  ($W$ boson helicity fractions, lepton angular distributions) at colliders. 
We include CPV effects in the angular distributions of the decay products of a single top  \cite{Aad:2015yem}, 
while we neglect CPV observables in 
$t \bar t$ and $t \bar t h$ production/decay~\cite{Bernreuther:1992be,Brandenburg:1992be,Bernreuther:1993hq,Choi:1997ie,Sjolin:2003ah,Antipin:2008zx,Gupta:2009wu,Gupta:2009eq,Hayreter:2015ryk, Mileo:2016mxg}  as these are not yet competitive. However, as the number of independent low-energy CPV observables is limited, future measurements of CPV observables at colliders can play an important complementary role.
Indirect probes involve top quarks in quantum loops, 
affecting both high-energy  (Higgs production and decay, precision electroweak tests) and  low-energy observables
($b \to s \gamma$  and EDMs).

The paper is organized as follows. 
In Section~\ref{sect:op-structure} we set up the operator analysis by discussing the high-scale top operators and identifying 
the relevant operators that are induced by operator mixing. 
In Section~\ref{sec:RGEs} we present the renormalization group equations (RGEs) needed to evolve the anomalous top couplings 
from high-scale to the energy scales associated to a variety of observables (ranging from collider to atomic EDMs). 
We discuss direct and indirect constraints from collider observables in Section~\ref{sect:collider}, 
and the impact of precision electroweak measurements in Section~\ref{sect:ew-precision}.
Indirect bounds from flavor physics and EDMs are discussed in Sections~\ref{sect:flav} and
\ref{sect:EDMs}, respectively. 
We then present  our fitting  strategy   (Section~\ref{analysis}) and results (Section~\ref{discussion}). 
We cast our analysis into the framework of Minimal Flavor Violation (Section~\ref{sect:MFV}) before presenting 
our conclusions (Section~\ref{sect:conclusion}).

\section{Operator structure}
\label{sect:op-structure}

In this work we  study chirality-flipping interactions involving the top quark and the Higgs boson. At dimension six,  five such interactions appear in the complete set of gauge-invariant operators \cite{Buchmuller:1985jz,Grzadkowski:2010es}. In their  $SU(3)_c\times SU(2)\times U(1)_Y$ invariant forms these operators consist of a hypercharge dipole, gluonic dipole, a non-standard Yukawa coupling, and two electroweak dipoles, 
\bea
\vL_{\rm eff}^{\rm BSM} &=& - \frac{g'}{\sqrt{2}}  \  \bar{q}_L  \simu B_{\mu\nu}   \, \Gamma_B^u  \, u_R \, \tilde  \varphi     
- \frac{g_s}{\sqrt{2}}  \  \bar{q}_L  \simu G^a_{\mu\nu} t^a  \, \Gamma_g^u  \, u_R \, \tilde  \varphi- \sqrt{2}  \varphi^\dagger \varphi  \  \bar{q}_L   \,  Y_u'  \, u_R  \, \tilde  \varphi \nn\\
&&-\frac{g}{\sqrt{2}}  \  \bar{q}_L  \simu W_{\mu\nu}^a \tau^a   \, \Gamma_W^u  \, u_R \, \tilde  \varphi 
- \frac{g}{\sqrt{2}}  \  \bar{q}_L  \simu W_{\mu\nu}^a \tau^a   \, \Gamma_W^d  \, d_R \,   \varphi +\text{h.c.}.
\eea
Here $q_L$ represents the left-handed quark doublet,  $u_R$ and $d_R$ are the right-handed quark singlets, and $\vp$ is the Higgs doublet ($\tilde \vp = i\tau^2 \vp^*$). 
$B_{\mu\nu},\, W^a_{\mu\nu}$, and $G^a_{\mu\nu}$ are the fields strengths of the $U(1)_Y$, $SU(2)$, and $SU(3)_c$ gauge groups, while  $g',\, g$, and $g_s$ denote their gauge couplings and $\tau^a/2$ and $t^a$ are the $SU(2)$ and $SU(3)_c$ generators, respectively. 
Our conventions are such that the covariant derivative is given by $D_\mu = \partial_\mu -ig_s t^a G^a_\mu -i \frac{g}{2}  \tau^aW^a_\mu -ig'Y B_\mu$, with $Y$ the hypercharge.
The couplings $Y_u'$ and  $\Gamma_{g,B,W}^{u,d}$ have mass dimension $-2$ and generally form $3\times 3$ matrices in flavor space.

Working in the unitary gauge, in which the Higgs doublet takes the form $\vp = (0,\, v+h)^T/\sqrt{2}$, one sees that to $\Or(h^0)$ the couplings $Y_u'$ then contribute to the up-type quark mass matrix.  After absorbing these contributions into the SM quark mass matrix, and moving to the physical (mass) basis, the resulting effective Lagrangian  encoding 
non-standard top couplings (neglecting flavor-changing neutral currents) at the high scale $\Lambda \gg v$ is~\footnote{Denoting the Standard Model 
Yukawa couplings by  ${\cal L}_Y \supset   - \sqrt{2} \bar{q}_L Y_u u_R \tilde{\varphi}$, 
the up-type quark mass  matrix  is  $m_{u} = v \left(Y_{u} - \frac{v^2}{2}  Y_{u}^\prime \right)$.  
Upon  expressing  ${\cal L}_Y + {\cal L}_{\rm eff}^{\rm BSM}$  
in terms of $m_{u}$ and $Y_{u}^\prime$,  
we get ${\cal L}_Y + {\cal L}_{\rm eff}^{\rm BSM} \supset - \frac{m_t}{v} \bar{t} t  h  +  (C_Y O_Y  + {\rm h.c.})$, 
with the relation between $C_Y$ and $[Y_u']_{33}$ given in  Table~\ref{tab:op-invariant}.
}
\begin{equation}
{\cal L}_{\rm top}   =
\sum_{\alpha \in \{ Y, g, \gamma, Wt,Wb\} }  \ C_\alpha  \, O_\alpha + {\rm h.c.} \qquad  \quad
\label{eq:Leff}
 \end{equation}
with complex couplings  $C_\alpha = c_\alpha + i \,   \tilde{c}_\alpha$  and
\begin{subequations}
\label{eq:operators}
\bea
O_\gamma &=&  - \frac{e Q_t}{2}  m_t  \, \bar{t}_L  \sigma_{\mu \nu} \left( F^{\mu \nu} - t_W  Z^{\mu \nu} \right) t_R \, \left(1 + \frac{h}{v}\right)   
\qquad 
\\
O_g &=&  - \frac{g_s}{2}  m_t  \, \bar{t}_L  \sigma_{\mu \nu} G^{\mu \nu} t_R \, \left(1 + \frac{h}{v}\right)   
\\
O_{Wt} \!\!   &=& -g m_t \bigg[  \frac{1}{\sqrt{2}}  \bar{b}_L'   \simu  t_R W_{\mu\nu}^- 
+  \bar t_L\simu t_R \bigg(\frac{1}{2c_W} Z_{\mu\nu}+i g W_\mu^-W_\nu^+\bigg)\bigg]\bigg(1+\frac{h}{v}\bigg) \qquad
\\
O_{Wb} \!\! &=& - g m_b   \bigg[\frac{1}{\sqrt{2}} \bar t_L'  \simu b_R  W_{\mu\nu}^+ 
- \bar b_L\simu b_R \bigg(\frac{1}{2c_W} Z_{\mu\nu}+i g W_\mu^-W_\nu^+\bigg)\bigg]\bigg(1+\frac{h}{v}\bigg) 
\\
O_Y  &=&   -  m_t  \bar{t}_L  t_R  \left( v h  + \frac{3}{2}  h^2 + \frac{1}{2}  \frac{h^3}{v}  \right)  ~, 
\label{eq:NSY}
\eea
\end{subequations}
where $Q_t = 2/3$,  $t_W = \tan \theta_W$,  $c_W = \cos \theta_W$,  with $\theta_W$ the Weinberg angle. 
The physical photon and $Z$ boson fields are given by  $Z_\mu = c_W W^3_\mu -s_W B_\mu$ and $A_\mu = c_W B_\mu +s_W W^3_\mu$.
Finally, the operators $O_{Wt,Wb}$  contain the combinations  $b' = V_{tb} b + V_{ts} s + V_{td} d$, and  $t' = V_{tb}^* t + V_{cb}^* c + V_{ub}^* u$.  
The relation between these operators in the quark mass basis and their  $SU(2)\times U(1)_Y$ invariant forms is given in Table \ref{tab:op-invariant}.

The couplings  $C_\alpha$ have mass dimension $[-2]$ and are related to  properties of the 
top quark, such as the electric and  magnetic dipole moments 
($d_t = (e m_t Q_t)  \tilde{c}_\gamma$ and  $\mu_t = (e m_t Q_t)  {c}_\gamma$), 
their non-abelian gluonic counterparts 
($\tilde d_t =  m_t  \tilde{c}_g$ and  $\tilde \mu_t =  m_t {c}_g$), 
and the Higgs-top, $W$-top, and $Z$-top couplings. 

We assume that at the high-scale  $\Lambda$ the largest non-standard  effects appear in the top sector, 
and hence set to zero all other couplings.  
We then evolve the  non-standard top couplings to lower scales through renormalization group 
flow and heavy SM particle thresholds.  The evolution induces  operators 
that impact a number of high-energy and low-energy phenomena (of which many do not involve the top quark directly), 
thus leading to  constraints  on non-standard top-Higgs couplings  at the scale $\Lambda$.

\subsection{Mixing Structure}

A consistent field-theoretical  analysis of the phenomenological implications of the effective Lagrangian \eqref{eq:Leff} requires  
extending the operator basis to include all the operators which $O_{\alpha }$  ($\alpha \in \{ Y, g, \gamma, Wt,Wb\}$) can mix into. 
We will consider the leading-logarithm contributions and include the effects induced by the Standard Model 
gauge couplings $g_s, g, g'$, and the top Yukawa coupling $y_t$. 

Let us first consider the dipole operators $O_{\gamma,g,Wt,Wb}$, 
that  belong to the $\psi^2 H X$  category in the notation of  Refs.~\cite{Grzadkowski:2010es,Jenkins:2013zja,Jenkins:2013wua, Alonso:2013hga}:

\begin{itemize}
\item At one loop,  $O_{\gamma,g,Wt,Wb}$   
mix into  gauge-Higgs operators ($X^2 H^2$  in the notation of  Refs.~\cite{Grzadkowski:2010es,Jenkins:2013zja,Jenkins:2013wua, Alonso:2013hga}) 
and into dipole operators $\psi^2 H X$ with different flavor structures.  In the latter group,  of particular phenomenological interest  
are the flavor diagonal light-flavor dipoles and the $b \to s$ dipoles. 
Note also that the gauge-Higgs operators ($X^2 H^2$) mix back into the dipoles of any flavor.

\item The top dipoles $O_{\gamma,g,Wt,Wb}$    mix into $O_Y$,  with their respective gauge-coupling strengths. 
We retain only the mixing proportional to the strong coupling, i.e.\ that of $O_g$   into  $O_Y$.

\item  $O_{\gamma,g,Wt,Wb}$ also mix into  four-fermion operators 
with chirality structure $(\bar{L}R) (\bar{L} R)$~\cite{Alonso:2013hga}, 
namely the four-quark  operators  $Q^{(1),(8)}_{quqd}$ and  the semi-leptonic operator  $Q^{(3)}_{lequ}$. 
These operators involve at least one third-generation quark (from the $O_{\gamma,g,Wt,Wb}$ vertex). 
While there are no strong phenomenological handles on  these four-fermion operators  
\footnote{The generated four-fermion operators involving only third-generations quarks  and leptons are the least suppressed by CKM factors and Yukawa couplings, 
but they are the hardest to constrain as they do not contribute to very sensitive observables. 
On the other hand, operators involving light fermion generations might mediate $t \bar{t}$  or single-top production  at colliders, 
but are   induced at a much suppressed level, proportional to the  light-fermion Yukawa couplings, thus leading to  
weak constraints on $C_{\gamma,g,Wt,Wb}$.},  
the semileptonic and four-quark operators  $Q^{(3)}_{lequ}$ and $Q^{(1),(8)}_{quqd}$ mix back into the lepton and quark dipoles, respectively~\footnote{The 
mixing of  $Q^{(1),(8)}_{quqd}$  into the dipole operators had been noticed in Refs.~\cite{Dekens:2013zca,Hisano:2012cc} and 
has been included in an updated version of Ref.~\cite{Jenkins:2013wua}. 
We thank Aneesh Manohar for confirming our results.}.
Thus,  $Q^{(3)}_{lequ}$ and $Q^{(1),(8)}_{quqd}$ feed into the electron, mercury, and neutron EDMs, 
and we therefore include them in our extended basis.

\item Finally,  at the top quark threshold $O_g$ and $O_Y$ induce
the Weinberg three-gluon operator $O_{\tilde{G}}$, which we therefore  include in our extended basis. 

\end{itemize}

The  operator $O_Y$   ($\psi^3 H^3$  in the notation of  Refs.~\cite{Grzadkowski:2010es,Jenkins:2013zja,Jenkins:2013wua, Alonso:2013hga}) 
mixes only into  $O_H = (\vp^\dagger \vp)^3$, which we do not include in our phenomenological analysis as it does not contribute to any precision 
observable useful to put constraints on $C_Y$. 
On the other hand,  $O_Y$ contributes  via threshold corrections  to  most of the operators mentioned above ($X^2 H^2$, $O_{\tilde{G}}$, and the light fermion dipoles).

\subsection{Extended Operator Basis}

Grouping the operators according to the processes they  contribute to,  we can write the 
 effective Lagrangian  at the high scale as
\begin{equation}
{\cal L}_{\rm eff} = {\cal L}_{\rm SM}  + 
{\cal L}_{\rm top}  + 
\vL_{\vp\vp XX} + 
\vL_{\vp\vp X\tilde X}  + 
\vL_{(\bar LR)(\bar LR)} +
\vL_{b \to s} + 
\vL_{\rm EDMs}
\label{eq:Leff-extended}
\end{equation}
where ${\cal L}_{\rm top}$ is given in \eqref{eq:Leff},  
\begin{subequations}
\bea
\vL_{\vp\vp XX}  &=& C_{\vp G}   O_{\vp G}   + C_{\vp B}   O_{\vp B} + C_{\vp W}   O_{\vp W} + C_{\vp W B}   O_{\vp WB}  
\label{eq:extendedLHC}
 \\
\vL_{\vp\vp X \tilde X}  &=& C_{\vp \tilde G}   O_{\vp \tilde G}   + C_{\vp \tilde B}   O_{\vp \tilde B} + C_{\vp \tilde W}   O_{\vp \tilde W} + C_{\vp \tilde W B}   O_{\vp \tilde W B}  
\label{eq:extendedEDMs2}
\\
\vL_{b \to s}  &=&   C_\gamma^{(bs)}  O_\gamma^{(bs)}  + C_g^{(bs)}  O_g^{(bs)}  
+\text{h.c.}
\label{eq:ExtendedFlavor}
\\
\vL_{(\bar LR)(\bar LR)}  &=&C_{lequ}^{(3)} O_{lequ}^{(3)}+C_{quqd}^{(1)} O_{quqd}^{(1)}+C_{quqd}^{(8)} O_{quqd}^{(8)}+{\rm h.c.}
\label{eq:ExtendedFourFermion}
\\
\vL_{\rm EDMs} &=&     \sum_{f=e,u,d,s,c,b}    \left(  C_\g^{(f)}  O_\g^{(f)}   +\text{h.c.} \right) +  
 \sum_{q=u,d,s,c,b}  \left(  C_g^{(f)}  O_g^{(f)}   +\text{h.c.} \right)    + C_{\tilde G}  O_{\tilde G}
 ~, 
\label{eq:ExtendedEDM1}
\eea
\end{subequations}
and the  operators  of the extended basis are  explicitly given  in Table \ref{tab:extended}. 
We will present  the anomalous dimensions  for the relevant mixing terms and the 
threshold corrections in Section \ref{sec:RGEs}.  
Eq.~\eqref{eq:Leff-extended}  will be the starting point of our phenomenological analysis, 
with all Wilson coefficients $C_\alpha (\Lambda)$ set to zero except for  
$\alpha \in \{ Y, g, \gamma, Wt,Wb\}$.

Note that the operators in the extended basis contribute to a large number of CP-even and CP-odd observables 
both at high- and low-energies  (see Tables~\ref{OverviewReal} and \ref{OverviewIm} for a synopsis)
that can thus be used to constrain the   chirality-flipping top-Higgs couplings  of Eq.~\eqref{eq:Leff}.
In particular:  
\begin{itemize}
\item 
The Higgs-gauge operators  in $\vL_{\vp\vp XX}$   affect electroweak precision tests and Higgs production and decay processes. 
In particular, they contribute to $h\to \g\g$ ($C_{\vp B,\, \vp W,\, \vp WB}$), $gg\leftrightarrow h$ ($C_{\vp G}$), 
and  the $S$ parameter ($C_{\vp WB}$), see Section \ref{sect:collider} for details.

\item 
The electromagnetic and strong dipole operators $O_{\g,g}^{(bs)}$   in ${\cal L}_{b \to s}$ contribute to $b\to s\g$ decays.
The connection of $C_{\g,g}^{(bs)}$ to the BR$(b\to s\g)$ and $A_{CP}(b\to s\g)$ are discussed in Section \ref{sect:rareB}.

\item 
Finally the EDMs of light fermions, the chromo-EDMs (CEDMs) of light quarks, and the three-gluon Weinberg operator, contained in ${\cal L}_{\rm EDMs}$, contribute to the EDMs of the electron, neutron,  and  diamagnetic atoms such as mercury.  
Although the operators involving the heavier quarks, $c$ and $b$, do not contribute directly, they do facilitate indirect contributions to EDMs 
via threshold contributions to the coefficient $C_{\tilde{G}}$.  The connection between the above  interactions and EDMs is discussed in Section \ref{sect:EDMs}.
 
It is important to note that the operators in ${\cal L}_{\rm EDMs}$ can be induced via mixing in two ways.
First,   the top-Higgs couplings in ${\cal L}_{\rm top}$  directly  mix into the quark (color-)EDMs, through one-loop diagrams. 
Second,  the CP-odd Higgs-gauge operators in  $\vL_{\vp\vp X \tilde X}$ and four-fermion operators in $\vL_{(\bar L R)(\bar LR)}$ mix into the light-fermion EDMs and light-quark chromo-EDMs.
In a leading logarithm analysis both effects have to be included, as the operators in ${\cal L}_{\rm top}$  mix into $\vL_{\vp\vp X \tilde X}$~\footnote{These 
interactions do not  contribute to electroweak precision tests and Higgs production and decay 
at the dimension-six level.} and $\vL_{(\bar L R)(\bar LR)}$.   This leads to a  two step path  ${\cal L}_{\rm top} \to  \vL_{\vp\vp X \tilde X,\,(\bar LR)(\bar LR)} \to {\cal L}_{\rm EDMs}$ connecting the 
top-Higgs electroweak dipoles to the light fermion EDMs,  which turns out to provide very powerful constraints. 
\end{itemize}

Having introduced all the operator structures relevant for our analysis, we discuss the renormalization group equations related to these operators in the next section.

\begin{table}[t]
\centering
$\begin{array}{|c|c|c|}
\hline 
\multicolumn{2}{|c|}{\rm Operator}   &  {\rm Coupling}   \\
\hline
\hline
- \sqrt{2}  \varphi^\dagger \varphi  \  \bar{q}_L   \,  Y_u'  \, u_R  \, \tilde  \varphi  &   O_Y  & 
y_t  C_Y  =   [Y_u']_{33} 
\\   \hline 
- \frac{g_s}{\sqrt{2}}  \  \bar{q}_L  \sigma \cdot G   \, \Gamma_g^u  \, u_R \, \tilde  \varphi &      O_g & 
y_t C_g =  [\Gamma_g^u]_{33}
\\  \hline \hline
- \frac{g'}{\sqrt{2}}  \  \bar{q}_L  \sigma \cdot B   \, \Gamma_B^u  \, u_R \, \tilde  \varphi  &    O_{\gamma,Wt}    & 
y_t  Q_t C_\gamma = -  [ \Gamma_B^u + \Gamma_W^u ]_{33} 
\\  
- \frac{g}{\sqrt{2}}  \  \bar{q}_L  \sigma \cdot W^a \tau^a   \, \Gamma_W^u  \, u_R \, \tilde  \varphi  &      &  
y_t C_{Wt} = [\Gamma_W^u]_{33} 
\\ \hline  \hline
- \frac{g}{\sqrt{2}}  \  \bar{q}_L  \sigma \cdot W^a \tau^a   \, \Gamma_W^d  \, d_R \,   \varphi  &    O_{Wb} & 
y_b C_{Wb} = [\Gamma_W^d]_{33} 
\\ \hline
\end{array}$
\caption{High-scale operators in $SU(2)\times U(1)$ invariant form~\cite{Buchmuller:1985jz,Grzadkowski:2010es} 
(left column) and mapping to the operators and couplings used in this work (center and right column).
$q_L$ represents the L-handed quark doublet, $\varphi$ is the Higgs doublet, and $\tilde \varphi = i \sigma_2 \varphi^*$.   
$g_s, g, g'$ denote the $SU(3)$, $SU(2)$,  and $U(1)$ gauge couplings, 
$y_{t,b} = m_{t,b}/v$,  
and  $\sigma \cdot X = \sigma_{\mu \nu} X^{\mu \nu}$.
The couplings $C_{\alpha}$ are related to the  $33$ components of the 
matrices $Y_u'$ and  $\Gamma_{g,B,W}^{u,d}$  in the quark mass basis. 
} \label{tab:op-invariant}
\end{table}

\begin{table}\begin{center}
\small
$\begin{array}{|c|}
\hline
\hline
O_{\varphi G} = g_s^2 \varphi^\dagger \varphi   G_{\mu \nu} G^{\mu \nu}  
\qquad 
O_{\varphi \tilde G}=- g_s^2\varphi^\dagger \varphi    G_{\mu \nu} \tilde G^{\mu \nu}  
\\ \hline
O_{\varphi W} = g^2 \varphi^\dagger \varphi   W^i_{\mu \nu} W^{i \mu \nu} 
\qquad
{O}_{\varphi \tilde W} = -g^2 \varphi^\dagger \varphi  \, \tilde W^i_{\mu \nu} W^{i \mu \nu} 
\\ \hline
O_{\varphi B} = g'^2 \varphi^\dagger \varphi   B_{\mu \nu} B^{\mu \nu} 
\qquad 
{O}_{\varphi \tilde B}= -g'^2\varphi^\dagger \varphi     \,  \tilde B_{\mu \nu} B^{\mu \nu} 
\\ \hline
O_{\varphi W B} = g g'\varphi^\dagger \tau^i  \varphi   W^i_{\mu \nu}  B^ {\mu \nu} 
\qquad 
{O}_{\varphi \tilde W B}=- g g' \varphi^\dagger \tau^i  \varphi     \,  \tilde W^i_{\mu \nu}  B^ {\mu \nu} 
\\ \hline 
O_{lequ}^{(3)}=(\bar l_L^I\simu e_R) \epsilon_{IJ}(\bar q^J_L\sigma_{\mu\nu}u_R)\\\hline
O_{quqd}^{(1)}=(\bar q_L^I u_R) \epsilon_{IJ} (\bar q_L^J d_R),\quad O^{(8)}_{quqd}=(\bar q_L^I\, t^a\, u_R) \epsilon_{IJ} (\bar q_L^J\, t^a\, d_R)
\\ \hline
O_{\tilde G} = (1/6)  g_s  f_{abc} \epsilon^{\mu \nu \alpha \beta}  G^{a}_{\alpha \beta}  G^{b}_{\mu \rho} G^{c \  \rho}_{\nu} 
\\ \hline
O_g^{(q)} =   O_g \vert_{t \to q} \qquad q =u,d,s 
\\ \hline
O_g^{(bs)} =   - (g_s/2)  \, m_b  \, \bar{s}_L \sigma_{\mu \nu} G^{\mu \nu}  b_R   \,  \left(1 + h/v \right) 
\\ \hline 
O_\gamma^{(f)} =   O_\gamma \vert_{t \to f}   \qquad f =  e, u,d,s  
\\ \hline
O_\gamma^{(bs)} =   - (Q_b e/2 \, ) m_b  \, \bar{s}_L \sigma_{\mu \nu}  \left( F^{\mu \nu}  - t_W  Z^{\mu \nu} \right) b_R  \,  \left(1 + h/v \right) 
\\\hline
\hline
\end{array}$
\caption{
Dimension-six operators induced by the  
top-Higgs interactions in Eq.~(\ref{eq:Leff}) via RG flow and threshold corrections.
We use the notation $\tilde{X}_{\mu \nu} \equiv \epsilon_{\mu \nu \alpha \beta}  X^{\alpha \beta}/2$ and $\epsilon^{0123}=+1$. Below the electroweak scale the same operator basis and naming scheme can be used, by simply replacing $\vp = (0,\, v+h)^T/\sqrt{2}$ and dropping terms involving the top quark, $Z^{\mu\nu}$, $W^{\mu\nu}$, and $h$.
} \label{tab:extended}\end{center}
\end{table}

\section{The renormalization group equations}\label{sec:RGEs}
To connect the top-Higgs couplings to observables,  Eq.\ \eqref{eq:Leff-extended} has to be evolved from the scale of new physics, $\Lambda$, to lower energies. For collider experiments the evolution to roughly the electroweak scale ($m_{t,h,Z}$) is sufficient, while for $b\to s\g$ measurements one has to lower the renormalization scale to $\mu\sim m_b$. Finally  the connection to EDMs will involve the evolution down to the QCD scale, $\MQCD\sim 1$ GeV, where QCD becomes strongly coupled and non-perturbative techniques are required. An overview of the leading contributions to observables induced in this way is presented in Table \ref{OverviewReal} and \ref{OverviewIm}.

The effects of lowering the energy scale on the real and imaginary parts of the top-Higgs couplings are determined by the renormalization group equations
\bea
\frac{d \,{\rm Re}\, \vec C_t}{d\ln \mu} = \g_t \cdot  {\rm Re}\, \vec C_t ,\qquad \frac{d \,{\rm Im} \, \vec{C}_t}{d\ln \mu} = \tilde\g_t \cdot {\rm Im}\, \vec C_t ,
\eea
where $\vec C_t = (C_\g,\, C_g,\, C_{Wt},\, C_{Wb},\, C_Y)^T$. The relevant anomalous dimensions are given by \cite{Degrassi:2005zd,Kaplan:1988ku,Grojean:2013kd,Bhattacharya:2015rsa}
\bea
\g_t = 
\frac{\al_s}{4\pi}\bma 
8C_F &-8C_F &\g_{Wt\to \g} & \g_{Wb\to \g} &0\\
\g_{\g\to g}
&16C_F-4N_c & \g_{Wt\to g}&0 &0\\
0 &2C_F &8C_F & 0&0\\
0 &0 &0 & 8C_F&0\\
0 &30 C_F y_t\sq &0 & 0&0
\label{eq:topRGEs}\ema
, \eea
where the electroweak contributions are given by 
\bea
\g_{Wt\to \g} &=& \frac{\al_{ em}}{s_W\sq\al_s}\bigg[-4\bigg[1+\frac{2T^3_t Q_b}{Q_t}\frac{T^3_t-2s_W\sq Q_t}{c_W\sq}\bigg]+\frac{2T^3_t}{Q_t}\bigg(\frac{m_t\sq}{m_W\sq}+2\sum_{q=d,s,b} |V_{tq}|\sq \frac{m_q\sq}{m_W\sq}\bigg)\bigg],\nn\\
\g_{Wb\to \g} &=& \frac{\al_{ em}}{s_W\sq\al_s}\frac{2T^3_t}{Q_t}\frac{m_b\sq}{m_W\sq}|V_{tb}|\sq ,\qquad \g_{Wt\to g}= \frac{\al_{ em}}{\al_s}\frac{4}{ s_W\sq}\bigg(1+2T^3_t\frac{T^3_t-2s_W\sq Q_t}{c_W\sq}\bigg),\nn\\
\g_{\g\to g}&=& -\frac{\al_{ em}}{\al_s}8Q_t\bigg(Q_t-\frac{T^3_t-2s_W\sq Q_t }{2c_W\sq} \bigg)\,,
\eea
with $\alpha_{ em} = e^2/(4\pi)$.
The anomalous dimensions for the imaginary parts of the couplings, $\tilde \g_t$, are equivalent, with the replacement $30\to 18$ in the $(5,2)$ element of Eq.\ \eqref{eq:topRGEs} \footnote{At one loop, the chromo-MDM and chromo-EDM operators induce, respectively,
a correction to the top mass (and dimension-4 top Yukawa) and a top pseudoscalar mass.
The pseudoscalar top quark mass term  is  not present in  the effective Lagrangian of Eq.~\eqref{eq:Leff-extended},
and  can be eliminated through an axial  transformation of the quark field, 
with the net effect of changing the  $(5,2)$ element of Eq.\ \eqref{eq:topRGEs}.
For all the observables we consider, the modification of the running of the top mass is effectively a dimension-eight effect, 
and thus beyond  our working accuracy.}.

Although the above equations can be used to evolve the top-Higgs couplings from $\Lambda$ to $m_t$, they are not sufficient to connect the top-Higgs couplings to experiment. In many cases important contributions arise from the inclusion of the additional operators of the previous section. To take these effects into account, we discuss how the additional operators are induced below, while a summary is presented in  Tables~\ref{OverviewReal} and \ref{OverviewIm}.

 --------------------------------------------------------------------------------

\subsection{RG equations for high-energy probes}\label{sec:RGLHC}
The operators appearing  
in ${\cal L}_{\vp \vp XX}$  (Eq.~\eqref{eq:extendedLHC})
are induced through the one-loop RGEs,
\bea
\frac{d \vec C_{\vp\vp XX}}{d\ln\mu}= \g_{t\to \vp\vp XX} \cdot {\rm Re}\, \vec C_t
\eea
where $\vec C_{\vp\vp XX}=(C_{\vp G},\,C_{\vp B},\, C_{\vp W},\, C_{\vp WB})^T$ and the anomalous dimensions are given by \cite{Alonso:2013hga,Elias-Miro:2013gya,Elias-Miro:2013mua},
\bea \g_{t\to \vp\vp XX}=
\frac{y_t\sq N_c}{(4\pi)\sq}\bma
0 & -C_F&0 &0&0\\
4Q_t(2Q_t-T^3_t) & 0&16Q_tT^3_t-2 &(16Q_bT^3_b-2)\big(\frac{y_b}{y_t}\big)^2 &0\\
0 & 0&-2 &-2\big(\frac{y_b}{y_t}\big)^2&0\\
-4Q_tT^3_t & 0 &16Q_t T^3_t-4&(16Q_b T^3_b-4)\big(\frac{y_b}{y_t}\big)^2&0
\ema .
\label{eq:HiggsGaugeRGE}\eea
Although $c_{Wb,\,Wt}$ do contribute to the Higgs-gauge operators, they do not induce the linear combination that contributes to $h\to\g\g$.
Up to one-loop in QCD, these operators do not undergo self-renormalization.

\subsection{RG equations for $b\to s\g$}\label{sec:RGFlavor}
The top-Higgs couplings contribute to $b\to s\g$ decays by inducing the flavor-violating operators of ${\cal L}_{b \to s}$ in Eq.~\eqref{eq:ExtendedFlavor}. 

The evolution and mixing of these two operators among themselves is the same as for $O_{\g}$ and $O_g$. In combination with the mixing with the top-Higgs couplings this gives rise to the following RGEs 
\bea
\frac{d\vec C_{(bs)}}{d\ln\mu} = \g_{(bs)} \cdot \vec C_{(bs)} + \g_{t\to (bs)}\cdot \vec C_t,
\eea
with $\vec C_{(bs)} = (C_\g^{(bs)},\,C_g^{(bs)})^T$ and \cite{Aebischer:2015fzz,Alonso:2013hga,Grzadkowski:2008mf}
\bea
 \g_{(bs)}  &=& \frac{\al_s}{4\pi} \bma 8C_F &-8C_F\\0& 16C_F-4N_c\ema
,\nn\\
\g_{t\to (bs)}&=& \frac{V_{tb}V_{ts}^*}{4\pi^2}y_t\sq\bma -\frac{1}{2}Q_t/Q_b&0 & -1/Q_b &2/Q_b&0\\ 0& -\frac{1}{2} &0 &0&0\ema.\label{eq:gammaFlavor}
\eea

\subsection{RG equations for EDMs}\label{sec:RGEDMs}
At low energies the experimental EDMs are determined by the quark (color-)EDMs,  the Weinberg operator, and the electron EDM, 
collected in  $\vec{C}_{\rm EDM}= (\tilde c_\g^{(q)},\,\tilde c_g^{(q)},\,C_{\tilde G}, \, \tilde c_\g^{(e)}  )^T$. 
The top-Higgs couplings induce these  Wilson coefficients  in two ways; either directly, or by generating an additional set of Higgs-gauge couplings, $\vec C_{\vp\vp X\tilde X}=(C_{\vp \tilde G},\,C_{\vp \tilde B},\, C_{\vp \tilde W},\, C_{\vp \tilde WB})^T$, and four-fermion operators, $\vec C_{\bar LR}= (C_{lequ}^{(3)},\, C_{q uqd}^{(1)},\,C_{quqd}^{(8)})^T$, which in turn contribute to   $\vec C_{\rm EDM}$.
Including the self-renormalization of the operators in Eq.\ \eqref{eq:ExtendedEDM1},  there are three relevant RG effects, described by the following RGEs,
\bea
\frac{d\vec C_{\rm EDM}   }{d\ln\mu} &=& \g_{\rm EDM} \cdot \vec C_{\rm EDM} + \g_{t\to {\rm EDM}}\cdot 
{\rm Im}\, \vec C_t\nn\\
&&+ \g_{\vp\vp X\tilde X\to {\rm EDM}}\cdot \vec C_{\vp\vp X\tilde X}+ \g_{\bar LR\to {\rm EDM}}\cdot {\rm Im}\, \vec C_{\bar LR}.\label{eq:EDMRGE}
\eea
The first term in Eq.\ \eqref{eq:EDMRGE} describes
the RG evolution of the quark (C)EDMs and Weinberg operator and the way they mix among themselves \cite{Weinberg:1989dx, Wilczek:1976ry, BraatenPRL}
(the electron EDM, $\tilde c_\g^{(e)}$ does not run  up to one-loop in QCD), 
\bea
\g_{\rm EDM}   = \frac{\al_s}{4\pi} \bma 8C_F &-8C_F&0 & 0 \\0& 16C_F-4N_c&2N_c & 0\\0&0& N_c+2n_f+\bt_0  & 0 \\
0 & 0 & 0 & 0 \ema~. 
\label{eq:DipoleRGE}\eea

The second term in Eq.\ \eqref{eq:EDMRGE}  
describes the direct contribution, while the third and fourth terms facilitate the two-step mechanism mentioned above.
We briefly discuss these terms below.

\subsubsection{Direct contribution}
The second term in Eq.\ \eqref{eq:EDMRGE}  represents the direct mixing of the top-Higgs interactions with the  quark (color-)EDMs. The anomalous dimensions are given by \cite{CorderoCid:2007uc},~\footnote{We neglected the contribution of $\tilde c_{Wb}$ to the b-quark EDM,   
because  $\tilde c_{Wb}$ is mainly constrained through its contribution to the b-quark CEDM and the Weinberg operator.}
\begin{align}
\g_{t\to {\rm EDM}} & = \label{eq:DirectEDM}\\   
 \frac{y_t\sq }{(4\pi)\sq} &\bma -2\frac{Q_t}{Q_q}|V_{tq}|\sq \delta_{d,s,b}^q&0 &4\frac{2T^3_q}{Q_q}|V_{tq}|\sq\delta_{d,s,b}^q& 4\frac{2T^3_q}{Q_q}|V_{qb}|\sq\frac{y_b\sq}{y_t\sq}\delta_{u,c}^q&0\\0&-2|V_{tq}|\sq\delta_{d,s,b}^q&0&16\frac{m_W\sq}{m_t\sq}\bigg[1+2T^3_b\frac{T^3_b-2s_W\sq Q_b}{c_W\sq}\bigg]\delta_{b}^q&0\\0&0&0&0&0
\\0&0&0&0&0
\ema .\nn
\end{align}
Note that the contributions to the light quark EDMs and CEDMs are always proportional to  the combination of  CKM elements, $|V_{td,ts,ub}|\sq$, and there are no direct one-loop contributions
to the electron EDM.

\begin{figure} \begin{center}
\includegraphics[width=.6\linewidth]{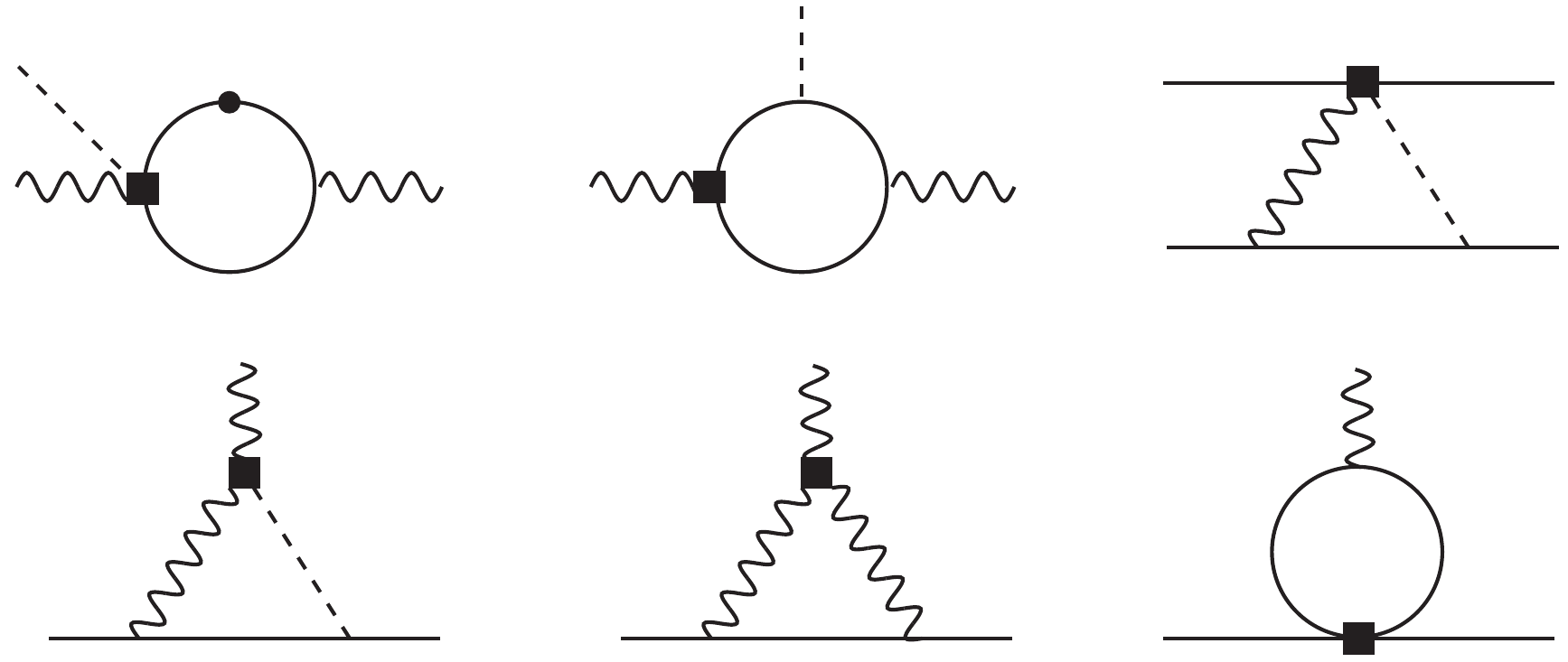}
\caption{
Representative diagrams contributing to the mixing of $C_\gamma$ into 
$C_{\varphi \tilde W, \varphi \tilde B, \varphi \tilde W B,quqd,lequ}$ (top panel), and 
the mixing of the latter into  light fermion  electroweak dipoles (bottom panel). 
The  square  (circle) represents an operator  (quark mass) insertion.
Solid, wavy, and dotted lines represent fermions,  electroweak gauge bosons, and the Higgs, respectively.  
}
\label{fig:diagrams}
\end{center}
\end{figure}

\subsubsection{Two-step mechanism}\label{sec:2step} The third and fourth terms in Eq.\ \eqref{eq:EDMRGE} are due to the two-step mechanism outlined above. There are two ways in which this mechanism can contribute to EDMs. In the first step, one induces the CP-odd Higgs-gauge interactions and four-fermion operators through the diagrams in the top panel of Fig.\ \ref{fig:diagrams}. In the second step these additional operators induce $\vec C_{\rm EDM}$ through the diagrams in the bottom panel of Fig.\ \ref{fig:diagrams}. We present the relevant anomalous dimensions for the two different paths, through the Higgs-gauge or four-fermion couplings, below.\\
\textbf{Path 1: $\vec C_t \to \vec C_{\vp\vp X\tilde X}\to \vec C_{\rm EDM}$}

In the first step of this path the CP-odd Higgs-gauge interactions are generated by the top-Higgs couplings in the same way as for their CP-even analogues,
\bea \frac{d\vec C_{\vp\vp X\tilde X}}{d\ln\mu}= \g_{t\to \vp\vp X\tilde X} \cdot {\rm Im}\, \vec{C}_t,\qquad  \g_{t\to \vp\vp X\tilde X} = \g_{t\to \vp\vp XX},\eea
where $\g_{t\to \vp\vp XX}$ is given in Eq.\ \eqref{eq:HiggsGaugeRGE},  while the Higgs-gauge interactions themselves do not run up to one-loop in QCD.~\footnote{
This is due to the fact that we include a factor of $g_s^2$ in the definition of  the operators $O_G$ and $O_{\tilde{G}}$.   
Moreover, we neglect self-renormalizations of $O(y_t^2)$ induced by the Higgs field wavefunction renormalization~\cite{Jenkins:2013zja,Jenkins:2013wua, Alonso:2013hga}.}

In the second step, the Higgs-gauge operators generate the quark (color-)EDMs and electron EDM through the third term in  Eq.\ \eqref{eq:EDMRGE}. The anomalous dimensions are \cite{Dekens:2013zca,Alonso:2013hga}
\bea
\g_{\vp\vp X\tilde X\to {\rm EDM}} =\frac{\al_{ em}}{4\pi}\bma 0 & 4\frac{2Q_q-T^3_q}{Q_q c_W\sq} & 4\frac{T^3_q}{Q_q s_W\sq}& 2\frac{t_W\sq (T^3_q-2Q_q)-3T^3_q}{Q_q s_W\sq}\\
-8\frac{\al_s}{\al_{ em}}&0&0&0\\0&0&0&0
\\
0 & 4\frac{2Q_e-T^3_e}{Q_e c_W\sq} & 4\frac{T^3_e}{Q_e s_W\sq}& 2\frac{t_W\sq (T^3_e-2Q_e)-3T^3_e}{Q_e s_W\sq}
\ema ~.
\eea
\\
\textbf{Path 2: $\vec C_t \to \vec C_{\bar LR}\to \vec C_{\rm EDM}$}

The generation and evolution of the four-fermion operators of Eq.~\eqref{eq:ExtendedFourFermion} in the first step is determined by the additional RGEs,
\bea
\frac{d\vec C_{\bar LR}}{d\ln\mu} = \g_{\bar LR} \cdot \vec C_{\bar LR} + \g_{t\to \bar LR}\cdot \vec C_t,
\eea
with $\vec C_{\bar LR} = (C_{lequ}^{(3)},\, C_{q uqd}^{(1)},\,C_{quqd}^{(8)})^T$. The QCD self-renormalization is determined by,
\bea
\g_{\bar LR} &=& \frac{\al_s}{4\pi}\bma
2C_F&0&0\\
0&-12C_F&\frac{4}{N_c}C_F\\
0&8&4C_F
\ema,\eea
while the anomalous dimensions relevant for the mixing are given by
\bea
\g_{t\to \bar LR}^{} &=& 
\frac{g\sq y_t}{(4\pi)\sq} \bma
-y_e t_W\sq Q_t(2Q_e-T^3_e)&0&-y_e(\frac{3}{2}+t_W\sq(2Q_e-T^3_e))&0&0\\
\frac{8}{N_c}y_{q}Q_t(2Q_{q}-T^3_{q})& -\frac{16}{N_c}\frac{g_s\sq}{g\sq}y_{q}C_F&\frac{8}{N_c}y_{q}(\frac{3}{2}+t_W\sq(2Q_{q}-T^3_{q}))&0&0\\
16y_{q}Q_t(2Q_{q}-T^3_{q})& \frac{16}{N_c}\frac{g_s\sq}{g\sq}y_{q}&16y_{q}(\frac{3}{2}+t_W\sq(2Q_{q}-T^3_{q}))&0&0
\ema.\nn\\
\eea
where the terms with $q=d,s,b$, contribute to the four-quark operators, $O_{q_{i}u_3q_3d_{i}}^{(1),(8)}$, with $i=1,2,3$, involving the first, second, or third generation down-type quarks, respectively. For example, the anomalous dimensions with $q=d$ generate the four-quark operators $O_{quqd}^{(1),(8)}$ with flavor structure $(\bar u_L t_R) (\bar b_L d_R)-(\bar d_L t_R) (\bar t_L d_R)$, while the flavor structure of the generated semi-leptonic operator of interest is $(\bar \nu_L t_R) (\bar b_L e_R)-(\bar e_L t_R) (\bar t_L e_R)$.

Finally, in the second step, the four-fermion operators generate the (C)EDMs of light quarks and the electron EDM through the following anomalous dimensions,
\bea
\g_{\bar LR\to {\rm EDM}}=
\frac{1}{(4\pi)\sq}\bma
0 &2\frac{Q_t y_t}{Q_{q}y_{q}} & 2C_F\frac{Q_t y_t}{Q_{q}y_{q}} \\
0 &-2\frac{ y_t}{y_{q}} & \frac{1}{N_c}\frac{ y_t}{y_{q}}\\
0&0&0\\
-16 N_c\frac{Q_t y_t}{Q_e y_e}&0&0
\ema,
\eea
where the terms with $q=d,s,b$  correspond to the contributions from four-quark operators involving the first, second, and third generation to the down, strange, and bottom (C)EDMs, respectively.

\subsection{RGE Summary}

In summary,  we can collect the Wilson coefficients of the extended operator basis in a single vector~\footnote{The real parts of the four-fermion couplings, $\vec C_{\bar LR}$, are generated in the same way as their imaginary parts.  However, since these do not contribute to EDMs, or any other sensitive observables, we neglect them in the basis, Eq.\ \eqref{eq:TotalBasis}. } 
\begin{equation}
\vec C =({\rm Re}\, \vec C_t,\,{\rm Im}\, \vec C_t,\,\vec C_{\vp\vp XX},\,\vec C_{\vp\vp X\tilde X},\, {\rm Im}\,\vec C_{\bar LR},\, {\rm Re}\, \vec C_{(bs)},\, {\rm Im}\,\vec C_{(bs)},\,  \vec C_{\rm EDM})^T
\label{eq:TotalBasis}\end{equation}
satisfying 
\begin{equation}
\frac{d \vec C}{d\ln\mu}=  \Gamma \cdot \vec{C}~,
\end{equation}
with 
\begin{equation}
\Gamma = \bma 
\g_t &- & - &-&-&-&-&-\\
-& \tilde \g_t&-&-&-&-&-&-\\
\g_{t\to \vp\vp XX}&-&-&-&-&-&-&-\\
-&\g_{t\to \vp\vp X\tilde X}&-&-&-&-&-&-\\
- &\g_{t\to \bar LR}&-&-&\g_{\bar LR}&-&-&-\\
\g_{t\to (bs)}&-&-&-&-&\g_{(bs)}&-&-\\
-&\g_{t\to (bs)}&-&-&-&-&\g_{(bs)}&-\\
-&\g_{t\to {\rm EDM}}&-&\g_{\vp\vp X\tilde X\to {\rm EDM}}&\g_{\bar LR\to {\rm EDM}}&-&-&\g_{\rm EDM}\\
\ema
\end{equation}
In our phenomenological analysis 
we will solve these RG equations with the initial condition at the 
high scale $\Lambda$ given by
$\vec{C} (\Lambda) = 
({\rm Re}\, \vec{C}_t (\Lambda), 
\, {\rm Im } \,\vec{C}_t (\Lambda), 
\, 0,
\, 0,
\, 0,
\, 0,\, 0,
\, 0)^T$.

\subsection{Evolution to $\mu = m_t$}

\begin{table}\footnotesize
\centering
$\begin{array}{c||cccccccc}
\Lambda = 1\, {\rm TeV}& c_\g(\Lambda) & c_g(\Lambda) &c_{Wt}(\Lambda)&c_{Wb}(\Lambda)&c_Y(\Lambda)
\\\hline\hline
c_\g(m_t^+)&0.86 &0.13& -9.2\Ex{-3}&-7.7\Ex{-6}&- \\
c_g(m_t^+)&2.8\Ex{-3}&0.87&-0.021&-1.2\Ex{-8}&-\\
c_{Wt}(m_t^+)&-5.4\Ex{-5}&-0.033&0.86& -&-\\
c_{Wb}(m_t^+)&-&-&-&0.86&- \\
c_{Y}(m_t^+)&-3.2\Ex{-4}&-0.20&2.4\Ex{-3}&-&1\\\hline
C_{\vp G}(m_t^+) &2.6\Ex{-5}&1.6\Ex{-2}&-2.0\Ex{-4} &-7.7\Ex{-11}&-\\
C_{\vp \g}(m_t^+)& -4.2\Ex{-4}&-3.1\Ex{-5} &2.1\Ex{-6} &1.8\Ex{-9} & -\\
C_{\vp WB}(m_t^+)& 1.6\Ex{-2}&1.5\Ex{-3} &-1.6\Ex{-2} &5.5\Ex{-6} &- 
\end{array}$ 
\caption{\small The contributions of the operators in Eq.\ \eqref{eq:operators} at $\Lambda = 1$ TeV, to the operators relevant for high-energy probes at $\mu = m_t$. A dash, $``-"$, indicates no, or a negligible, contribution.
\label{tab:mtLHC} }
\end{table}

As discussed above, for collider observables, it is mainly the mixing among the top-Higgs couplings themselves and the Higgs-gauge interactions, Eq.\ \eqref{eq:extendedLHC}, that is relevant. The  RGEs of section \ref{sec:RGLHC} can be used to first run the couplings down to $\mu=m_t$, where the top-Higgs couplings are integrated out. To evaluate the RGEs, we employ  input on the gauge couplings,  $W$, $Z$, and quark masses, and CKM elements from 
Ref.~\cite{Agashe:2014kda}. In particular, we used the values of the quark masses in the $\overline{\textrm{MS}}$ scheme.  
The resulting  Wilson coefficients, relevant for high-energy probes, are collected in Table \ref{tab:mtLHC}.

Apart from the RG effects,  additional threshold effects appear, as a result of integrating out the top-quark. At this scale, the top Yukawa induces contributions to the Higgs-gauge operators, 
\bea
C_{\vp G}(m_t^-) &= & C_{\vp G}(m_t^+) +\frac{1}{48\pi\sq}\big(1+c_Y(m_t^+)\big),\nn\\
C_{\vp\g}(m_t^-) &=  &C_{\vp\g}(m_t^+) +\frac{1}{4\pi\sq}\bigg[\frac{N_c Q_t\sq}{6}\big(1+c_Y(m_t^+)\big)-\frac{21}{24}A(\tau_W)\bigg],\nn\\
A(\tau) &=& \frac{1}{7} \left(  2 + 3 \tau + 3 \tau (2 -\tau) \arcsin\left( \frac{1}{\sqrt{\tau}} \right)^2 \right),\qquad \tau_W=4\frac{m_W\sq}{m_h\sq},
\label{eq:Afunction}
\eea
where the constant terms arise from the SM top and $W^\pm$ loops, while $C_{\vp\g}$ represents the coupling of $e^2 vhF_{\mu\nu}F^{\mu\nu}$, which corresponds to the linear combination $C_{\vp B}+C_{\vp W}-C_{\vp WB}$.

\subsection{Evolution to $\mu =\mu_b$}

\begin{table}\footnotesize
\centering
$\begin{array}{c||cccccccc}
\Lambda = 1\, {\rm TeV}& C_\g(\Lambda) & C_g(\Lambda) &C_{Wt}(\Lambda)&C_{Wb}(\Lambda)&C_Y(\Lambda)
\\\hline\hline
C^{(bs)}_\g(m_t^+)&6.2\Ex{-4} &-1.2\Ex{-5}& 1.9\Ex{-3}&-3.7\Ex{-3}&- \\
C^{(bs)}_g(m_t^+)&-5.2\Ex{-7}&-3.2\Ex{-4}&3.9\Ex{-6}&1.5\Ex{-12}&-\\\hline
C^{(bs)}_\g(\mu_b)&3.4\Ex{-4} &-1.2\Ex{-4}& 1.0\Ex{-3}&-2.0\Ex{-3}&- \\
C^{(bs)}_g(\mu_b)&-3.0\Ex{-7}&-1.9\Ex{-4}&2.3\Ex{-6}&9.0\Ex{-13}&-
\end{array}$ 
\caption{\small The contributions of the operators in Eq.\ \eqref{eq:operators} at $\Lambda = 1$ TeV, to the operators relevant for $b\to s\g$ transitions at $\mu = m_t$ and $\mu=\mu_b\approx 2$ GeV. A dash, $``-"$, indicates no, or a negligible, contribution.
\label{tab:mtFlavor} }
\end{table}

For the $b\to s\g$ observables, the main contributions follow from the mixing onto the flavor-violating operators, $O^{(bs)}_{\g,\,g}$. In this case we employ the RGEs of section \ref{sec:RGFlavor}, to run to the scale $\mu=m_t$.  Below this threshold the top-Higgs operators are integrated out, such that the top-Higgs couplings no longer contribute below this scale. The  operators $O^{(bs)}_{\g,\,g}$ can then  straightforwardly be evolved down to $\mu =\mu_b$  by use of the same RGEs with the mixing terms set to zero, $\g_{t\to  (bs)}\to 0$. As we will employ expressions for the $b\to s\g$ observables in terms of the couplings at $\mu = m_t$ and $\mu=\mu_b$, we present the values of $C^{(bs)}_{\g,\,g}$ at both scales in Table \ref{tab:mtFlavor}.

\subsection{Evolution to $\mu =\MQCD$}
Evaluating the contributions to EDMs is somewhat more involved. At low energies, {around $\Lambda_\chi$},
 the light-quark (C)EDMs, the Weinberg operator, and the electron EDM contribute to EDMs, while the charm- and bottom-quark CEDMs facilitate indirect contributions. As a result, the mixing of the  original top-Higgs operators in Eq.~\eqref{eq:Leff} 
with the additional operators in Eq.\ \eqref{eq:ExtendedEDM1} determines the contribution to EDMs. Apart from the mixing, the matching corrections at the different thresholds are relevant as well. 

First the RGE of Eq.\ \eqref{eq:EDMRGE} is used to run the operators from $\mu= \Lambda $ to $\mu=m_t$, where we integrate out the top quark and the Higgs boson. This implies that  the  top-Higgs and the additional Higgs-gauge couplings and their corresponding operators are removed from the EFT  below  
$\mu=m_t$.  This gives rise to several threshold corrections to the operators in Eq.\ \eqref{eq:ExtendedEDM1}. The Yukawa interaction, $\tilde c_Y$, contributes to the (C)EDMs \cite{Barr:1990vd,Gunion:1990iv,Abe:2013qla,Jung:2013hka,Dekens:2014jka} and the Weinberg operator \cite{Weinberg:1989dx,Dicus:1989va} through Barr-Zee diagrams, while the top CEDM gives rise to a one-loop threshold contribution to the Weinberg operator \cite{BraatenPRL,Boyd:1990bx}. In total we have the following matching conditions,
\bea\label{mtThreshold}
\tilde c^{(f)}_\g(m_t^-) &=& \tilde c^{(f)}_\g(m_t^+)+ 24\frac{\alpha_{ em}}{(4\pi)^3} Q_{t}\sq f(x_{t}) \tilde c_Y(m_t^+),\nn\\
\tilde c_g^{(q)}(m_t^-) &=& \tilde c_g^{(q)}(m_t^+)-4 \frac{\alpha_s}{(4\pi)^3} f(x_{t})\tilde c_Y(m_t^+),\nn\\
C_{\tilde G}(m_t^-)&=&C_{\tilde G}(m_t^+)-\frac{\al_s}{8\pi } \tilde c_g(m_t^+)-4 \frac{\al_s}{(4\pi)^3}h(m_t,m_h)\, \tilde c_Y(m_t^+) ,
\eea
where $m_t^+$ ($m_t^-$) indicates a scale just above (below) $m_t$, $x_t\equiv \frac{m_t\sq}{m_h\sq}$, and the functions $f$, and $h$ are given by,
\bea
&f(z) \equiv \frac{z}{2}\int_0^1 dx \frac{1-2x(1-x)}{x(1-x)-z}\ln \frac{x(1-x)}{z}\,\,, \qquad h(m,M) = \frac{m^4}{4}\int_0^1 dx\int_0^1 du \frac{u^3x^3(1-x)}{[m\sq x(1-ux)+M\sq(1-u)(1-x)]\sq}\,\,\,.\nn
\eea

Below $\mu=m_t$, our basis consists of the operators in Eq.\ \eqref{eq:ExtendedEDM1}. The  anomalous dimensions in Eq.\ \eqref{eq:DipoleRGE} 
control the running  down to $\mu=m_b$ and subsequently to $\mu=m_c$. At these thresholds the bottom and charm quarks and their (C)EDMs are integrated out, which results in additional threshold corrections to the Weinberg operator,
\bea
C_{\tilde G}(m_{c,b}^-)&=&C_{\tilde G}(m_{c,b}^+)-\frac{\al_s}{8\pi } \tilde c_g^{(c,b)}(m_{c,b}^+) .
\label{eq:dWthreshold}
\eea
After the charm threshold the remaining operators can be evolved to $\MQCD$ using Eq.\ \eqref{eq:DipoleRGE}. The numerical result of this analysis is presented in Table \ref{tab:mtEDM} for $\Lambda=1$ TeV. 

An overview of the effects of the running and threshold contributions to observables presented in Table \ref{OverviewReal} for $c_\al$ and in Table \ref{OverviewIm} for $\tilde c_\al$.

\begin{table}\footnotesize
\centering
$\begin{array}{c||ccccc}
\Lambda =1\,  \text{TeV} & \tilde{c}_{\gamma }(\Lambda) & \tilde{c}_g(\Lambda) & \tilde{c}_{{Wt}}(\Lambda) & \tilde{c}_{{Wb}}(\Lambda) & \tilde{c}_Y(\Lambda) \\\hline\hline
\tilde c_\g^{(e)}(\Lambda_\chi ) & 3.8\Ex{-4} & 1.4\Ex{-5} & -4.4\Ex{-4} & 2.3\Ex{-8} & 4.0 \Ex{-5} \\
\tilde c_\g^{(u)}(\Lambda_\chi ) & 1.4\Ex{-4} & 6.3\Ex{-4} & -1.2\Ex{-4} & -2.9\Ex{-6} & -6.1\Ex{-5} \\
\tilde c_g^{(u)}(\Lambda_\chi ) & 3.9\Ex{-6} & 1.1\Ex{-3} & -1.9\Ex{-5} & -1.7\Ex{-5} & -1.0 \Ex{-4}\\
\tilde c_\g^{(d)}(\Lambda_\chi ) & 2.0\Ex{-4} & 8.6\Ex{-4} & -9.1\Ex{-4} & -2.9\Ex{-6} & -6.1\Ex{-5} \\
\tilde c_g^{(d)}(\Lambda_\chi ) & 2.9\Ex{-6} & 1.3\Ex{-3} & -2.4\Ex{-5} & -1.7\Ex{-5} & -1.0\Ex{-4} \\
\tilde c_\g^{(s)}(\Lambda_\chi )& 1.9\Ex{-4} & 8.6\Ex{-4} & -9.5\Ex{-4} & -2.9\Ex{-6} & -6.1\Ex{-5} \\
\tilde c_g^{(s)}(\Lambda_\chi ) & 2.9\Ex{-6} & 1.3\Ex{-3} & -2.4\Ex{-5} & -1.7\Ex{-5} & -1.0\Ex{-4} \\
 C_{\tilde G}(\Lambda_\chi ) & -2.8\Ex{-6} & -8.8\Ex{-4} & 2.2\Ex{-5} & 7.8\Ex{-5} & -8.1\Ex{-7} \\
\end{array}
$
\caption{\small The contributions of the operators in Eq.\ \eqref{eq:operators} at $\Lambda = 1$ TeV, to the operators relevant for EDMs  at $\mu = \Lambda_\chi$.
\label{tab:mtEDM} }\end{table}

\begin{table}\centering\footnotesize
\renewcommand{\arraystretch}{2}
$\begin{array}{c|c||c cccc}
&{\rm \textbf{Obs.}}& c_\g & c_g&c_{Wt}&c_{Wb}&c_Y
\\\hline\hline
&t& $\ding{55} $&$\ding{55} $ & \large \checkmark& $\ding{55} $&$\ding{55} $ \\ 
&
t\bar t& $\ding{55} $&\large\checkmark & $\ding{55} $ &$\ding{55} $ & $\ding{55} $\\\rotatebox{90}{\rlap{\textbf{Direct}}}
&t\bar t h& $\ding{55} $& \large\checkmark& $\ding{55} $& $\ding{55} $&\large \checkmark\\
 &F_0,\, F_L,\,\delta^-& $\ding{55} $& $\ding{55} $&\large\checkmark & $\ding{55} $&$\ding{55} $\\
  \hline
&gg\leftrightarrow h& $\ding{55} $&\g_{t\to \vp\vp XX}^{(1,2)} &$\ding{55} $ &$\ding{55} $ & {\rm \textbf{Threshold} \, \eqref{eq:Afunction} }\\
\rotatebox{90}{\rlap{\hspace{-.2cm} \textbf{Indirect}}}
&h\to \g\g&\begin{tabular}{c}$\g_{t\to \vp\vp XX}^{(2,1),\,(4,1)}$\end{tabular}& $\ding{55} $& $\ding{55} $&$\ding{55} $ &{\rm \textbf{Threshold}\, \eqref{eq:Afunction} } \\
\hline
\rotatebox{90}{\rlap{\textbf{S}}}&S& \g_{t\to \vp\vp XX}^{(4,1)}  & $\ding{55} $&\g_{t\to \vp\vp XX}^{(4,3)} &  $\ding{55} $&$\ding{55} $ \\
\hline
&&&\\
\rotatebox{90}{\rlap{\hspace{-.4cm}$\boldsymbol{b\to s \g}$}}
&{\rm BR},\, A_{CP} &
\g_{t\to (bs)}^{(1,1)} & $\ding{55} $& \g_{t\to (bs)}^{(1,3)}& \g_{t\to (bs)}^{(1,4)}& $\ding{55} $\\&&&
\end{array}$ 
\caption{An overview of the dominant contributions of the real parts of the anomalous top-Higgs couplings to high- and low-energy observables. $\checkmark$ indicates a direct (tree-level) contribution, \ding{55} a negligible contribution,  $\gamma_{t \rightarrow X}$ a contribution induced by the RG flow, and \textbf{Threshold} a threshold contribution with the appearing numbers indicating the corresponding equations. 
}\label{OverviewReal}
\end{table}

\begin{landscape}

\begin{table}\hspace{-0.75cm}\footnotesize
$\begin{array}{c|c||ccccc}
&{\rm \textbf{Obs.}}& \tilde c_\g &\tilde  c_g&\tilde c_{Wt}&\tilde c_{Wb}&\tilde c_Y
\\\hline\hline
&t& $\ding{55} $&$\ding{55} $ &$\ding{55} $ &$\ding{55} $ & $\ding{55} $\\ 
&
t\bar t&$\ding{55} $ &$\ding{55} $ &  $\ding{55} $&$\ding{55} $ & $\ding{55} $\\\rotatebox{90}{\rlap{\textbf{Direct}}}
&t\bar t h& $\ding{55} $& $\ding{55} $&$\ding{55} $ & $\ding{55} $& $\ding{55} $\\
 &F_0,\, F_L,\,\delta^-& $\ding{55} $&$\ding{55} $ &\large\checkmark &$\ding{55} $&$\ding{55} $ \\ \hline
\rotatebox{90}{\rlap{\hspace{-.4cm}$\boldsymbol{b\to s \g}$}}
&{\rm BR},\, A_{CP}& \begin{tabular}{c} \\
$\g_{t\to (bs)}^{(1,1)}$\\ \\  
\end{tabular} &$\ding{55} $ & \g_{t\to (bs)}^{(1,3)}& \g_{t\to (bs)}^{(1,4)}& $\ding{55} $\\\hline
\rotatebox{90}{\rlap{\hspace{-1.2cm}\textbf{EDMs}}}
&d_{\rm ThO} & \g_{ t\to \vp\vp X\tilde X, \bar LR}^{(i,1)} \g_{\vp\vp X\tilde X,\bar LR\to {\rm EDM}}^{(4,i)}
&  \begin{array}{c}  \g_{t}^{(1,2)}\g_{t\to \vp\vp X\tilde X,\bar LR}^{(i,1)} \g_{\vp\vp X\tilde X,\bar LR\to {\rm EDM}}^{(4,i)}\\
 \g_{t}^{(3,2)}\g_{t\to \vp\vp X\tilde X,\bar LR}^{(i,3)} \g_{\vp\vp X\tilde X,\bar LR\to {\rm EDM}}^{(4,i)}
\end{array}
& \g_{t\to \vp\vp X\tilde X,\bar LR}^{(i,3)}\g_{\vp\vp X\tilde X,\bar LR\to {\rm EDM}}^{(4,i)}& $\ding{55} $& {\rm \textbf{Threshold} \,\eqref{mtThreshold}} 
\\[8ex]  
&d_n,\, \dHg & \g_{t\to \vp\vp X\tilde X,\bar LR}^{(i,1)}\g_{\vp\vp X\tilde X, \bar LR\to {\rm EDM}}^{(1-3,i)} &{\rm \textbf{Threshold} \,\eqref{mtThreshold}}& \g_{t\to \vp\vp X\tilde X,\bar LR}^{(i,3)}\g_{\vp\vp X\tilde X,\bar LR\to {\rm EDM}}^{(1-3,i)}&\g_{t\to {\rm EDM}}^{(2,4)}, {\rm \textbf{Threshold} \,\eqref{eq:dWthreshold}} 
&{\rm \textbf{Threshold}\,\eqref{mtThreshold}} 
\end{array}$ 
\caption{
Similar to Table~\ref{OverviewReal} but now for the imaginary parts of the anomalous top-Higgs couplings. Contributions that are generated at the two- or three-loop level are represented by entries involving a combination of anomalous dimensions and/or threshold contributions. For example, the leading contribution of $\tilde c_{Wb}$ to the mercury and neutron EDMs is due to the b-quark CEDM, $\tilde c_g^{(b)}$, which is RG-induced through $\g_{t\to {\rm EDM}}^{(2,4)}$ (Eq.\ \eqref{eq:DirectEDM}) and subsequently generates the Weinberg operator, $C_{\tilde G}$, through the threshold contribution of \eqref{eq:dWthreshold}.}\label{OverviewIm}\end{table}
\end{landscape}

\newpage
\section{Collider constraints}
\label{sect:collider}
In this section we discuss the constraints that collider experiments set on the top couplings introduced in Eq.~\eqref{eq:Leff}.
The top quark was discovered at the Tevatron \cite{Abe:1995hr,Abachi:1995iq} and is abundantly produced at the LHC. 
This makes it possible to directly probe the properties of the top quark by measuring the cross sections of processes with top final states, and the subsequent top decays. In the former category, we consider single top, $t\bar t$, and associated Higgs $t\bar t$ production cross sections that are sensitive to anomalous top-gluon ($O_g$), top-W ($O_{Wt}$ and $O_{Wb}$), and top-Higgs ($O_Y$) couplings. In the latter category we study the helicity fractions of $W$ bosons that are produced in top quark decays.  These fractions are sensitive to the top-bottom-W operators,  $O_{Wt}$
and, to a lesser extent, $O_{Wb}$.  In addition to its contribution to single top production, the operator $O_{Wb}$ generates a dipole coupling of the $Z$ boson to $b \bar b$ pairs.   This coupling affect the branching ratio $Z \rightarrow b \bar b$, which was precisely measured at LEP.   
Since the bounds turn out to be weak,  we do not further discuss this observable.

The couplings of the top can be probed indirectly, by studying observables that do not have a top quark in the final state, but instead receive sizable corrections from top loops. We consider corrections to precision EW observables, in particular the $S$ parameter \cite{Peskin:1990zt,Peskin:1991sw,Barbieri:2004qk}, and the Higgs boson production and decay signal strengths. 
In the SM, the main Higgs production mechanism is gluon fusion and proceeds through a top loop. This process is therefore quite sensitive to modifications of the top Yukawa, $C_Y$, and to the top chromo-dipole moment $C_g$. In a similar way, the SM decay process $h \rightarrow \gamma \gamma$ is loop induced, and can be used to constrain $C_Y$ and  $C_\gamma$.
We do not include corrections to Higgs production and decay mechanisms that are tree level in the SM, like vector boson fusion (VBF), $h \rightarrow W W^*$, or $h \rightarrow b \bar b$. Contributions to these processes from the operators in Eq. \eqref{eq:Leff} are loop suppressed such that any resulting constraints are weak and can be neglected.

Finally we comment on the contributions from the anomalous couplings that we include. Our EFT approach is based on an expansion in $Q/\Lambda$ where $Q$ is a low-energy scale that can be identified with the typical energy in a process, the Higgs vev, or the mass of a SM particle. We always present our results as functions of the dimensionless combinations $v^2 C_\alpha$. The most important contributions are linear in this combination and appear at the dimension-six level, $\sim \mathcal O(1/\Lambda^2)$. For production and decay cross sections such terms arise from the interference between SM and anomalous amplitudes. As we do not consider explicit dimension-eight operators, to be consistent we should truncate our expansion at $\mathcal O(1/\Lambda^2)$ and not consider terms that depend quadratically on $v^2 C_\alpha$. This means for example that for most of the collider observables under investigation, the imaginary parts of the Wilson coefficients do not contribute.

That being said, in the expressions given below we do give quadratic contributions. We will use these terms mainly as a diagnostic tool,
to check whether the EFT is working satisfactorily, in which case quadratic terms should only affect the results of the fits by a small amount.
This is particularly important for the global fits we present in Section \ref{global}.
A more detailed discussion on dimension-eight effects in the framework of the SM EFT is given in Refs.~\cite{Berthier:2015oma,Berthier:2015gja,Contino:2016jqw}.
We  stress that the dimension-eight contributions given below are not complete. In the cross sections and branching ratios we include dimension-eight effects from two insertions of the operators $C_\alpha$
in tree-level diagrams, but never consider the insertion of genuine dimension-eight operators, or the mixing of two dimension-six operators onto dimension-eight operators.
The terms we do include should be thought of as a rough probe of the impact of higher-order corrections.

\subsection{Direct constraints}\label{Direct}

\subsubsection{$t \bar t$ production}\label{ttbar}

\begin{table}
\center
\begin{tabular}{||c| c |c |c| c |c c||}
\hline \hline
process     & $\sqrt{S} $ (TeV) & \multicolumn{3}{c|}{$\sigma$ (pb)}  & \multicolumn{2}{c||}{ Experiment}   \\
\hline 
$t \bar t $ &  1.96  & \multicolumn{3}{c|}{$7.6 \pm 0.4$}  & CDF, D0  &  \cite{Aaltonen:2013wca} \\
	    &  8     & \multicolumn{3}{c|}{$242 \pm 10$}   & ATLAS 	 &  \cite{Aad:2014kva} \\
	    &	     & \multicolumn{3}{c|}{$239 \pm 13$}   & CMS	 &  \cite{Chatrchyan:2013faa} \\
\hline	  
\hline
process     & $\sqrt{S} $ (TeV) & $t$ & $\bar t$ & $t$ + $\bar t$ & \multicolumn{2}{c||}{ Experiment}  \\
\hline
single top  		& 7   & $46 \pm 6$&   $23 \pm 3$ &  $68 \pm 8$ & ATLAS &  \cite{Aad:2014fwa}\\
$t$-channel		&     & --        & -- & $67 \pm 7 $  & CMS &  \cite{Chatrchyan:2012ep}\\
			& 8   & --        & -- &   $83 \pm 12$            & ATLAS & \cite{ATLAS-CONF-2014-007} \\
			&     &  $54  \pm 5$&  $28 \pm 4$ & $84 \pm 8 $ & CMS & \cite{Khachatryan:2014iya}\\
			& 13  & $133 \pm 26$& $96 \pm 24$ & $229 \pm 48$ & ATLAS &  \cite{ATLAS-CONF-2015-079}\\
	&   & $142 \pm 23$& $81 \pm 15$ & $228 \pm 34$ & CMS & 		\cite{CMS-PAS-TOP-16-003}\\
\hline
\end{tabular}
\caption{$t\bar t$ and single top total cross sections, measured at CDF, D0, ATLAS and CMS.}\label{TabTop}
\end{table}

The cross section induced by the top chromo-magnetic dipole moment (CMDM), $c_g$, and CEDM, $\tilde c_g$, was computed in Refs.~\cite{Atwood:1994vm,Haberl:1995ek} and is given by
\begin{eqnarray}\label{ttbar.0}
\frac{\sigma_{t \bar t }(1.96 \, \textrm{TeV})}{\rm pb} &=& (7.45 \pm 0.44) - (10.8 \pm 0.6)   (v^2 c_g) + (7.1 \pm 0.7)  (v^2 c_g)^2  \nonumber   + (2.5 \pm 0.5)   (v^2 \tilde{c}_g)^2  \nonumber\\
\frac{\sigma_{t \bar t }(8\, \textrm{TeV})}{\rm pb} &=& (252.9 \pm 20) - (333 \pm 28)   \left(v^2 c_g \right) + (476 \pm 44)   \left( v^2 c_g\right)^2   + (336 \pm 33)  \left( v^2 \tilde c_g\right)^2. \label{ttbar.1} \nonumber\\
\end{eqnarray}
The SM $t\bar t$ cross section has been computed using the program TOP++ \cite{Czakon:2011xx}. It includes next-to-next-to-leading order (N${}^2$LO) corrections \cite{Czakon:2013goa} and soft gluon resummation. 
The cross section and  the couplings $c_g$ and $\tilde c_g$ are evaluated at the renormalization scale $\mu = m_t$.
The theoretical uncertainties on the SM cross section arises from PDF and scale variations. 
As the contribution of $C_g$ is only included at LO, the theoretical errors on terms proportional to $c_g$ and $\tilde c_g$ in Eq. \eqref{ttbar.0} only include PDF and $\alpha_s$ uncertainties, which are obtained by following the recipe of the PDF4LHC working group \cite{Botje:2011sn} with the three PDF sets CT10 \cite{Lai:2010vv}, MSTW08 \cite{Martin:2009iq}, and NNPDF2.3 \cite{Ball:2012cx}. In the SM, NLO and N${}^2$LO corrections to the $t \bar t$ cross section are large \cite{Czakon:2013goa} suggesting the need to include NLO corrections for the dipole operators as well \cite{Franzosi:2015osa}. We have not included these corrections here.

Our fits include data from the Tevatron experiments \cite{Aaltonen:2013wca} at $\sqrt{S} = 1.96$ TeV 
and from the ATLAS and CMS experiments at $\sqrt{S} = 8$ TeV \cite{Aad:2014kva,Chatrchyan:2013faa}. The experimental results are summarized in Tab.~\ref{TabTop}.

\subsubsection{Associated production of a Higgs boson and a $t \bar t$ pair}\label{ttbarH}

The $t\bar t h$ cross section receives contributions from the  anomalous Yukawa coupling, $C_Y$, and from the dipole operator, $C_g$.
In Ref.~\cite{Chien:2015xha} we  presented
the $t \bar t h$ cross section induced by $\tilde c_g$ and $\tilde c_Y$
at LO in QCD. Here we extend the calculation with contributions from the real part of the couplings, $c_Y$ and $c_g$, which interfere with the SM and induce genuine $\mathcal O(1/\Lambda^2)$ effects.
The $t\bar t h$ cross section in the SM is known at NLO in QCD \cite{Beenakker:2001rj,Beenakker:2002nc,Reina:2001sf,Dawson:2002tg}. 
The contribution of the pseudoscalar Yukawa couplings is also known at NLO  \cite{Frederix:2011zi}, while the effects of $C_g$ has been considered at tree level
in Refs. \cite{Degrande:2012gr,Hayreter:2013kba}. Our results agree with Refs. \cite{Degrande:2012gr,Hayreter:2013kba}.

The observable we study is the ratio, $\mu_{t\bar t h}$, of the production cross section with and without dimension-six operators,
\begin{equation}
\mu_{t\bar t h} = \frac{\sigma_{p p \rightarrow t \bar t h}}{\sigma^{SM}_{p p \rightarrow t \bar t h}}.
\end{equation}
At center-of-mass energy of  8 and 14 TeV, we find
\begin{eqnarray}
\mu_{ t \bar t h}(8\, \textrm{TeV}) &=& \left(1 + v^2 c_Y\right)^2 + (0.33 \pm 0.02) ( v^2 \tilde c_Y)^2  
- (7.11 \pm 0.02)  (v^2 c_g) \nonumber
\\
& & + (52 \pm 5)   (v^2 c_g)^2   + (44 \pm 4)  (v^2 \tilde{c_g})^2  \nonumber \\ & &  
- (11.0 \pm 0.1)  \left(v^2 c_g \right)  (v^2 c_Y) 
- (0.12 \pm 0.16)  \left(v^2 \tilde{c}_g \right)  (v^2 \tilde c_Y) \\ 
\mu_{ t \bar t h}(14 \, \textrm{TeV}) &=& \left(1 + v^2 c_Y \right)^2  
+ (0.42 \pm 0.01) (v^2 \tilde c_Y)^2
- (7.57 \pm 0.03) \left(v^2 c_g \right) \nonumber
\\ && 
+ (80 \pm 5)  \left(v^2 c_g \right)^2  
+ (72 \pm 5)  \left(v^2 \tilde c_g \right)^2  \nonumber 
\\ & & 
 - (11.5 \pm 0.1)   \left(v^2 c_g \right)  (v^2 c_Y)
 - (0.79 \pm 0.06)  \left(v^2 \tilde c_g \right)  (v^2 \tilde c_Y)
  \nonumber .
\end{eqnarray}
The theoretical error only includes PDF and $\alpha_s$ variations. The cross section, and the couplings $C_Y$ and $C_g$, are evaluated at the scale $\mu = m_t$.
The ATLAS and CMS measurements of  the signal strength $\mu_{t\bar t h}$ are given in Table \ref{HiggsSS}. 

\subsubsection{Single top production}\label{ST}

\begin{figure}
\center
\includegraphics[width=8cm]{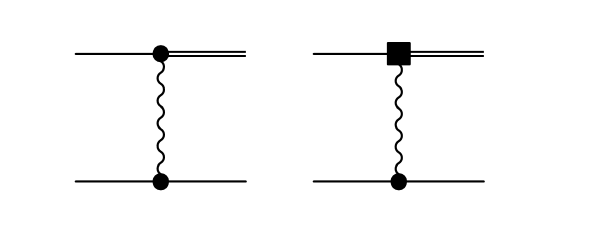}
\caption{Contribution of $C_{Wt}$ and $C_{Wb}$ to $t$-channel single top production. Solid lines denote light quarks, double lines the top quark, and wavy lines the $W$ boson. SM vertices are denoted by a dot, while an insertion of a $C_{Wt}$ or $C_{Wb}$ by a square.}\label{SingleTop}
\end{figure}

The weak dipole operators $C_{W t}$ and $C_{W b}$ provide tree-level corrections  
to single top production cross sections and to the $t\rightarrow Wb$ decay. 

The largest SM contribution to single top production is through the $t$-channel exchange of a $W$ boson. Smaller contributions arise from the associated production of a top and a $W$ boson and by $s$-channel $W$ exchange. 
 Production via the $t$-channel was first observed at the Tevatron \cite{Abazov:2009ii,Aaltonen:2010jr}. 
ATLAS published the measurement of the inclusive and differential  cross section  at $\sqrt{S} =$ 7 TeV with luminosity of 5 fb$^{-1}$  \cite{Aad:2014fwa}, and 
preliminary results at $\sqrt{S} = 8$ and $13$ TeV, with luminosity of, respectively,  20 fb$^{-1}$ and $3.2$ fb$^{-1}$ , are also available \cite{ATLAS-CONF-2014-007,ATLAS-CONF-2015-079}.
CMS published results at $\sqrt{S} = 7$ and $\sqrt{S} = 8$ TeV with luminosity of 1.56 and 20 fb$^{-1}$, respectively \cite{Chatrchyan:2012ep,Khachatryan:2014iya}.
Preliminary results at $\sqrt{S} = 13$ TeV are given in Ref. \cite{CMS-PAS-TOP-16-003}.
The associated production of a top and a $W$ boson has also been observed by both ATLAS and CMS \cite{Aad:2015eto,Chatrchyan:2014tua}, while the first observation of $s$-channel single top production
has been recently announced by the Tevatron experiments \cite{CDF:2014uma}.
In our analysis, we include only $t$-channel production, which gives the strongest bounds at the moment.

The total and differential SM cross sections are known at NLO in QCD  \cite{Bordes:1994ki,Stelzer:1997ns,Harris:2002md,Campbell:2009ss},
both in the 5 flavor scheme, in which the $b$ quark is considered massless and appears in the initial state, and in the 4 flavor scheme, which 
keeps into account $m_b$ effects. A detailed comparison of the two schemes is discussed in Ref. \cite{Campbell:2009ss}. 
We computed the corrections of the operators $C_{W t}$ and $C_{W b}$ to the $t$-channel single top cross section in the 5 flavor scheme,
including NLO QCD effects. The tree-level diagrams are displayed in Fig. \ref{SingleTop}. 
$c_{W t}$ interferes with the SM through terms proportional to $m_t$ and contributes to the total cross section at $\mathcal O(1/\Lambda^2)$. The interference of $c_{W b}$ with the SM is proportional to $m_b$ 
and vanishes in the 5 flavor scheme.
The imaginary couplings $\tilde{c}_{W t}$ and $\tilde{c}_{W b}$ only contribute to the total cross section at $\mathcal O(1/\Lambda^4)$. 

At NLO in QCD, the inclusive $t$-channel single top  cross section in the presence of the operator $C_{W t}$ is  
\begin{eqnarray}
\frac{\sigma_{t}( 7\; \textrm{TeV}) }{\rm pb} & = & 
(41.9 \pm 1.8) 
- \left( 9.4  \pm 0.3\right) \,  v^2 c_{Wt}    
+   ( 15.6 \pm 0.2 ) \,   (  (v^2 c_{W t}) ^2 + ( v^2\tilde{c}_{W t})^2)\ ,   \nonumber  \\ 
\frac{\sigma_{\bar t}( 7\; \textrm{TeV}) }{\rm pb} & = & 
(22.7 \pm 1.0) - \left( 0.3 \pm 0.1\right) v^2 c_{Wt}    + 
 (5.5 \pm 0.2 ) \, (  (v^2 c_{W t}) ^2 + ( v^2\tilde{c}_{W t})^2)   \ ,  \nonumber \\
\frac{\sigma_{t}( 8\; \textrm{TeV})  }{\rm pb} &=&
(56.4 \pm 2.4) - \left( 11.7 \pm 0.3 \right) \,  v^2 c_{Wt} +
 ( 21.0 \pm 0.5 ) \,  (  (v^2 c_{W t}) ^2 + ( v^2\tilde{c}_{W t})^2)     \ ,    \nonumber \\
\frac{\sigma_{\bar t}( 8\; \textrm{TeV})}{\rm pb} & = &  
(30.7 \pm 1.3)  - \left( 0.5 \pm 0.2\right)  \,  v^2 c_{Wt} + 
 (7.7 \pm 0.2) \, (  (v^2 c_{W t}) ^2 + ( v^2\tilde{c}_{W t})^2) \ ,
    \nonumber  \\
\frac{\sigma_{t}( 13\; \textrm{TeV})}{\rm pb}  &=&  (136{\color{white}.} \pm 5.4) - \left( 26.2 \pm  0.4 \right) \,  v^2 c_{Wt} + 
  (57.0 \pm 1.0 )  \, (  (v^2 c_{W t}) ^2 + ( v^2\tilde{c}_{W t})^2)     \ ,    \nonumber \\
\frac{\sigma_{\bar t}( 13\; \textrm{TeV}) }{\rm pb} & = & (81.0 \pm 4.1)
- \left( 2.6 \pm 0.4\right) \,  v^2 c_{Wt}  +  (24.7 \pm 1.0) \, (  (v^2 c_{W t}) ^2 + ( v^2\tilde{c}_{W t})^2)\ .      \nonumber  \\
\end{eqnarray}
The cross sections were obtained by setting the factorization and renormalization scales  to $\mu = m_t$. The couplings $c_{W t}$
and $\tilde{c}_{W t}$ are also evaluated at this scale and are related to the couplings at $\mu = \Lambda$ by the RGE discussed in Sec. \ref{sec:RGLHC}. The scale uncertainty was estimated by varying the factorization and renormalization scales between $\mu = m_t/2$ and $\mu = 2 m_t$. In the evaluation of the scale uncertainties,
we accounted for the running of $C_{W t}$ between the central scale and $\mu = m_t/2$ or  $\mu =2 m_t$.
We obtained the PDF and $\alpha_s$ uncertainties following the PDF4LHC recipe, using the three PDF sets CT10 \cite{Lai:2010vv}, MSTW08 \cite{Martin:2009iq} and NNPDF2.3 \cite{Ball:2012cx}. 
PDF uncertainties turn out to dominate the theoretical uncertainty. 
Corrections to the single top cross section from $C_{Wt}$ and $C_{Wb}$ have been considered  in Ref. \cite{Zhang:2010dr} at LO, and, recently, the NLO QCD corrections have been included \cite{Zhang:2016omx}.
Our results are in agreement with Refs.  \cite{Zhang:2010dr,Zhang:2016omx}.

In Table \ref{SingleTop1} we give the differential cross section $d\sigma/d p_T$, where $p_T$ is the $t$ or $\bar t$ transverse momentum,
in the $p_T$ bins relevant to the analysis of Ref.~\cite{Aad:2014fwa}. 
The different $p_T$ bins are correlated and we included the experimental correlations given in Ref.~\cite{Aad:2014fwa} when constructing the $\chi^2$. We neglected any correlations of theoretical uncertainties.

The contribution of $C_{W b}$ to single top production is suppressed by the factor of $m_b$ appearing in the definition of the operator. We find    	 	
\begin{eqnarray}
(\sigma_t + \sigma_{\bar t}) (7 \, \textrm{TeV}) &=&   (0.11  \pm 0.01)  \left(  (v^2 c_{W b}) ^2 + ( v^2\tilde{c}_{W b})^2 \right)  \, {\rm pb}  \ ,\nonumber \\
(\sigma_t + \sigma_{\bar t}) (8 \, \textrm{TeV}) &=&   (0.15  \pm 0.01) \left( (v^2 c_{W b}) ^2 + ( v^2\tilde{c}_{W b})^2   \right) \, {\rm pb}   \ ,\nonumber \\
(\sigma_t + \sigma_{\bar t}) (13 \, \textrm{TeV}) &=&   (0.44  \pm 0.01) \left(  (v^2 c_{W b}) ^2 + ( v^2\tilde{c}_{W b})^2 \right) \, {\rm pb}\  .  \label{STcwb}
\end{eqnarray}
In the 5 flavor scheme, $C_{W b}$ does not interfere with the SM, since the $b$ quark in the initial state is taken to be massless. We do not expect the interference to be significantly larger than the terms in Eq. \eqref{STcwb}.    
These results imply that the single top cross section is sensitive to $v^2 C_{W b} \sim \mathcal O(10)$. Such large values of $v^2 C_{W b}$ are forbidden by flavor processes such as $b \rightarrow s \gamma$. For this reason, we do not include Eq.~\eqref{STcwb} in our analysis.

\begin{table}
\center
\begin{tabular}{||c |c|l l l ||}
\hline
 &  \multicolumn{4}{c||}{$d\sigma(t)/d p_T$  ($10^{-3}$ pb/GeV)}    \\
\hline
bins (GeV)          &   exp. & \multicolumn{3}{c||}{theory}  \\
\hline
0-45         &   $440 \pm 65$           & $(373\pm 22)$ & $   -  (71  \pm 5)  \,v^2 c_{W t} $  & $+  (48\pm 4) \, (  (v^2 c_{W t})^2 + (v^2 \tilde{c}_{W t})^2 )   $        \\ 
45-75        &   $370 \pm 56$           & $(382\pm 15)$ & $   -  (85  \pm 1)  \,v^2 c_{W t} $  & $+  (96\pm 7)\, (  (v^2 c_{W t})^2 + (v^2 \tilde{c}_{W t})^2 )   $     	\\
75-110       &   $250 \pm 37$           & $(207\pm 8) $ & $   -   (52  \pm 5)  \,v^2 c_{W t} $  & $+  (87\pm 4) \, (  (v^2 c_{W t})^2 + (v^2 \tilde{c}_{W t})^2 )   $    	\\
110-150      &   $133 \pm 26$           & $(90\pm 3 ) $ & $   -   (26  \pm 4)  \,v^2 c_{W t} $  & $+  (66\pm 2) \, (  (v^2 c_{W t})^2 + (v^2 \tilde{c}_{W t})^2 )   $     	\\
150-500      &   $7.8 \pm 1.4$             & $(7.7\pm 0.4) $ & $ -   (2.2 \pm 0.5)\,v^2 c_{W t} $  & $+  (13\pm 1)  \, (  (v^2 c_{W t})^2 + (v^2 \tilde{c}_{W t})^2 )   $        \\
\hline
\hline
 &  \multicolumn{4}{c||}{$d\sigma(\bar t)/d p_T$  ($10^{-3}$ pb/GeV)}    \\
\hline
          bins (GeV)&   exp. & \multicolumn{3}{c||}{theory}  \\
\hline
0-45         &   $190 \pm 52$           & $(210\pm 14)  $ & $   -   (10.2    \pm 4.7) \, v^2  c_{W t} $ & $+  (14 \pm 2)  \, (  (v^2 c_{W t})^2 + (v^2 \tilde{c}_{W t})^2 )    $        \\ 
45-75        &   $230 \pm 43$           & $(202\pm 10)  $ & $   -   (4.7   \pm 1.3) \, v^2  c_{W t} $ & $+  (33 \pm 3)  \, (  (v^2 c_{W t})^2 + (v^2 \tilde{c}_{W t})^2 )    $     	\\
75-110       &   $97 \pm 26$            & $(102\pm 4)   $ & $   +   (0.7  \pm 4.7)  \, v^2  c_{W t} $ & $ +  (34 \pm 2) \, (  (v^2 c_{W t})^2 + (v^2 \tilde{c}_{W t})^2 )    $    	\\
110-150      &   $13 \pm 9$              & $(42 \pm 2)   $ & $   +   (2.7    \pm 1.3)  \, v^2  c_{W t} $ & $ +  (25 \pm 1) \, (  (v^2 c_{W t})^2 + (v^2 \tilde{c}_{W t})^2 )    $     	\\
150-500      &   $1.4 \pm 0.8$         & $(3.2\pm 0.2) $ & $   +   (0.4    \pm 0.2) \, v^2 c_{W t} $ & $+  (5.0 \pm 0.2)\, (  (v^2 c_{W t})^2 + (v^2 \tilde{c}_{W t})^2 )    $        \\
\hline

\end{tabular}
\caption{Single top differential cross section induced by $C_{Wt}$ at $\sqrt{S} = 7$ TeV. }\label{SingleTop1}
\end{table}

\subsubsection{$W$ helicity fractions}\label{HelFrac}

\begin{table}
\center
\begin{tabular}{||c|c|c|c|c c||}
\hline
\hline
$F_0$            	  & $F_L$  & $F_R$  & $\delta_-/\pi$    			& \multicolumn{2}{c||}{experiment} \\
\hline
$0.72 \pm 0.08$    &  $0.31 \pm 0.09$          & $-0.03 \pm 0.04$  & --	& CDF \& D0 &\cite{Aaltonen:2012rz}\\
$0.67 \pm 0.07 $   &  $0.32 \pm 0.04$          & $ 0.01 \pm 0.04$  & --	& ATLAS &\cite{Aad:2012ky} \\
$0.68 \pm 0.04$    &  $0.31 \pm 0.03$	       & $ 0.01 \pm 0.01$  & --	& CMS &\cite{Chatrchyan:2013jna} \\
$0.72 \pm 0.06 $   &  $0.30 \pm 0.04$	       & $-0.02 \pm 0.02 $ & --	& CMS &\cite{Khachatryan:2014vma} \\
	--	   &  $0.37 \pm 0.07$ - $F_R$  &	--	   & $ -0.014 \pm 0.036  $& ATLAS & \cite{Aad:2015yem} \\
\hline
\end{tabular}
\caption{$W$ helicity fractions measured at CDF, D0, ATLAS, and CMS. }\label{Tab:HelFrac}
\end{table}

The helicity fractions of $W$ bosons produced from top quark decays are sensitive to the operator $C_{W t}$ and, to a lesser extent, $C_{W b}$. We consider three helicity fractions: $F_0$, denoting the fraction of longitudinally-polarized W bosons, and $F_{L,R}$, denoting the fraction of left/right-handed transversely-polarized W bosons. 
These helicity fractions have been measured at the Tevatron and LHC \cite{Aaltonen:2012rz,Aad:2012ky,Chatrchyan:2013jna,Aad:2015yem,Khachatryan:2014vma} and in Table~\ref{Tab:HelFrac} we summarize the results used in our analysis. The experimental error is obtained by combining in quadrature the statistical and systematic errors reported by the experimental collaborations,
and in the $\chi^2$ function we consider correlations between $F_0$ and $F_{L,R}$.
In addition to the helicity fraction, the ATLAS collaboration has measured the phase, $\delta^-$,
between the amplitudes for the longitudinally-  and transversely-polarized $W$ bosons, recoiling against a left-handed $b$ quark, in the decay of a single top \cite{Boudreau:2013yna,Aad:2015yem}. This phase is sensitive to the imaginary parts of the dimension-six operators.

The SM helicity fractions have been computed at N${}^2$LO in QCD \cite{Czarnecki:2010gb}. They are a function of the ratio $x = m_W/m_t$, and, for the top pole mass, $m_t = 173$ GeV, and $m_W = 80.4$ GeV,  
the SM helicity fractions are $F_0 = 0.687$, $F_L = 0.311$, and $F_R = 0.0017$. The theoretical uncertainty is very small, at the permil level, and is negligible compared to the experimental error. 
The phase $\delta^-$ vanishes at tree level in the SM, but receives non-vanishing contributions from electroweak loops in which the internal $W$ boson and $b$ quark go on-shell   \cite{GonzalezSprinberg:2011kx}. However, these contributions are negligible with respect to the current experimental uncertainty. 

The corrections to the helicity fractions induced by the operators  $C_{W t}$ and $C_{W b}$ have been computed at NLO in QCD in Ref. \cite{Drobnak:2010ej}. 
Also in this case, the contribution of  $C_{W b}$ is suppressed by the bottom Yukawa coupling and gives bounds that are not competitive with those from flavor physics.
The LO correction of $C_{W t}$ to $F_0$ and $F_L$ is \cite{Drobnak:2010ej}
\begin{eqnarray} \label{hf1}
F_0 = \frac{ 1 - 4 y_t^2 x^2  (v^2 c_{W t}) + 4 x^4 y_t^4 \left(    (v^2 c_{W t})^2 + (v^2 \tilde{c}_{W t})^2   \right) }{ 
	    (1 + 2 x^2) - 12 y_t^2 x^2 (v^2 c_{W t})  + 4 x^2 (2 + x^2) y_t^4  \left(    (v^2 c_{W t})^2 + (v^2 \tilde{c}_{W t})^2   \right)  } \\
F_L = \frac{ 2 x^2 \left( 1 - 4 y_t^2   (v^2 c_{W t}) + 4  y_t^4 \left(    (v^2 c_{W t})^2 + (v^2 \tilde{c}_{W t})^2   \right) \right)}{ 
(1 + 2 x^2) - 12 y_t^2 x^2 (v^2 c_{W t})  + 4 x^2 (2 + x^2) y_t^4 \left(    (v^2 c_{W t})^2 + (v^2 \tilde{c}_{W t})^2   \right) }
\label{hf2}.
\end{eqnarray}
At tree level, the SM and $C_{Wt}$ contributions to  $F_R$ vanish. They are not zero at one loop \cite{Drobnak:2010ej}. In our analysis, we used the NLO expressions of Ref.  \cite{Drobnak:2010ej}.
Similar to the single top cross section, $c_{W t}$ interferes with the SM and gives rise to a genuine dimension-six effect, while the imaginary part of $C_{W t}$ corrects the helicity fractions at $\mathcal O(1/\Lambda^4)$.

The phase $\delta_-$ is linear in $\tilde c_{W t}$ and, at tree level, is given by \cite{Boudreau:2013yna,Aad:2015yem}
\begin{equation}\label{hf4}
\delta_- = V_{tb}^2 \; \textrm{arg} \left( (x - g_R)( 1 - x g_R)^* \right), \qquad \textrm{with} \quad g_R = 2 \frac{m_W}{v} y_t (v\sq c_{W t} + i\, v\sq \tilde{c}_{W t}).
\end{equation}

\subsection{Indirect constraints}

\subsubsection{Higgs production and decay}
\label{subsect:higgs-prod-decay}

The Higgs production cross section and branching ratios are sensitive probes of couplings of the top quark
to the Higgs boson, gluon, and photons. We already discussed the associated production of a Higgs boson and a $t\bar t$ pair, which provides a direct probe of the Higgs-top Yukawa coupling and the top chromo-dipole operator $C_g$.
In the SM the dominant Higgs production mechanism is gluon fusion and proceeds via a top loop. Similarly, the Higgs boson decay into two photons, the Higgs discovery channel, gets a sizable contribution from a top loop.
We can thus expect Higgs production and decay  processes to be very sensitive to anomalous top couplings, in particular  to the modification of the top Yukawa, $C_Y$, and to the top electromagnetic and color dipoles $C_\gamma$ and $C_g$. 

Through mixing onto $C_{\varphi W}$ and $C_{\varphi W B}$, the operators $C_{W t}$ and $C_{W b}$ also affect 
important  Higgs production and decay mechanisms such as vector boson fusion (VBF), associated production of a Higgs boson and a $W$ or $Z$ boson (WH and ZH), and the 
$W W^*$ and $Z Z^*$ decay channels. However, in this case the contribution of the dimension-six operators is suppressed by one electroweak loop with respect to the SM contribution, which arises at tree level, such that the resulting bounds turn out to be negligible.

The observables we consider are the Higgs signal strengths, which are observed to be compatible with the SM \cite{Khachatryan:2014jba,Aad:2015gba}. 
For a given Higgs production mechanism, $i \rightarrow h$, followed by the decay of the Higgs to the final state $f$,  
the signal strength is defined as 
\begin{equation}\label{signal}
\mu_{i \rightarrow h \rightarrow f}  =    \frac{ \sigma^{}_{i \rightarrow h} }{\sigma^{SM}_{i \rightarrow h}}     \frac{\Gamma^{}_{h \rightarrow f}}{\Gamma^{SM}_{h \rightarrow f} } \biggm/  \frac{\Gamma^{}_{\textrm{tot}}}{\Gamma^{SM}_{\textrm{tot}}
},
\end{equation}
where  $\sigma^{}_{i \rightarrow h}$ and $\sigma^{SM}_{i \rightarrow h}$ are, respectively, the production cross section in the channel $i$ including the effects of dimension-six operators, 
and the production cross section in the SM. $\Gamma^{}_{h \rightarrow f}$ and $\Gamma^{SM}_{h \rightarrow f}$ are the decay widths in the channel $f$, and $\Gamma^{}_{\textrm{tot}}$ and $\Gamma^{SM}_{\textrm{tot}}$ the Higgs total width, 
with and without the inclusion of dimension-six operators. For $m_h = 125$ GeV, the SM Higgs total width 
is $\Gamma^{SM}_{\textrm{tot}} = 4.07 \pm 0.16$ MeV \cite{Heinemeyer:2013tqa}.

The only production channel which is significantly affected by the operators we consider is gluon fusion.
The gluon fusion cross section can be computed in terms of the effective operators 
$\mathcal O_{\varphi G}$ and  $\mathcal O_{\varphi \tilde{G}}$ and is known at  N${}^2$LO in $\alpha_s$  \cite{Harlander:2002wh,Anastasiou:2002yz,Ravindran:2003um,Anastasiou:2002wq,Harlander:2002vv}.
The scalar and pseudoscalar top Yukawa couplings $c_Y$ and $\tilde{c}_Y$ induce threshold corrections to $C_{\varphi G}$ and $C_{\varphi \tilde{G}}$ at the scale $m_t$,
while the top chromo-dipole moments mix onto $\mathcal O_{\varphi G}$ and  $\mathcal O_{\varphi \tilde{G}}$, with the anomalous dimension \eqref{eq:HiggsGaugeRGE}.
As discussed in Ref. \cite{Chien:2015xha}, the NLO and N${}^2$LO corrections, and the theoretical uncertainties, mostly cancel 
in the ratio of the production cross section induced by $\mathcal O_{\varphi G}$ and  $\mathcal O_{\varphi \tilde{G}}$ and the SM. We therefore use the tree-level expression, which, in the limit of $m_t \rightarrow \infty$, is given by
\begin{equation}\label{ggF}
\frac{\sigma_{ggF}}{\sigma^{SM}_{ggF}} =  \bigg(1 + v^2 c_Y(m_t^+) + 48 \pi^2 v^2  C_{\varphi G}(m_t^+)  \bigg)^2 +
 \bigg(48 \pi^2 v^2 C_{\varphi \tilde{G}}(m_t^+) + \frac{3}{2} v^2 \tilde c_Y(m_t^+)  \bigg)^2\ ,
\end{equation}
and neglect the small theoretical uncertainties. $C_{\varphi G}(m_t^+)$ is given in Table~\ref{tab:mtLHC} and in Eq.~\eqref{ggF} we explicitly show the threshold corrections induced by the anomalous scalar and pseudoscalar Yukawa couplings.

The decay channels most affected by the operators under consideration are  $h \rightarrow g g$ and $h \rightarrow \gamma \gamma$.
The gluonic decay of the Higgs is not directly observed at the LHC because of the large QCD background. However, it affects  the signal strength by modifying the Higgs total width. 
In the infinite top mass limit, the Higgs decay width into gluons is  \cite{Spira:1997dg},   
\begin{equation}
\Gamma_{h \rightarrow g g} =  \frac{\alpha_s^2 m_h^3}{72 v^2 \pi^3} \left(   \bigg(1 + v^2 c_Y + 48 \pi^2 v^2 C_{\varphi G} \bigg)^2 + 
\bigg(48 \pi^2 v^2 C_{\varphi \tilde{G}} 
+ \frac{3}{2} v^2 \tilde{c}_Y  \bigg)^2    \right) ,
\end{equation}
where the couplings are evaluated at $m_t^+$.

The Higgs decay into photons in the infinite top mass limit becomes \cite{Spira:1997dg}
\begin{eqnarray}
\frac{\Gamma_{h \rightarrow \gamma \gamma}}{\Gamma^{SM}_{h \rightarrow \gamma \gamma}}  &=&  \frac{  \left( N_c Q_t^2 (1 + v^2 c_Y) + 24 \pi^2 (v^2 C_{\varphi \gamma}) - \frac{21}{4} A(\tau_W)  \right)^2 + \left( \frac{3}{2} N_c Q^2_t \, v^2 \tilde{c}_Y  + 24 \pi^2 (v^2 C_{\varphi \tilde{\gamma}})\right)^2  }{ \left( N_c Q_t^2  - \frac{21}{4} A(\tau_W)  \right)^2 }\ , \nonumber \\
\end{eqnarray}
where the couplings are evaluated at $m_t^+$, $\tau_W = 4 m_W^2/m_h^2$, and the loop function $A(\tau)$ is given in Eq.~\eqref{eq:Afunction}.
The $\gamma \gamma$ channel gives negligible corrections to the total width. 
NLO QCD corrections to the SM decay widths $\Gamma^{SM}_{h \rightarrow \gamma \gamma}$ and $\Gamma^{SM}_{h \rightarrow gg}$ are known \cite{Spira:1997dg}, and we included them in our analysis.

\begin{table}
\center
\begin{tabular}{||c|ccccc||}
\hline \hline
ATLAS              & $g g$ & VBF  & WH & ZH & $t\bar t h$ \\ 
\hline
$\gamma \gamma$   &  $1.3 \pm 0.4$ &  $0.8 \pm 0.7$ & $1.0 \pm 1.6$ & $0.1^{+3.7}_{-0.1}$ &  $1.3^{+2.6}_{-1.7}$ \\ 
$Z Z^*$ 	  &  $1.7 \pm 0.5$ &  $\left(0.3^{+1.6}_{-0.9}\right)^\dagger$  &  --           &  --           &   --    \\
$W W^*$ 	  &  $1.0 \pm 0.3$ &  $1.3 \pm 0.5$ & $3.0 \pm 1.6^*$ & --            &  $2.1\pm 1.4$   \\
$\tau \tau$       &  $2.0 \pm 1.5$ &  $1.2 \pm 0.6^\dagger$ & --            & --            & --  \\
$b\bar b$         &   --           &   --           & $1.1 \pm 0.6$ & $0.05 \pm 0.5$ & $1.5 \pm 1.1$ \\
\hline \hline
CMS              & $g g$ & VBF  & WH & ZH & $t\bar t h$ \\ 
\hline 

$\gamma \gamma$   &  $1.1 \pm 0.4$ &  $1.2 \pm 0.6^\dagger$ & -- & -- &  $2.7 \pm 2.6$  \\ 
$Z Z^*$ 	  &  $0.8 \pm 0.5$ &  $1.7 \pm 2.2^\dagger$  &  --           &  --           &   --    \\
$W W^*$ 	  &  $0.7 \pm 0.2$ &  $0.6 \pm 0.4$ & $0.4 \pm 2.0^*$ & --            &  $4.0 \pm 1.8$   \\
$\tau \tau$       &  $0.3 \pm 0.9$ &  $0.9 \pm 0.4^\dagger$ & --            & --            & --  \\
$b\bar b$         &   --           &  -- & $0.8 \pm 0.4^*$           & --   & $0.7 \pm 1.9$ \\
\hline \hline
\end{tabular}
\caption{Higgs signal strength in various production and decay channels, as measured by the ATLAS and CMS collaborations \cite{Aad:2015gba,Khachatryan:2014jba}. 
$\dagger$ denotes entries for which the signal strengths  is given in the combined bosonic production modes (VBF + WH + ZH),
$*$ denotes entries for which the combined WH + ZH signal strength is given.    }\label{HiggsSS}
\end{table}

The signal strengths as measured by the ATLAS and CMS collaborations \cite{Aad:2015gba,Khachatryan:2014jba} are given in Table \ref{HiggsSS}.
In our fits, we symmetrized the uncertainties in Table \ref{HiggsSS}, and treated them as statistical errors.

\section{Electroweak precision tests}
\label{sect:ew-precision}
The corrections from BSM physics to the self energies of the SM gauge bosons
can be described by three parameters (up to terms linear $q\sq$) \cite{Peskin:1990zt,Peskin:1991sw,Barbieri:2004qk},
\bea
\frac{\al_{ em}}{4s_W\sq c_W\sq}S &= &\Pi_{ZZ}'(0)-\frac{c_W\sq-s_W\sq}{c_Ws_W}\Pi_{\g Z}'(0)-\Pi_{\g\g}'(0),\nn\\
\al_{ em} T &=& \frac{\Pi_{WW}(0)}{m_W\sq}-\frac{\Pi_{ZZ}(0)}{m_Z\sq},\nn\\
\frac{\al_{ em}}{4s_W\sq}U &=& \Pi_{WW}'(0)-c_W\sq\Pi_{ZZ}'(0)-2s_Wc_W \Pi_{\g Z}'(0)-s_W\sq\Pi_{\g\g}'(0),
\eea
where $\Pi_{XY}$ denotes the self energy of the vector bosons $X$ and $Y$, while the primes indicate differentiation with respect to $q\sq$. Two of these parameters, $S$ and $T$, are generated by dimension-six operators, namely $O_{\vp WB}$ and $O_{HD}=\big|\vp^\dagger D_\mu \varphi\big|\sq$ \cite{Grinstein:1991cd,Barbieri:2004qk,Alonso:2013hga},
\begin{equation}
 S = 16\pi v^2  C_{\vp WB}(m_t),\qquad T = -\frac{2\pi}{e\sq} v\sq C_{HD}(m_t).
\end{equation}
In contrast, the $U$ parameter receives its first contribution at the dimension-eight level. As a result, when considering leading-log effects, the top-Higgs couplings only generate the $S$ parameter, and do not contribute to $C_{HD}$ or the dimension-eight operator responsible for $U$. The contributions to $S$ arise through the RG mixing with $O_{\vp WB}$ described in section \ref{sec:RGLHC}, which coincides with the divergent parts of the loop contributions discussed in \cite{Zhang:2010dr,Greiner:2011tt}. 

To derive the constraints resulting from the $S$ parameter, we expres $C_{\vp WB}$ in terms of the Higgs-top couplings (Table \ref{tab:mtLHC}) and employ the result of a fit to experimental data, $S=-0.03\pm 0.10$ \cite{Agashe:2014kda}~\footnote{This constraint results from a fit in which $T$ and $U$ are allowed to vary. Since we do not consider contributions to these parameters, we could take $T=U=0$ to be a prediction of our scenario and force $U$ and $T$ to zero in the fit. This would lead to a more stringent constraint on $S$, with the SM point, $S=0$, excluded at $90\%$. We therefore prefer to allow $T$ and $U$ to vary and employ the more conservative constraint  $S=-0.03\pm 0.10$.}~\footnote{It should be noted that the extraction of $S$ could be complicated in case there are additional interactions that significantly modify the fermion-$Z$ vertex \cite{Elias-Miro:2013gya,Elias-Miro:2013mua}. Such deviations from the SM could be induced by operators of the form $(\bar f_{L,R} \g^\mu f_{L,R})(i\vp^\dagger (\overleftarrow D_\mu -\overrightarrow D_\mu)\vp)$, however, since these 
interactions only receive mixing contributions from the top-Higgs interactions at the three-loop level, we neglect these effects here.}.

\section{Flavor physics}\label{sect:flav}
Flavor physics offers a large number of processes that can, in principle,  receive contributions from the anomalous top interactions. However, we find that most of these observables only give rise to fairly weak constraints. We briefly discuss here flavor observables that are not very sensitive to these couplings, after which we turn to the $b\to s \g$ transitions which do lead to significant limits.

Starting in the $B$ meson sector, the coupling $c_{Wt}$ induces flavor-changing four-quark operators at one loop which can contribute to $\bar B_{d,s}- B_{d,s}$ mixing \cite{Drobnak:2011wj}. Comparing the resulting limits \cite{Drobnak:2011wj} to those coming from the rare $B$ decays discussed below, we find that the $B$ mixing constraints are weaker and do not affect our bounds significantly. We therefore neglect the $\bar B_{d,s}- B_{d,s}$ mixing observables in what follows.

In the kaon sector, there are potential constraints from similar observables as in the $B$-meson sector, although the uncertainties from long-distance effects are generally larger. Here, $c_{Wt}$ induces a flavor-changing four-quark operator, analogous to the $B$-mixing operator, which contributes to CP violation in kaon mixing, $\epsilon_K$. In addition, the top-Higgs couplings induce flavor-changing dipole operators. A gluonic dipole ($s\to d g$) is mainly generated by $C_g$ and affects direct CP violation, $\epsilon'$, while $C_{\g,Wt}$ contribute to the photonic dipole operator ($s\to d \g$) which induces rare kaon decays such as $K_L\to \pi^+\pi^-\g$ \cite{Mertens:2011ts}. 
We employ the expressions in Refs.\ \cite{Bertolini:2014sua}, \cite{Lenz:2010gu}, and \cite{Mertens:2011ts}, to estimate the constraints from $\epsilon_K,\, \epsilon'$, and rare decays, respectively. 
We find an $\Or(1)$ constraint on $v\sq c_{Wt}$ in the case of $\epsilon_K$, while the remaining observables give rise to weaker bounds. Thus, none of the above flavor observables leads to competitive constraints, and we focus on the limits coming from $b\to s\g$ transitions to be discussed below.

\subsection{Rare B decays}
\label{sect:rareB}
To study the effects of the top-Higgs interactions on the $b\to s\g$ observables, namely the branching ratio and  CP asymmetry, we closely follow the procedure outlined in Refs.\ \cite{Altmannshofer:2012az,Altmannshofer:2011gn}. 
The branching ratio and CP asymmetry are mainly affected by the $C_{\g,g}^{(bs)}$ interactions induced by the top-Higgs operators. These couplings are related to the couplings $C_{7,8}$ that are more commonly used in the literature \cite{Altmannshofer:2012az,Altmannshofer:2011gn,Misiak:2006zs,Lunghi:2006hc,Kamenik:2011dk,Benzke:2010tq}, by the relations
\bea
C_\g^{(bs)} = \frac{V_{tb}V_{ts}^*}{4\pi\sq Q_b } \frac{C_7}{v\sq},\qquad C_g^{(bs)} =- \frac{V_{tb}V_{ts}^*}{4\pi\sq } \frac{C_8}{v\sq}\ .
\eea
The dimension-six contributions to these operators can be read off from Table~\ref{tab:mtFlavor}.

\subsubsection{$\text{BR}\,(B\to s\g)$}\label{sec:BRFlavor}
We describe the branching ratio by the expression derived in Refs.~\cite{Kagan:1998ym,Lunghi:2006hc}, rescaled to the SM prediction of \cite{Misiak:2006zs,Misiak:2015xwa,Czakon:2015exa},
\begin{equation}
\text{BR}\,(B\to s\g) = 10^{-4}\frac{3.36}{2.98}\bigg[2.98 +4.743 |C_7^{\rm NP}|\sq + 0.789 |C_8^{\rm NP}|\sq
\qquad \qquad \qquad  \qquad \qquad \qquad 
\nonumber 
\end{equation}
\begin{equation}
\qquad +\text{Re}\, \bigg((-7.184+0.612 i)C_7^{\rm NP}+(-2.225-0.557 i)C_8^{\rm NP}+(2.454-0.884 i)C_7^{\rm NP}C_8^{\rm NP *}\bigg)\bigg] ,
\end{equation}
where $C_{7,8}^{\rm NP}$ stand for the non-standard contributions to $C_{7,8}$, which are to be evaluated at the top scale, $\mu = 160\, \text{GeV}$. This expression should be compared with the current experimental world average  \cite{Agashe:2014kda}, 
\bea\text{BR}\,(B\to s\g)=(3.43\pm 0.21\pm0.07)\times 10^{-4}.
\eea 
In order to derive constraints we  follow  Refs.\ \cite{Altmannshofer:2012az,Altmannshofer:2011gn} and use the relative uncertainty on the SM prediction as our theoretical error, $\sigma = \frac{0.23}{3.36}{\rm BR}(b\to s\g) $. This theoretical uncertainty is then added in quadrature to the experimental one.
\subsubsection{$A_{CP}(B\to s\g)$}
For the CP asymmetry we follow Refs.~\cite{Altmannshofer:2012az,Altmannshofer:2011gn,Kamenik:2011dk} and employ the expression derived in Ref.~\cite{Benzke:2010tq},
\bea
\frac{A_{CP}(B\to s\g)}{\pi}&\equiv & \frac{1}{\pi}\frac{\Gamma(\bar B\to X_s\g)-\Gamma(B\to X_{\bar s}\g)}{\Gamma(\bar B\to X_s\g)+\Gamma(B\to X_{\bar s}\g)} \nn\\&\approx & \bigg[\bigg(\frac{40}{81}-\frac{40}{9}\frac{\Lambda_c}{m_b}\bigg)\frac{\al_s}{\pi}+\frac{\Lambda_{17}^c}{m_b}\bigg]\text{Im}\,\frac{C_2}{C_7}-\bigg(\frac{4\al_s}{9\pi}+4\pi\al_s\frac{\Lambda_{78}}{3m_b}\bigg)\text{Im}\,\frac{C_8}{C_7}\nn\\
&&-\bigg(\frac{\Lambda_{17}^u-\Lambda_{17}^c}{m_b}+\frac{40}{9}\frac{\Lambda_c}{m_b}\frac{\al_s}{\pi}\bigg)\text{Im}\,\bigg(\frac{V_{ub}V_{us}^*}{V_{tb}V_{ts}^*}\frac{C_2}{C_7}\bigg)\ ,
\eea 
where $C_2$ is a four-quark operator, $\sim (\bar d_L\g^\mu b_L)(\bar s_L\g_\mu d_L)$ \cite{Buras:1998raa}, which, along with $C_{7,8}$, should to be evaluated at the factorization scale $\mu_b\simeq 2$ GeV. We employ the following SM values for these coefficients at $\mu_b=2\, \text{GeV}$  \cite{Benzke:2010tq},
\bea
C_2^{\rm SM} = 1.204,\qquad C_7^{\rm SM} = -0.381,\qquad C_8^{\rm SM} = -0.175.\label{BcoeffSM}
\eea
The contributions from the top-Higgs couplings at $\mu=m_t$  and $\mu=2$ GeV can again be read from Table \ref{tab:mtFlavor}. 
In addition, the CP asymmetry depends on the scale, $\Lambda_c\simeq 0.38\, \text{GeV}$,  and on three hadronic parameters that are estimated to lie in the following ranges \cite{Benzke:2010tq},
\bea
\Lambda_{17}^u\in [-0.33,\, 0.525]\, \text{GeV},\qquad \Lambda_{17}^c\in [-0.009,\, 0.011]\, \text{GeV},\qquad \Lambda_{78}\in [-0.017,\, 0.19]\, \text{GeV}.\label{lambdas}
\eea
We deal with these rather large uncertainties by using the R-fit procedure \cite{Charles:2004jd}; we  vary the $\Lambda$ parameters in their allowed ranges, selecting the values which produce the smallest $\chi\sq$. The final ingredient we require is the current experimental value of the CP asymmetry given by \cite{Amhis:2014hma}
\bea
A_{CP}(B\to s\g) = 0.015\pm 0.02.
\eea

\section{Electric dipole moments}
\label{sect:EDMs}

\begin{table}[t]
\begin{center}\small
\begin{tabular}{||c||ccc|ccc||}
\hline
&$d_e$ & $d_n$& $\dHg$ &$d_{p,D}$  & $\dXe$ & $\dRa$\\
\hline
\rule{0pt}{3ex}
current limit &$8.7 \cdot 10^{-29} $ &$ 3.0 \cdot 10^{-26}$  & $6.2 \cdot 10^{-30}$  & x  & $5.5 \cdot 10^{-27}$  , & $4.2\cdot 10^{-22}$  \\
expected limit &$5.0 \cdot 10^{-30}  $  &$ 1.0 \cdot 10^{-28} $ &$6.2 \cdot 10^{-30}$&$ 1.0 \cdot 10^{-29}$  & $5.0 \cdot 10^{-29}$ & $1.0 \cdot 10^{-27}$\\
\hline
\end{tabular}
\end{center}
\caption{\small Current limits on the electron\cite{Baron:2013eja}, neutron \cite{Baker:2006ts,Afach:2015sja}, and mercury \cite{Griffith:2009zz,Graner:2016ses} EDMs in units of $e$ cm ($90\%$ confidence level). We also show an indication of their prospective limits  \cite{Kumar:2013qya,Chupp:2014gka} as well as those of the proton, deuteron, xenon \cite{PhysRevLett.86.22}, and radium \cite{Parker:2015yka} EDMs, which could provide interesting constraints in the future.}
\label{tab:EDMexps}  
\end{table}

Permanent EDMs of leptons, nucleons, nuclei, atoms, and molecules probe flavor-diagonal CP violation with essentially no SM background. CP violation from the CKM mechanism predicts EDMs that are orders of magnitude below current experimental sensitivities. The only SM background then arises from the QCD vacuum angle, the so-called theta term, which, in principle, induces large EDMs of hadrons and nuclei. The absence of an experimental signal for the neutron and ${}^{199}$Hg EDMs leads to the strong constraint $\theta < 10^{-10}$ \cite{Baluni:1978rf}. This smallness begs for an explanation that can be provided by the Peccei-Quinn mechanism~\cite{Peccei:1977hh}, 
which dynamically relaxes the vacuum angle to zero at the cost of a, so far, unmeasured axion. In this work, we assume the Peccei-Quinn mechanism to be at work such that the bare theta term is removed from our EFT. However, in the presence of dimension-six sources of CP violation, the Peccei-Quinn mechanism does not completely remove the theta term. Instead, the vacuum angle is relaxed to a finite value proportional to the coefficients of the dimension-six CPV operators. The contribution from the induced vacuum angle is taken into account in our analysis by the value of the hadronic matrix elements~\cite{Pospelov_review}.
Recent developments in lattice QCD \cite{Guo:2015tla, Bhattacharya:2015esa, Shindler:2015aqa, Alexandrou:2015spa, Shintani:2015vsx} and chiral effective field theory \cite{deVries:2012ab, Bsaisou:2014oka}
have improved the description of hadronic and nuclear EDMs. If future experiments detect nonzero EDMs, their precise pattern could potentially disentangle a nonzero theta term from BSM sources of CP violation \cite{Dekens:2014jka}. For now, however, we apply the Peccei-Quinn mechanism to essentially remove the theta term from our analysis.

At present, the most stringent  constraints come from  measurements of the neutron,  ${}^{199}$Hg atom,  and ThO   molecule  EDMs. Here we give a brief overview of our analysis of these EDMs and refer to Ref.~\cite{Chien:2015xha} for more details. 

We begin with ThO measurement \cite{Baron:2013eja}, which, for the set of dimension-six operators under discussion, can be interpreted as a measurement of the electron EDM\footnote{
Apart from the electron EDM, the ThO EDM also receives contributions from semi-leptonic four-fermion interactions, which can be generated by the top-Higgs couplings at loop-level. However, these induced semi-leptonic interactions are always negligible due to suppression by small Yukawa couplings and/or CKM elements.}
\bea
d_e = e \,Q_e m_e \tilde c_\g^{(e)}(\MQCD)&\leq &8.7\cdot 10^{-29}\,e \,\mathrm{cm} \quad (90\%\, {\rm C.L.}) \,\,\, .
\eea
This rather clean theoretical interpretation in terms of  the electron EDM involves an estimated $\Or(15\%)$ uncertainty \cite{Skripnikov,Fleig:2014uaa}. As this error estimate only affects the bound by an overall factor (it does not allow for cancellations) and it is far below the uncertainties related to the hadronic/nuclear EDMs, we neglect it here.

The neutron and proton EDMs are plagued by much larger hadronic uncertainties. They can be expressed in terms of the operators of Eq.~\eqref{eq:ExtendedEDM1} via the relations
\bea
d_n&=&
-(0.22\pm0.03)\,e\,Q_u m_u \tilde c_\g^{(u)}+(0.74\pm0.07 )\,e\,Q_d m_d \tilde c_\g^{(d)}+(0.0077\pm0.01)\,e\,Q_s m_s \tilde c_\g^{(s)}\nn\\
&&-(0.55\pm0.28)\,e\, m_u \tilde c_g^{(u)}-(1.1\pm0.55)\,e\,m_d \tilde c_g^{(d)} \pm(50\pm40)e\,g_sC_{\tilde G},\nn\\
d_p&=&
(0.74\pm0.07)\,e\,Q_u m_u \tilde c_\g^{(u)}-(0.22\pm0.03 )\,e\,Q_d m_d \tilde c_\g^{(d)}+(0.0077\pm0.01)\,e\,Q_s m_s \tilde c_\g^{(s)}\nn\\
&&+(1.30\pm0.65)\,e\, m_u \tilde c_g^{(u)}+(0.60\pm0.30)\,e\,m_d \tilde c_g^{(d)} \mp(50\pm40)e\,g_sC_{\tilde G},
\eea
where all coefficients should be evaluated at $\mu=\MQCD$. Because of recent lattice calculations \cite{Bhattacharya:2015esa, Bhattacharya:2015wna}, the contributions from the up- and down-quark EDMs in this expression are known to $\Or(15\%)$, while the strange contribution is still highly uncertain. The up- and down-quark CEDM contributions have an estimated $50\%$ uncertainty based on QCD sum-rule calculations \cite{Pospelov_qCEDM, Pospelov_deuteron, Pospelov_review, Hisano1}, while the Weinberg operator appears with the largest uncertainty, $\Or(100\%)$, based on a combination of QCD sum-rules \cite{Pospelov_Weinberg} and naive dimensional analysis estimates \cite{Weinberg:1989dx}. 
The magnitude of the strange-quark CEDM contribution is currently unresolved and is often assumed to vanish in the Peccei-Quinn scenario. We do so here as well, but point out that this assumption might be unwarranted \cite{Hisano2}.

Finally, the ${}^{199}$Hg  EDM receives contributions from the nucleon EDMs as well as from the CP-odd isoscalar and isovector pion-nucleon couplings\footnote{A potential third contribution from a CP-odd isotensor pion-nucleon interaction is negligible for all operators in Eq.~\eqref{eq:ExtendedEDM1} \cite{deVries:2012ab}.}, $\bar g_0$ and $\bar g_1$ (here we use the conventions of Ref.\ \cite{Chien:2015xha}). The induced nucleon EDMs are given above, while the pion-nucleon couplings are generated by the quark CEDMs \cite{Pospelov_piN},
\bea
\bar g_0 = (5\pm 10)(m_u\tilde c^{(u)}_g + m_d\tilde c^{(d)}_g)\, \mathrm{fm}^{-1}\,\,\,,\qquad
\bar g_1 = (20^{+40}_{-10})(m_u\tilde c^{(u)}_g- m_d\tilde c^{(d)}_g)\,\, \mathrm{fm}^{-1}
\,\,\,.
\eea
Combining the contributions of the nucleon EDMs \cite{Dmitriev:2003sc} with those of the pion-nucleon couplings \cite{deJesus:2005nb,Ban:2010ea,Dzuba:2009kn,Engel:2013lsa}, then gives the following expression for the ${}^{199}$Hg EDM,
\bea
\dHg = -(2.8\pm 0.6)\Ex{-4}\bigg[(1.9\pm0.1)d_n +(0.20\pm 0.06)d_p+\bigg(0.13^{+0.5}_{-0.07}\,\bar g_0 + 0.25^{+0.89}_{-0.63}\,\bar g_1\bigg)e\, {\rm fm}\bigg],
\eea
where the small number in front of the main brackets is the Schiff screening factor. The large nuclear uncertainties appearing in the dependencies on $\bar g_{0,1}$ dilute the constraining power of $\dHg$. 

We summarize the current and prospective limits in Table \ref{tab:EDMexps}. The table also shows the limits on systems which are not yet competitive, but could provide interesting constraints in the future. EDM experiments on ${}^{225}$Ra and ${}^{129}$Xe atoms have already provided limits \cite{PhysRevLett.86.22, Parker:2015yka} and are quickly improving. We use the following expressions for these EDMs \cite{Engel:2013lsa}
\bea
d_{\mathrm{Xe}} &=& (0.33\pm 0.05)\Ex{-4}\left(-0.10^{+0.037}_{-0.53}\,\bar g_0 - 0.076^{+0.038}_{-0.55}\,\bar g_1\right)e\, {\rm fm}\ , \\
d_{\mathrm{Ra}} &=& -(7.7\pm 0.8)\Ex{-4}\left(-19^{+6.4}_{-57}\,\bar g_0 + 76^{+227}_{-25}\,\bar g_1\right)e\, {\rm fm}\, .
\eea
We point out that these expressions do not contain the dependencies on the single-nucleon EDMs as these have, as far as we know, not been calculated. The associated nuclear uncertainties are still significant but smaller than for $\dHg$. $d_{\mathrm{Ra}}$ has the additional benefit of a smaller screening factor and a large dependence on $\bar g_{0,1}$ due to the octopole deformation of the nucleus (see Ref.~\cite{Engel:2013lsa} and references therein). 

Plans exist to measure the EDMs of charged nuclei in electromagnetic storage rings \cite{Eversmann:2015jnk}. Here we consider the impact of a deuteron EDM measurement. Light nuclei have the advantage that the theoretical calculations can be performed accurately within a controlled power counting scheme \cite{deVries2011b,Bsaisou:2014zwa}. The deuteron EDM can be expressed as
\bea
d_D &=& (0.94\pm 0.01)(d_n + d_p) + (0.18 \pm 0.02) \bar g_1\, e \, {\rm fm}\,\ ,
\eea
which, as this is a measurement of a nuclear EDM, has no Schiff screening factor. EDMs of other light nuclei, such as ${}^3$He , ${}^6$Li , and ${}^{13}$C  have been investigated along similar lines \cite{Stetcu:2008vt, deVries2011b, Song:2012yh,Bsaisou:2014zwa,Yamanaka:2015qfa,Yamanaka:2016itb} but are not considered here.

\subsection{Lepton anomalous magnetic moments}
The same mechanisms that generate the electron dipole moment $\tilde{c}^{(e)}_{\gamma}$ also induce the  magnetic moments of charged leptons. As the magnetic moments of the electron \cite{Hanneke:2008tm} and muon \cite{Bennett:2006fi} are measured to very high accuracy and have precise SM predictions, we briefly discuss whether they lead to significant constraints on the top-Higgs couplings.
The magnetic moment is defined as 
\begin{equation}
\vec{M}_{l} =  \frac{e}{2 m_{l}} g_{l} \vec{S},
\end{equation}
where $l=(e,\,\mu,\, \tau)$, $\vec S$ is the spin of the charged lepton, and $g_{l} = 2$ at tree level in the SM.
Loop effects in the SM lead to corrections to $g_l$, thereby inducing  anomalous magnetic moments,
\begin{equation}
a_{l} = \frac{g_l - 2}{2}.
\end{equation}
Due to the small uncertainties in the measurement of $a_e$ \cite{Hanneke:2008tm}, it can be used to obtain the most precise value of the electromagnetic fine-structure constant \cite{Aoyama:2012wj}. To instead compare the measurement to  the SM value of $a_e$, the fine-structure constant has to be extracted from an independent experiment. Currently, the most precise determination (apart from $a_e$) comes from a measurement of the ratio of Planck's constant and the mass of the ${}^{87}$Rb atom \cite{Bouchendira:2010es}. Employing the obtained fine structure constant and comparing the SM predictions for $a_{e,\,\mu}$ with the experimental results gives  \cite{Agashe:2014kda}, 
\bea
\Delta a_e &=& a^\textrm{exp}_e - a^{\textrm{SM}}_e =   -1.05(0.82)\cdot 10^{-12}\, ,\nn\\
\Delta a_\mu &=& a^\textrm{exp}_\mu - a^{\textrm{SM}}_\mu = 2.88 (0.63) (0.49) \cdot 10^{-9}, \label{eq:g-2}
\eea
where $a_\mu$ is in some tension with the SM prediction while $a_e$ is consistent with the SM.

The real parts of the top couplings in Eq.\ \eqref{eq:Leff} induce corrections to the lepton magnetic moments in the same way as the imaginary parts induce the electron EDM. For $c_Y$ this occurs through Barr-Zee diagrams, Eq.\ \eqref{mtThreshold}, while for $c_{\g,\,g,\, Wt,\,Wb}$ the main contributions arise through the two-step mechanism explained in Section \ref{sec:RGEDMs}.
Extending Eq. \eqref{eq:ExtendedEDM1} to include the fermion magnetic dipole operators, we can parametrize the corrections to 
$a_{l}$ as
\begin{equation}
\Delta a_{l} = -  2  \frac{m^2_{l}}{v^2} Q_l  \, \left(v^2 c_\gamma^{(l)}(\Lambda_\chi) \right).
\end{equation}
Since the running of the real and imaginary part of the operators  $C_{\alpha}$ in Eq. \eqref{eq:Leff} onto the lepton magnetic and electric dipole operators   
is identical, the values of $c_\gamma^{(l)}(\Lambda_\chi)$ as a function of 
the top couplings at the scale $\Lambda = 1$ TeV can be read from the first line of Table   
\ref{tab:mtEDM}, giving
\bea
\Delta a_e &=&  \left( 3.3 \, (v^2  c_\gamma) + 0.12 \, (v^2  c_g) - 3.8\, (v^2  c_{W t}) + 0.35\, (v^2  c_{Y})     \right)  \cdot 10^{-15}\ , \nn\\
\Delta a_\mu &=&  \left( 13.8\, (v^2  c_\gamma) + 0.51 \, (v^2  c_g) - 16.0\, (v^2  c_{W t}) + 1.5\, (v^2  c_{Y})     \right)  \cdot 10^{-11},
\eea
where the contribution of $c_{Wb}$ is negligible.

Comparing with Eq.\ \eqref{eq:g-2} we see that the uncertainty on $\Delta a_e$ is  a factor $\Or(10^{3})$ smaller than on $\Delta a_\mu$, while the latter  is more sensitive to the top-Higgs couplings by a factor $m_\mu\sq/m_e\sq\sim 4\Ex{4}$. Despite this sensitivity, large values of the couplings, $v^2\, c_\alpha \sim \mathcal O(10) -\mathcal O(100)$, are needed to explain the observed tension with the SM. Furthermore, the determination of  $\Delta a_e$  leads to $\Or(100)$ constraints on $v^2 \, c_\al$. As we discuss in Section \ref{discussion}, such large values are already excluded by other direct and indirect observables. This implies that the $g_\mu-2$ anomaly cannot be due to the dimension-six operators we investigate, and $\Delta a_e$ does not give competitive constraints. 
We therefore do not include the electron and muon anomalous magnetic moments in our analysis.

\section{Analysis strategy}\label{analysis}

\subsection{The $\chi\sq$ functions}
To set constraints on the top-Higgs couplings using a given observable we construct a $\chi\sq$ in the usual way,
\bea
\chi_i\sq = \bigg(\frac{\Or_i^{\rm th}-\Or_i^{\rm exp}}{\sigma_i}\bigg)\sq .
\eea
Here $\Or_i^{\rm exp}$ stands for the experimentally measured value of the observable $i$, $\Or_i^{\rm th}$ is its theoretical expression, and $\sigma_i$ is the related experimental uncertainty.\footnote{There is one exception to this in the case of BR$(b\to s\g)$. As described in section \ref{sec:BRFlavor}, in this case we treat the theory error as statistical and add it to the experimental one in quadrature, i.e.\ $\sigma\sq = (\sigma^{\rm th})\sq+(\sigma^{\rm exp})\sq$.}
As there is a large number of observables to consider, we combine them in a number of ways. For the collider observables, we differentiate between direct and indirect constraints,
\bea
\chi_{\rm direct}\sq = \sum_{i=t,\,t\bar t, \,t\bar th,\, F_0,\,F_L,\, F_R,\, \delta_-}\chi_i\sq,\qquad \chi_{\rm indirect}\sq = \sum_{i,j}\chi_{i\to h(W,Z)\to j}\sq,
\eea
where the indirect constraints include all the Higgs production and decay channels, mentioned in Section \ref{sect:collider}. The direct constraints include $t$, $t\bar t$, and $t\bar t h$ production, as well as the $W$ helicity fractions, while the constraint from the electroweak precision tests are simply captured by $\chi\sq_{S}$.

For the rare B decays we combine the constraints from both observables into a single constraint, 
\bea\chi_{b\to s\g}\sq = \chi_{\text{BR}}\sq+\chi_{A_{CP}}\sq. \eea

Finally the EDM constraints are combined into a single $\chi$-squared as follows,
\bea
\chi\sq_{\rm EDMs} = \sum_{i = d_n,\, d_{\rm Hg},\, d_{\rm ThO},\, d_p,\, d_D,\, d_{\rm Ra},\, d_{\rm Xe}} \chi\sq_i,
\eea
where the final four observables are typically only relevant when considering future constraints. The combined $\chi\sq$, taking into account all observables, is then given by,
\bea
\chi\sq_{\rm Total} = \chi\sq_{\rm direct}+\chi\sq_{\rm indirect}+\chi\sq_{S}+\chi\sq_{b\to s\g}+\chi\sq_{\rm EDMs}.
\eea

\subsection{Theoretical uncertainties}\label{rfit}
Through $\Or_i^{\rm th}$ the above $\chi\sq$ functions depend on both the top-Higgs couplings (at the scale $\Lambda $), as well as parameters which have theoretical uncertainties. In the case of high-energy probes these `parameters' are the theory prediction for the SM and BSM contributions to cross sections and signal strengths, while for the $b\to s\g$ observables and EDMs the hadronic and nuclear matrix elements play the role of these parameters. We deal with these theoretical uncertainties in two different ways:
 
\begin{itemize}
\item \textbf{Central:} Here we neglect theoretical uncertainties in the hadronic and nuclear matrix elements entering $d_n$, $d_{\textrm{Hg}}$ and $A_{CP}$. Instead, for collider observables, where the uncertainties are  under  better control and generally smaller, we apply the R-fit procedure explained below.
\item \textbf{R-fit:}
Here we vary all theoretical uncertainties, appearing in $d_n$, $\dHg$, $A_{CP}$, and collider observables, within the allowed ranges assuming a flat distribution, and minimize the total $\chi\sq$. This method corresponds to the Range-fit (R-fit) procedure defined in Ref. \cite{Charles:2004jd}. It always gives the weakest (= most conservative) constraint   as it allows for cancellations between different contributions. 

\end{itemize}

\section{Discussion}\label{discussion}

\subsection{Single coupling analysis}\label{singlecoupling}

\begin{figure}[t!]
\center
\includegraphics[width=0.47\linewidth]{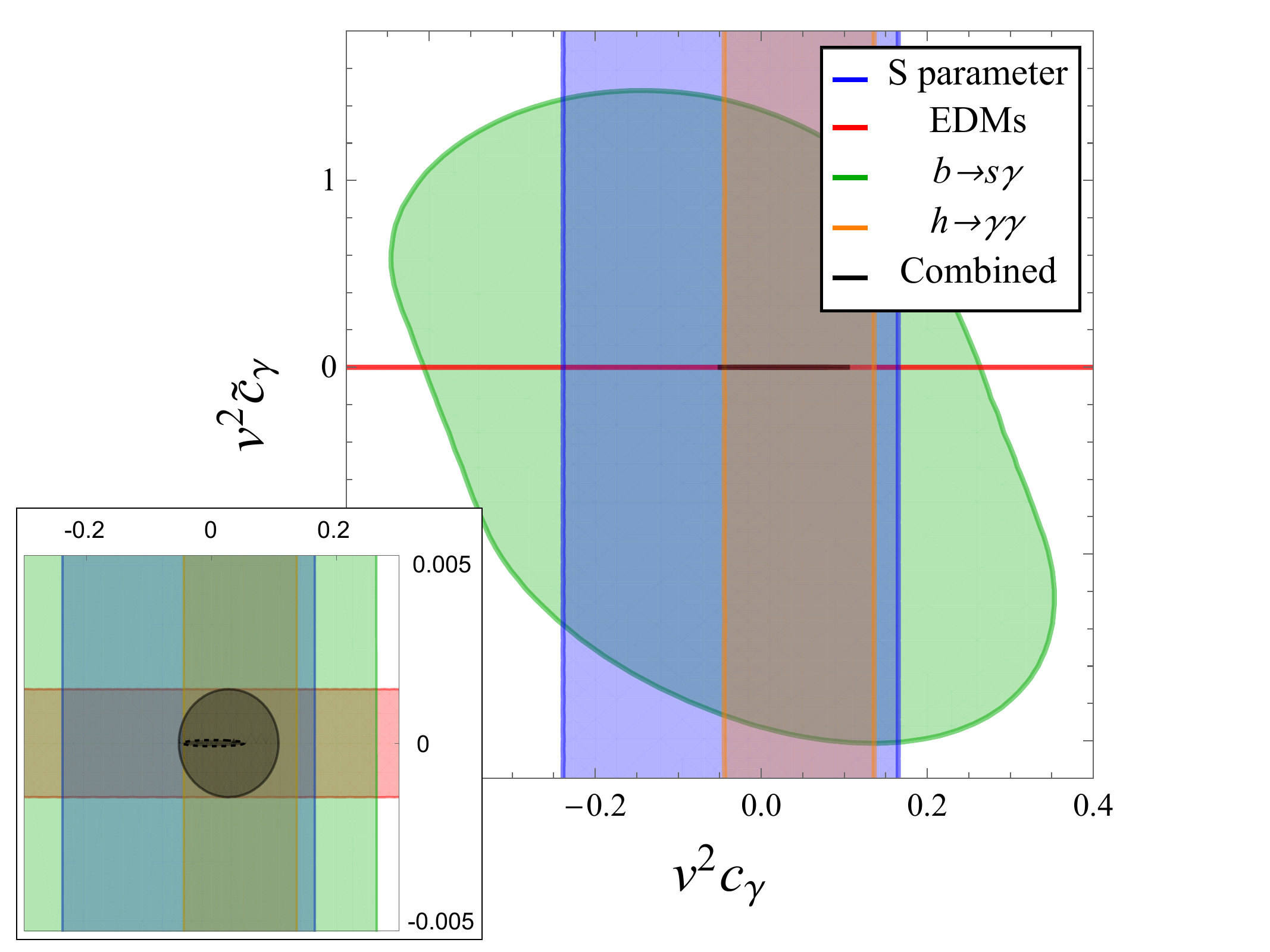}
\includegraphics[width=0.47\linewidth]{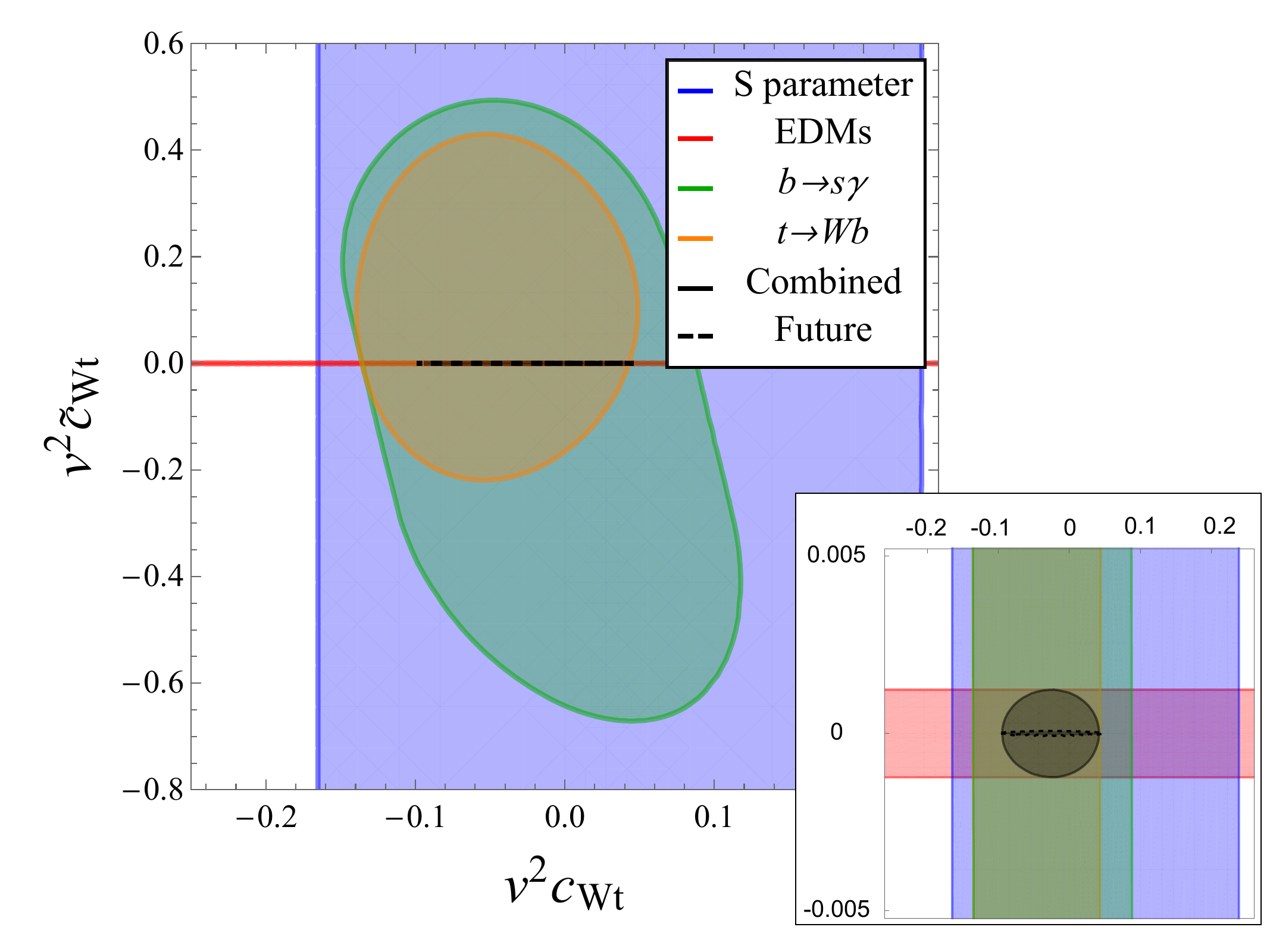}
\caption{90\% C.L.\  allowed regions in the   $v^2 c_{\gamma} - v^2  \tilde{c}_{\gamma}$   (left panel) 
and $v^2 c_{Wt} - v^2  \tilde{c}_{Wt}$  planes (right panel), with couplings evaluated at   $\Lambda = 1$~TeV.
In both cases, the inset  zooms into the current combined allowed region and shows projected future sensitivities. 
}\label{fig:Cgamma}
\end{figure}

\begin{figure}[t]
\center
\includegraphics[width=0.32\linewidth]{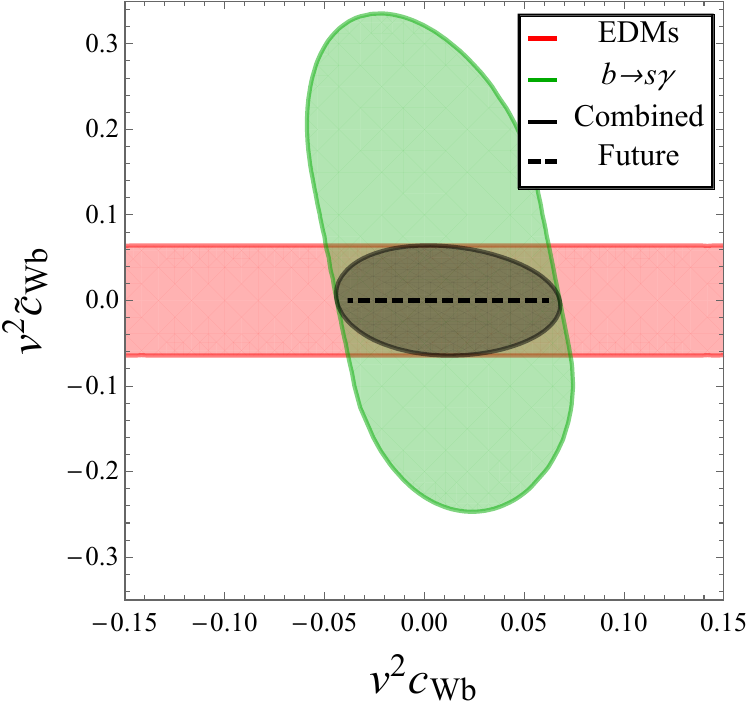}
\includegraphics[width=0.32\linewidth]{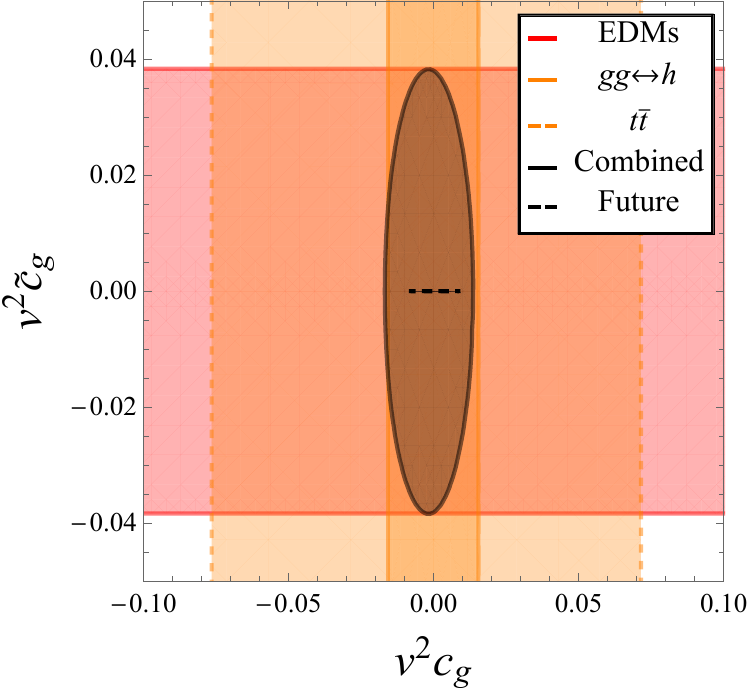}
\includegraphics[width=0.32\linewidth]{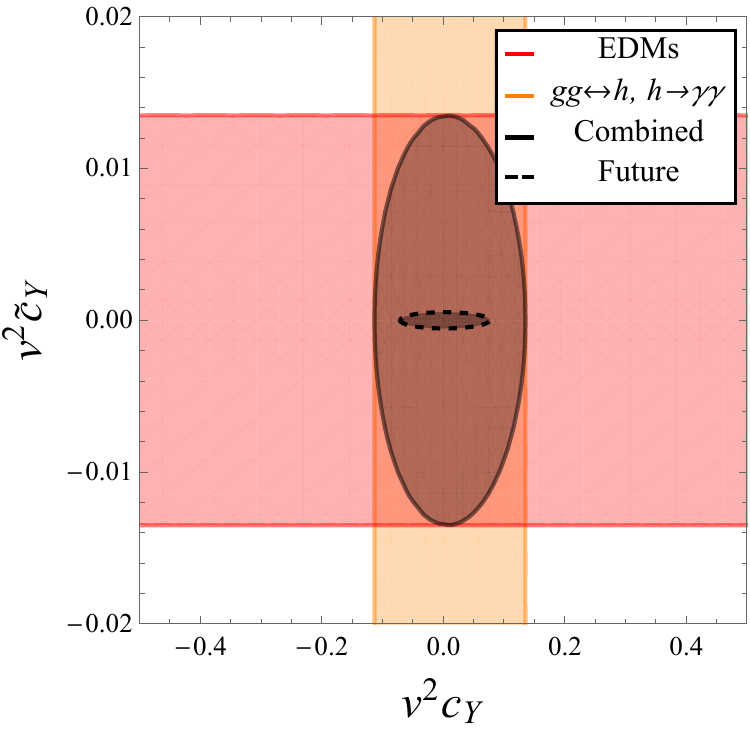}
\caption{90\% C.L.\  allowed regions in the   $v^2 c_{Wb} - v^2  \tilde{c}_{Wb}$   (left panel) 
and $v^2 c_{g} - v^2  \tilde{c}_{g}$  (center panel) 
and $v^2 c_{Y} - v^2  \tilde{c}_{Y}$  (right panel) planes,
with couplings evaluated at   $\Lambda = 1$~TeV.
}\label{fig:CWb}
\end{figure}

We first focus on the case in which a single operator dominates at the high scale.
In Figs. \ref{fig:Cgamma} and \ref{fig:CWb} we show the 90\% C.L. allowed region  for the complex couplings $C_{\gamma}$,
$C_{W t}$, $C_{W b}$, $C_g$ and $C_Y$, obtained under the assumption that only one coupling is  nonzero at a scale $\Lambda =$ 1 TeV (the bounds on the dimensionless parameters  do not significantly change if larger values of $\Lambda$ are applied). 
For each coupling we show the combined allowed region (black), 
and the  most constraining bounds coming from EDMs (red), flavor physics (green),  electroweak precision observables (blue), and  direct and indirect collider searches (orange).
Theoretical uncertainties play a large role for the constraints arising from EDMs and flavor observables. The plots in Fig. \ref{fig:Cgamma} and \ref{fig:CWb} are obtained with the R-fit procedure described in Section \ref{analysis} which particularly affects the constraints on the imaginary parts of the couplings \cite{Chien:2015xha}.

The dashed black contour denotes future sensitivities, considering improvements in EDM experiments~\cite{Kumar:2013qya},
super-B factory measurements~\cite{Bona:2007qt,Nishida:2011dh}, and collider searches~\cite{CMS:2013xfa,ATL-PHYS-PUB-2014-016}.
For the electron and neutron EDM we use the expected limits in Table \ref{tab:EDMexps}, which project  
an improvement on the electron and neutron EDMs by a factor of 10 and 100, respectively. 
We assume the bound on $d_{\textrm{Hg}}$ to stay at the current level.
Future super-B factory experiments can reduce the statistical error on the 
$b \rightarrow s \gamma$ branching ratio to about 3\%, and  
improve the error on $A_{CP}$ by a factor of 5 \cite{Bona:2007qt,Nishida:2011dh}. 
For the Higgs signal strengths, we use the projected uncertainties of Ref.~\cite{CMS:2013xfa,ATL-PHYS-PUB-2014-016} 
for the LHC Run 2, with $\sqrt{S} = 14$ TeV, and integrated luminosity of 300 fb$^{-1}$.
We assume a central value of $1$ (SM prediction) in every production and decay channel. The projected uncertainty on
the gluon fusion channel with decay in $\gamma \gamma$ and $Z Z^*$,
$\mu_{gg \rightarrow h \rightarrow \gamma \gamma}$ and $\mu_{gg \rightarrow h \rightarrow Z Z^*}$,
is about 10\%, while uncertainties in other relevant channels range from 20\% on  $\mu_{gg \rightarrow h \rightarrow W W}$ and $\mu_{VBF \rightarrow h \rightarrow \gamma \gamma}$,
to 30\%-40\% on   $\mu_{h \rightarrow b \bar b}$
and   $\mu_{t\bar t h}$.
For all observables, we do not assume improvement in the theoretical uncertainties, but stress that improvements on hadronic/nuclear matrix elements could have a large impact on EDM constraints \cite{Chien:2015xha}.

In the single operator analysis, EDMs put extremely strong bounds on the imaginary parts of the coefficients $C_\alpha$.
This is true in particular for $\tilde{c}_\gamma$ and $\tilde c_{Wt}$. As shown in Fig. \ref{fig:Cgamma}, the mixing of these operators  into the electron EDM leads to constraints that are a factor of $10^3$ stronger \cite{Cirigliano:2016njn} than constraints from the $A_{CP}$ asymmetry in  $b \rightarrow s\gamma$  or from the phase $\delta^-$ measured in top decays.
The current bound on the electron EDM limits $c_{\gamma}$ to be $|v^2 \tilde{c}_{\gamma}| < 1.4 \cdot 10^{-3}$. 
The real part of the coupling, $c_{\gamma}$, can be larger and is mainly constrained by the $S$ parameter and by  Higgs decay into photons. We find the allowed region for $c_{\gamma}$ to be  $-0.05 < v^2 c_\gamma < 0.11$ ($90\%$ C.L.). 
Projected experimental improvements on the electron EDM can improve the bounds on  
$\tilde{c}_{\gamma}$ by a factor of 10, while the LHC Run 2 has the possibility of improving the bound on the real part by a factor of $2$. Additional direct information on the real part can be obtained by studying additional observables at LHC Run 2 in processes such as $\bar t t + \gamma,\, \bar t t +Z$ \cite{Bylund:2016phk,Schulze:2016qas}.

The situation is similar for $C_{W t}$. In the single operator analysis, the imaginary part of $C_{W t}$ is extremely well constrained by the electron EDM,
$|v^2 \tilde{c}_{W t}  | < 1.2 \cdot 10^{-3}$. Bounds from $A_{CP}$ and $\delta^-$ are more than a hundred times weaker. In this case, the real part of the coupling receives competitive constraints from $b \rightarrow s \gamma$, the $S$ parameter, single top production, and the $W$ boson helicity fractions. The combined allowed region for $c_{W t}$ is $ -0.10 < v^2 c_{W t} < 0.04$ ($90\%$ C.L). 

The left panel of Fig. \ref{fig:CWb} shows the allowed region for the coupling $C_{Wb}$. The EDM bound is dominated by the neutron EDM and is five time stronger than the limit extracted from $A_{CP}$, even with the conservative treatment of the theoretical errors that we are adopting. We find $|v^2 \tilde{c}_{Wb} | < 0.06 $.
The real part, $c_{Wb}$, is mainly constrained by the $b \rightarrow s \gamma$ branching ratio yielding $ -0.04 <  v^2 c_{W b} < 0.07$. 
Collider observables, like $Z \rightarrow b \bar b$ or single top production, and indirect observables like the $S$ parameter, give much weaker bounds,  $v^2 c_{Wb} \sim O(1) - \mathcal O(10)$.
A neutron EDM bound at the level of  $10^{-28}$ $e$ cm, which should be in reach of the next generation of EDM experiments, 
would strengthen the constraint on $\tilde c_{W b}$ by a factor of 100 even without improvements on the hadronic matrix elements. 
The impact of future super-B factory  experiments appears to be more limited and only slightly affect the bound on $c_{Wb}$.

The center and right panels of Fig. \ref{fig:CWb} show, respectively, the allowed regions for $C_g$ and $C_Y$. As shown in Table \ref{tab:mtEDM}, $\tilde c_g$ contributes to both the neutron and the electron EDM.
In the R-fit procedure, the neutron EDM is subject to cancellations between  the contributions of the Weinberg operator and those of the light quark (C)EDMs. This severely weakens the neutron EDM constraint, such that the strongest bound comes from the electron EDM and we obtain $| v^2 \tilde{c}_g| < 0.038$. This constraint can be significantly improved with a better understanding of the effect of the Weinberg operator on the neutron EDM  \cite{Chien:2015xha}.
The top chromo-magnetic dipole moment, $c_g$, strongly affects the gluon fusion Higgs production channel and the decay width $h \rightarrow g g$, resulting in a very strong bound $-0.017 < v^2 c_g < 0.014$. 
In the center panel of Fig. \ref{fig:CWb} we compare the bound on $c_g$ from gluon fusion and the direct bound from the 
$t \bar t$ production cross section. Notwithstanding the sizable experimental uncertainties on the Higgs signal strengths, the indirect bound is already five time stronger than the direct bound from $t \bar t$.
$c_g$ also contributes to the associated production of a Higgs boson and a  $t\bar t$ pair. At the moment, the bound from $t\bar t h$ is not competitive with gluon fusion or $t \bar t$.
Data from the LHC Run 2 will improve the bound on $c_g$ to the sub-percent level, while new neutron and electron EDM experiments are projected to improve the bound on $\tilde{c}_g$ by one to two orders of magnitude.
 
In the single operator scenario, the pseudo-scalar Yukawa, $\tilde{c}_Y$, receives its strongest bound from the electron EDM, $| v^2 \tilde{c}_Y| < 0.013$. The real part of the anomalous Yukawa coupling, $c_Y$, affects the Higgs gluon fusion production cross section and the decays $h\rightarrow gg$ and $h\rightarrow \gamma \gamma$. $c_Y$ can also be probed directly by studying the associated $t\bar t h$ production. With the current experimental data, the bound is dominated by the Higgs signal strengths and is at the 10\% level, $-0.12 < v^2 c_Y < 0.14$. The bound from $t\bar t h$ is noticeably weaker (and has a preference for positive values of $c_Y$), $ -0.1 <  v^2 c_Y < 1.1$  and is not shown in Fig. \ref{fig:CWb}.

\begin{table}
\center
\begin{tabular}{||c | c| |c|c| c ||}
\hline \hline
Real &\textbf{R-fit}   & Imaginary &\textbf{R-fit} & \textbf{Central}    \\
\hline
$v^2\, c_\gamma$ & $[-5.3,11] \cdot 10^{-2}$  & $v^2\,\tilde{c}_\gamma$& $[-1.4,1.4]\cdot 10^{-3}$ & $[-1.4,1.4]\cdot 10^{-3}$  \\
$v^2\,c_{W t}$  & $[-9.5,4.2]  \cdot 10^{-2}$ & $v^2\,\tilde{c}_{Wt}$& $[-1.2,1.2]\cdot 10^{-3}$ &  $[-1.2,1.2]\cdot 10^{-3}$  \\
$v^2\,c_{W b}$  & $[-4.4,6.7]  \cdot 10^{-2} $  & $v^2\,\tilde{c}_{Wb}$& $[-6.4,6.4] \cdot 10^{-2}$  & $[-4.2,4.4]\cdot 10^{-3}$ \\
$v^2\,c_g$ 	  & $[-1.7,1.4] \cdot 10^{-2}$  & $v^2\,\tilde{c}_g$& $[-3.8,3.8] \cdot 10^{-2}$  &$[-2.9,2.9]\cdot 10^{-4}$  \\
$v^2\,c_Y$      & $[-12,14]\cdot 10^{-2}$  & $v^2\,\tilde{c}_Y$ & $[-1.3,1.3] \cdot 10^{-2}$	& $[-1.3,1.3] \cdot 10^{-2}$ \\
\hline \hline
\end{tabular}
\caption{Allowed region ($90\%$ C.L.) for the couplings $C_\alpha$, with the assumption that one complex coupling is  turned on at the scale $\Lambda = 1$ TeV. Constraints in the second and third columns  are obtained by using the R-fit strategy of Sect.~\ref{rfit}, while the constraints on the imaginary couplings in the fourth column are based on central matrix elements. }
\label{Tab:Individual}
\end{table}

\subsubsection{Summary}
The constraints shown in Figs.\ \ref{fig:Cgamma} and \ref{fig:CWb} are summarized in Table \ref{Tab:Individual}. EDM limits provide the most stringent constraints for all the imaginary parts of the top-Higgs couplings, which are therefore generally constrained to be smaller than the real parts by one order of magnitude or more. Exceptions are $C_g$ and $C_{Wb}$ where the imaginary parts can still be of the same order as the real parts. In part, these exceptions are due to the large hadronic uncertainties related to the Weinberg operator, which in the case of $\tilde c_g$ allows the contributions to the neutron EDM to cancel completely. Clearly, a better understanding of the relevant matrix element would lead to improvement of these constraints. To illustrate this, in the  fifth  column of Table \ref{Tab:Individual} we also give the bounds that can be set on the imaginary couplings if we ignore the uncertainties in the matrix elements and simply use central values. This has a large impact on the couplings $\tilde c_{Wb}$ and $\tilde c_g$ illustrating the importance of improving the theoretical understanding of CPV operators in nucleons and nuclei. 

The real parts of the couplings are constrained by a more diverse set of observables. Higgs production and decay processes provide the most stringent constraints on $c_\g$ ($h\to \g\g$), $c_{g}$ ($gg\to h$), and $c_Y$ (both $gg\to h$ and $h\to\g\g$). The weak dipole operator $c_{Wb}$ is constrained purely by its contribution to the $b\to s\g$ transition, while for $c_{Wt}$ the $W$ helicity fractions give rise to slightly stronger constraints. Finally, the $S$ parameter provides competitive constraints in the case of $c_\g$ and $c_{Wt}$. We find that the constraints are not significantly affected by theoretical uncertainties and find only small differences when using central matrix elements. 

The constraints in Table \ref{Tab:Individual} were derived truncating the expansion of observables at $\mathcal O(v^2/\Lambda^2)$, including only genuine dimension-six effects.
We explicitly checked that  dimension-eight effects in the collider cross sections and signal strengths, and in $b\rightarrow s \gamma$,
do not significantly impact the bounds.

Finally, we notice that for $c_\gamma,$ $c_{Wt}$, $c_{Wb}$,  $c_Y$,  $\tilde{c}_Y$, $\tilde c_g$ and the bounds on $c_g$ from $t\bar t$ production, 
our results are compatible with the existing literature (for example, Refs. \cite{Kamenik:2011dk,Brod:2013cka,Aguilar-Saavedra:2014iga,deBlas:2015aea,Hioki:2015env,Aguilar:2015vsa,Birman:2016jhg,Buckley:2015lku,Buckley:2015nca}).
For $\tilde{c}_\gamma$, $\tilde{c}_{Wt}$, $\tilde{c}_{Wb}$, we find that EDMs provide stronger bounds than  previously realized.  
For $c_g$, the strongest constraint comes from the contribution to Higgs production through gluon fusion.

\begin{figure}
\center
\includegraphics[width=0.47\linewidth]{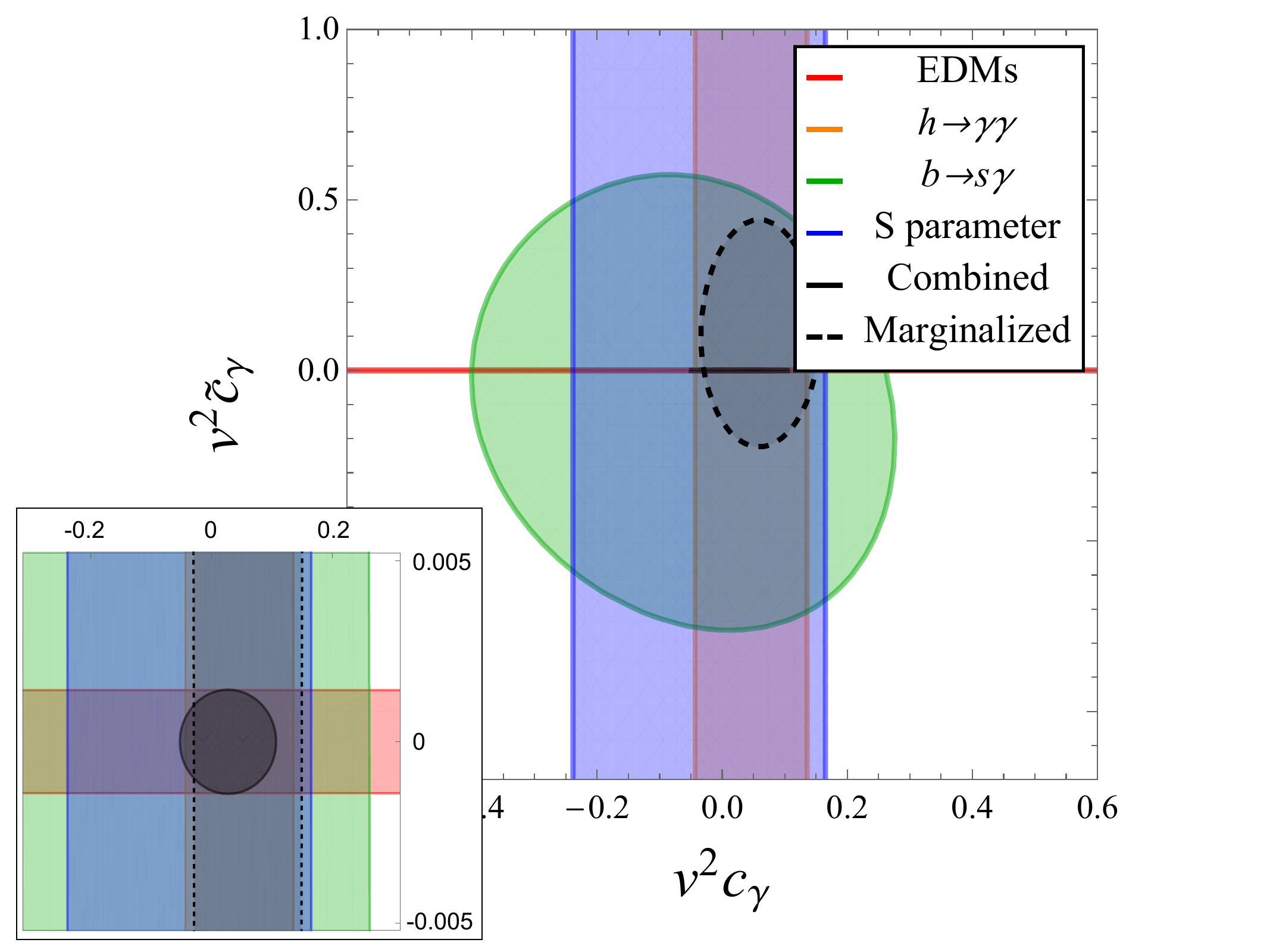}
\includegraphics[width=0.47\linewidth]{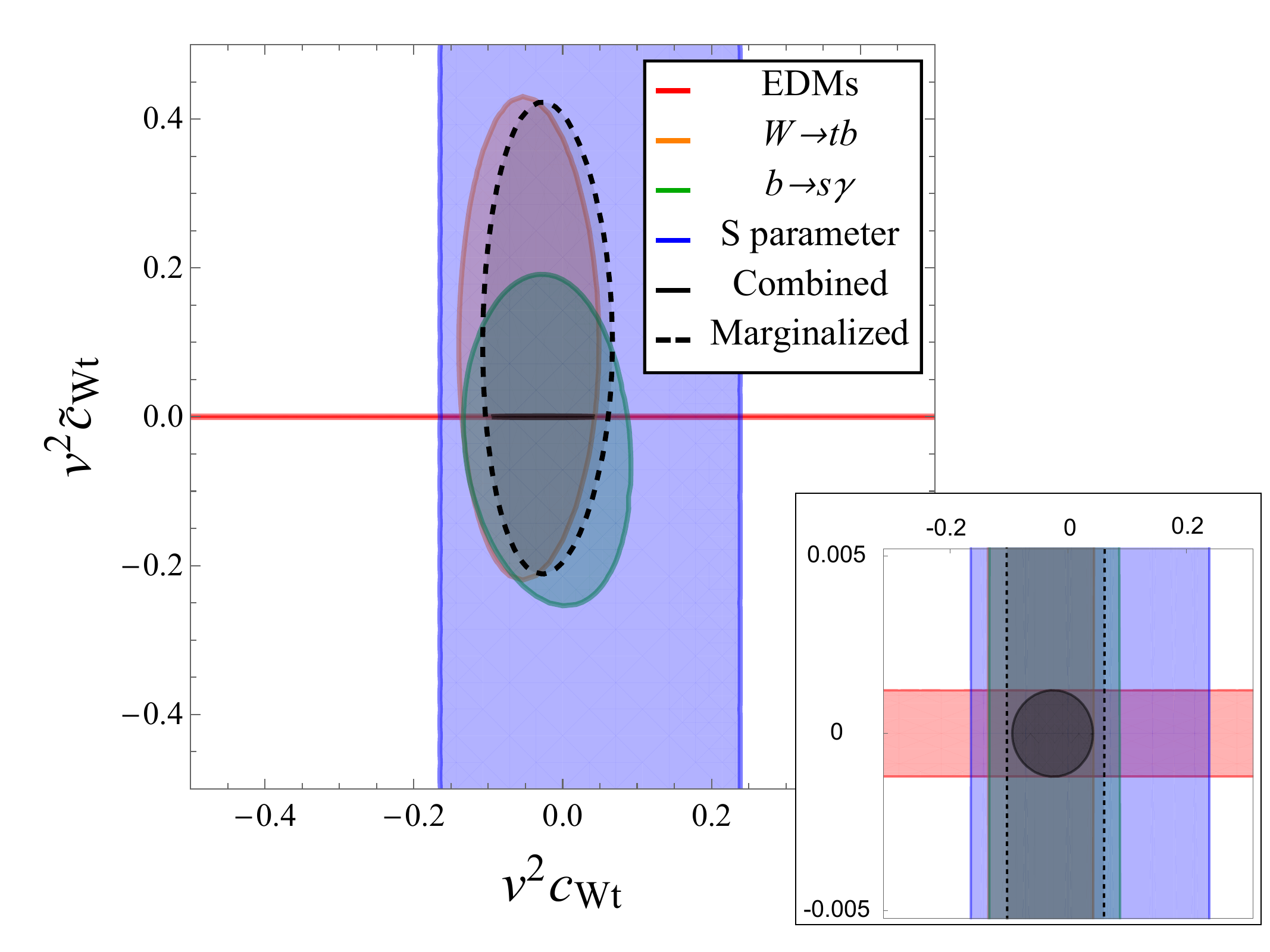}
\caption{Same as Fig.\ \ref{fig:Cgamma}, showing the 90\% C.L.\  allowed regions in the   $v^2 c_{\gamma} - v^2  \tilde{c}_{\gamma}$   (left panel) 
and $v^2 c_{Wt} - v^2  \tilde{c}_{Wt}$  planes (right panel), but now assuming central values for the relevant nuclear and hadronic matrix elements. Both the allowed regions in the single coupling case (solid lines) and marginalized case (dashed lines) are shown.}\label{marginalized1}
\end{figure}

\subsection{Global fit}\label{global}

We now investigate the scenario in which new physics generates all the operators in Eq. \eqref{eq:Leff} at the high scale $\Lambda = 1$ TeV, with arbitrary coefficients. This scenario allows us to quantify the effects of possible cancellations between contributions from various top couplings to direct and indirect observables, and to test the robustness of the strong EDM bounds discussed in Section \ref{singlecoupling}.

The large theoretical uncertainties of the hadronic and nuclear matrix elements entering the mercury EDM cause the bound from $d_{\textrm{Hg}}$ to effectively disappear in the R-fit approach, reducing the number of CP-odd observables to four (electron and neutron EDMs, $A_{CP}$ in $b\rightarrow s \gamma$, and the phase $\delta^-$ in top decays). As we investigate five anomalous couplings, this gives rise to free directions for the imaginary parts leading to unbound $\tilde{c}_\alpha$ for all $\al$ apart from $\tilde c_{Wt}$ which remains constrained by the $W$ helicity fractions discussed in Section \ref{HelFrac}. This situation is certainly unrealistic and requires an unmotivated cancellation between various couplings and matrix elements. Furthermore, the free directions can be removed by including less sensitive observables which we have neglected so far, or by including dimension-eight effects such as contributions of $\tilde{c}_{\alpha}$
to CPC total cross sections and decay rates, which become relevant for $v^2 \tilde{c}_\alpha \sim O(1)$ (of course, this does not protect us from further cancellations against possible dimension-eight BSM operators).  The  latter possibility is, however,  at the limit of validity of our assumption that  the leading effects of BSM physics are captured by non-renormalizable operators of lowest canonical dimension. Finally, future EDM measurements on systems such as the proton, deuteron, or radium can also remove unconstrained directions \cite{Chien:2015xha}.

In the rest of this Section we study one case in which the $C_\alpha$ can be bound, that is if we neglect theoretical uncertainties in the hadronic and nuclear matrix elements entering $d_n$, $d_{\textrm{Hg}}$ and $A_{CP}$. Although this might seem rather wishful at the moment, relatively modest improvements from both lattice QCD and nuclear many-body theory regarding various matrix elements (see the discussion in Ref.~\cite{Chien:2015xha}) would be sufficient to make this a realistic scenario.

\begin{figure}
\center
\includegraphics[width=0.3\linewidth]{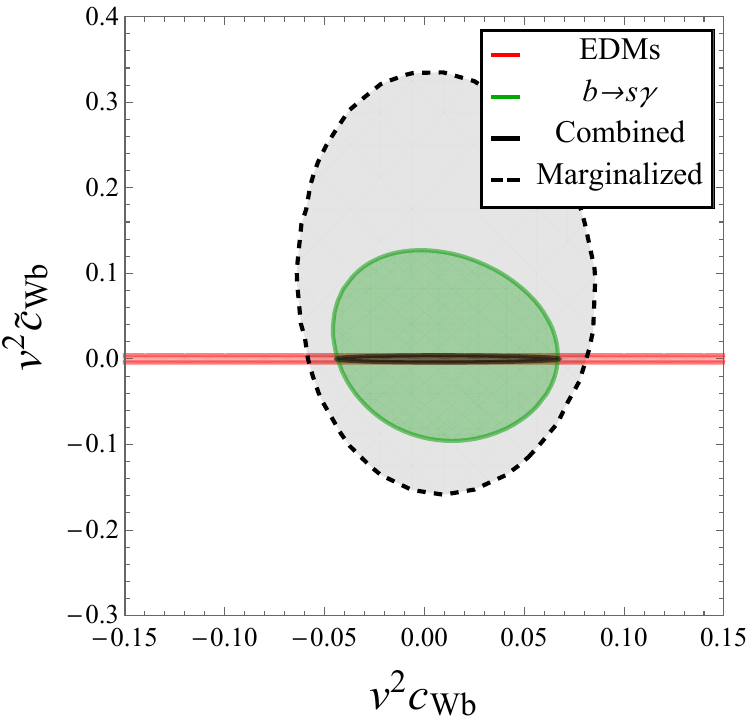}
\includegraphics[width=0.3\linewidth]{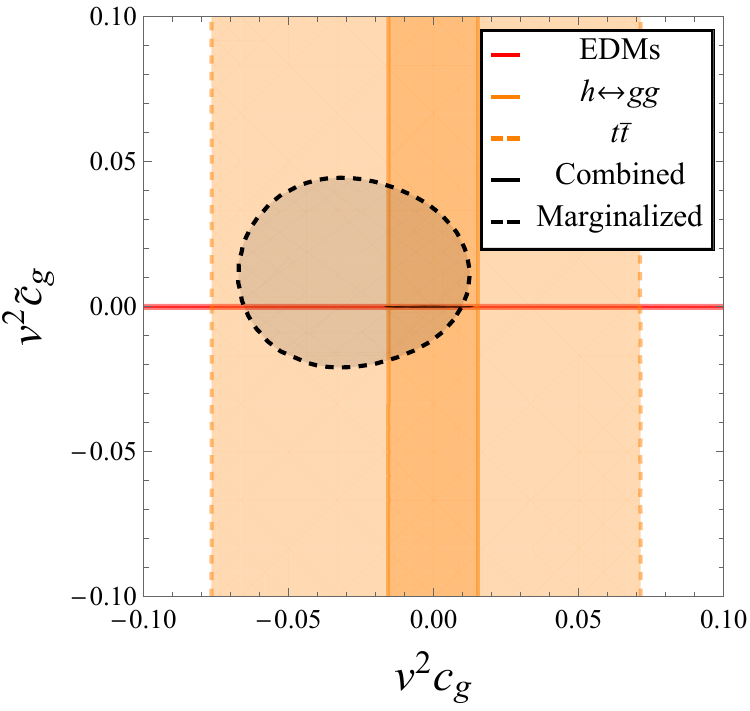}
\includegraphics[width=0.3\linewidth]{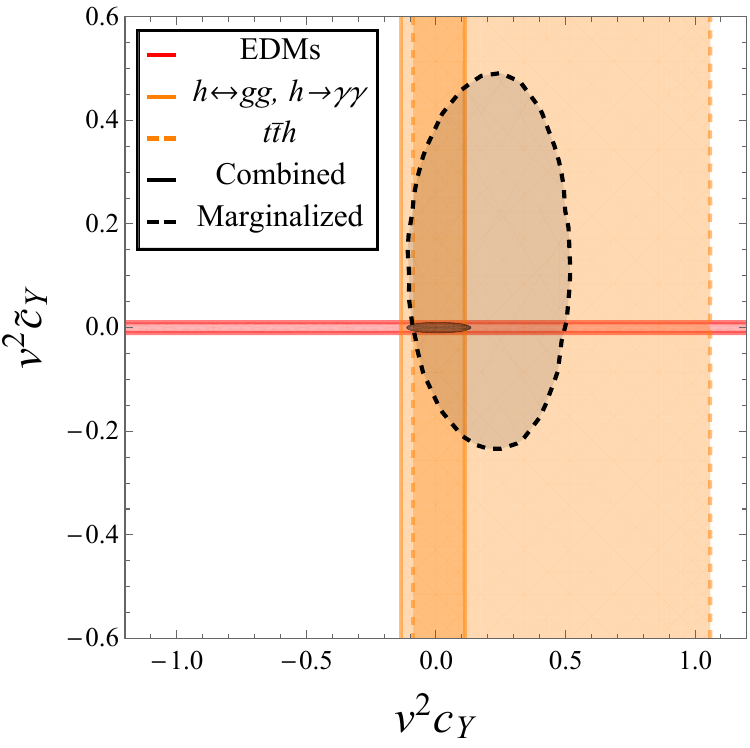}
\caption{
Same as Fig.\ \ref{fig:CWb}, showing the 90\% C.L.\  allowed regions in the   
$v^2 c_{Wb} - v^2  \tilde{c}_{Wb}$   (left panel) 
and $v^2 c_{g} - v^2  \tilde{c}_{g}$  (center panel) 
and $v^2 c_{Y} - v^2  \tilde{c}_{Y}$  (right panel) planes,  but now assuming central values for the relevant nuclear and hadronic matrix elements. Both the allowed regions in the single coupling case (solid lines) and marginalized case (dashed lines) are shown.}\label{marginalized3}
\end{figure}

\begin{table}
\center\begin{tabular}{||c |c| c| |c|c| c ||}
\hline \hline
Real &\textbf{Individual} &\textbf{Marginalized}  & Imaginary &\textbf{Individual} &\textbf{Marginalized}  \\
\hline
$v^2\, c_\gamma$ & $[-5.3,11]  \cdot 10^{-2}$  &$[-3.4,15]       \cdot 10^{-2}$ & $v^2\,\tilde{c}_\gamma$& $[-1.4,1.4]\cdot 10^{-3}$ & $[-22,44] \cdot 10^{-2}$  \\
$v^2\,c_{W t}$   & $[-9.5,4.2]  \cdot 10^{-2}$  &$[-11,6.8]  \cdot 10^{-2}$& $v^2\,\tilde{c}_{Wt}$      & $[-1.2,1.2]\cdot 10^{-3}$ & $[-21,42] \cdot 10^{-2}$  \\
$v^2\,c_{W b}$   & $[-4.4,6.7]  \cdot 10^{-2}$  & $[-6.4,8.5] \cdot 10^{-2}$ & $v^2\,\tilde{c}_{Wb}$     & $[-4.2,4.4]\cdot 10^{-3}$ & $[-16,34] \cdot 10^{-2}$       \\
$v^2\,c_g$ 	 & $[-1.7,1.4] \cdot 10^{-2}$  &$[-6.7,1.1]  \cdot 10^{-2}$  & $v^2\,\tilde{c}_g$      & $[-2.9,2.9]\cdot 10^{-4}$ & $[-2.0,4.4] \cdot 10^{-2}$  \\
$v^2\,c_Y$       & $[-12,14]   \cdot 10^{-2}$  & $[-11,52] \cdot 10^{-2}$ & $v^2\,\tilde{c}_Y$         & $[-1.3,1.3] \cdot 10^{-2}$& $[-24,50] \cdot 10^{-2}$ \\
\hline \hline
\end{tabular}\\
\caption{\small Allowed regions ($90\%$ C.L.) for the couplings $C_\alpha$, at the scale $\Lambda = 1$ TeV, while employing the `central' strategy outlined in Section \ref{global}. The second and fifth columns show the constraints when assuming only a single complex coupling is generated at the high scale, while in the third and sixth columns we assume all $C_\al$ are present at the scale of new physics and marginalize over the remaining couplings.}
\label{Tab:Margin}
\end{table}
\subsubsection{Global analysis: central values of the hadronic matrix elements}
\begin{figure}
\center
\includegraphics[width=7cm]{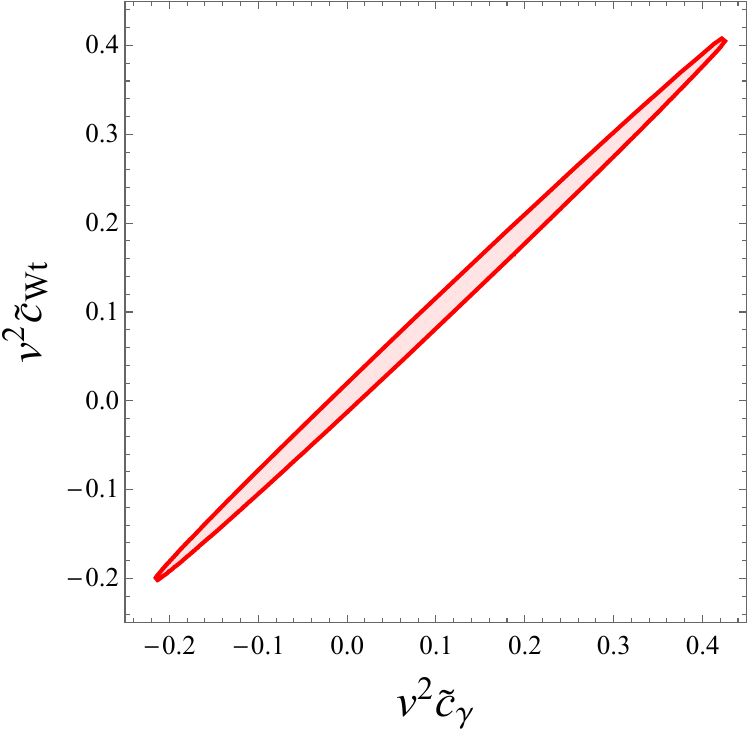}
\caption{90\% C.L.\  allowed regions in the   $v^2 \tilde{c}_{\gamma} - v^2  \tilde{c}_{Wt}$ plane. We marginalized over the remaining couplings and assumed central values for the relevant nuclear and hadronic matrix elements.}\label{edmvswdm}
\end{figure}

Figs. \ref{marginalized1} - \ref{marginalized3} show the marginalized constraints as well as those resulting from the single-coupling analysis (at 90\% C.L.), using the central procedure in both cases.
We immediately notice that the limits on $\tilde c_\alpha$ weaken considerably because the imaginary parts of the couplings are strongly correlated.
The bounds on  $v^2 \tilde{c}_{\gamma}$  and $v^2 \tilde{c}_{W t}$ deteriorate from the few permil level to about 40\%. This can be understood from the fact that the electron EDM, which provides the strongest bound in the single operators analysis, is sensitive to the combination $\sim \tilde{c}_\gamma - \tilde{c}_{W t}$,
leaving the orthogonal linear combination unconstrained. This is illustrated in Fig. \ref{edmvswdm}. The required orthogonal constraint comes from $\delta^-$, which only receives contributions from $v\sq \tilde c_{Wt} $ and is therefore unaffected by marginalizing. The combination of the two then provides the $\sim 40\%$ bounds on $\tilde c_{\g,Wt}$. The resulting constraint on $\tilde c_\gamma$ is still somewhat stronger than the bound from $b \rightarrow s \gamma$. On the other hand, in the marginalized case there are not enough EDM measurements to constrain $\tilde{c}_{W t}$, and the bound becomes almost identical to that of the direct observable $\delta^-$.

The couplings $\tilde c_{Wb}$ and $\tilde c_{g}$ also exhibit strong correlations, because they mainly contribute to the neutron EDM. In the marginalized case, due to the possible cancellations, the bound on $\tilde c_{Wb}$ is then mainly determined by $A_{CP}$, while $d_n$ and $d_{\textrm{Hg}}$ 
set strong constraints on $\tilde c_{g}$.  Similarly, the bound on $ \tilde{c}_Y$ is weakened and now allows 
for a large top pseudoscalar Yukawa coupling,  up to  $50\%$ of the SM top Yukawa.
This further motivates direct searches for CP-odd effects, for instance in the measurements of triple correlations in $t\bar t h$ production \cite{Mileo:2016mxg}, which can become sensitive to $v^2 \tilde c_{Y}\sim \mathcal O(0.3)$ at the LHC Run II. 
The current situation could also be improved by additional EDM experiments. For instance, a proton or deuteron EDM measurement at the level of $d_{p,D}\leq 3.0\Ex{-26} e$ cm, would significantly shrink the allowed region in the $\tilde c_g - \tilde c_Y$ plane \cite{Chien:2015xha}.

\begin{figure}
\center
\includegraphics[width=0.45\linewidth]{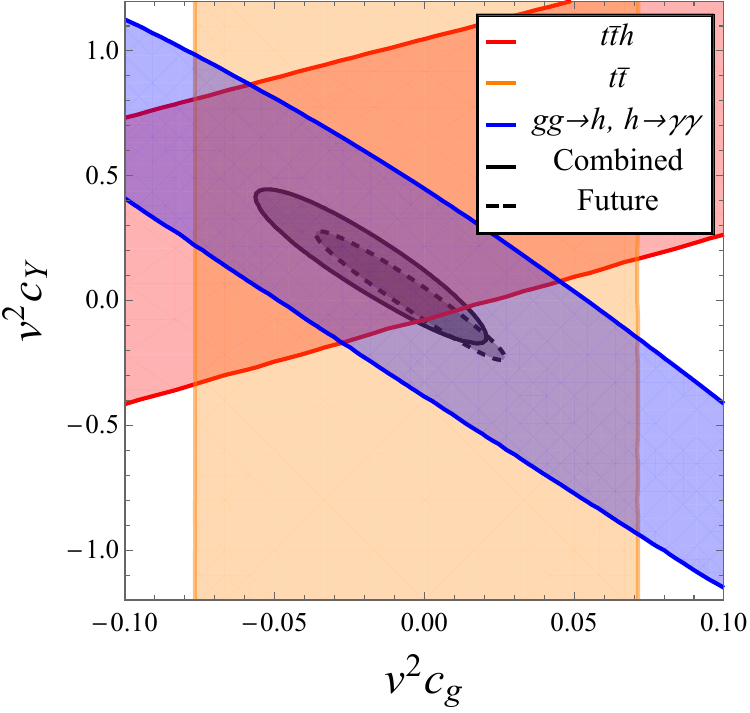}\qquad 
\caption{90\% C.L.\  allowed regions in the   $v^2 c_{g} - v^2  {c}_{Y}$     plane, with couplings evaluated at   $\Lambda = 1$~TeV. 
We assume that only $C_g$ and $C_Y$ are generated at the high scale and the theoretical uncertainties are dealt with using the R-fit procedure.}\label{Fig:Cg-Cy}
\end{figure}

The real parts of the coefficients $C_\alpha$ are much less affected by considering multiple operators at the new physics scale $\Lambda$.
The bounds on $c_{\gamma}$, $c_{W t}$ and $c_{W b}$, which are respectively dominated by $h \rightarrow \gamma \gamma$, the W boson helicity fractions and $b \rightarrow s \gamma$,
are barely changed, and these couplings are nearly uncorrelated. 
In the single coupling analysis, gluon fusion provides the strongest constraints on both $c_g$ and $c_Y$. Turning on both couplings therefore allows for cancellations that weaken the bound. The center and right panels of Fig. \ref{marginalized3} 
show that the marginalized bounds on $c_g$ and $c_Y$ from gluon fusion are still better than the individual direct bounds from $t\bar t$ and $t\bar t h$. The bound on $c_g$ also remains strong in the marginalized case,
while cancellations between $c_g$ and $c_Y$ allow for large corrections to the top Yukawa, up to 50\%. As is illustrated in Fig.\ \ref{Fig:Cg-Cy}, an improved direct measurement of $t\bar t h$ at the LHC Run 2, with uncertainties reduced to the 30\% -
40\% level, would improve the upper bound on $c_Y$ by a factor $\sim 2$.

The $90\%$ C.L.\ limits resulting from the marginalized central analysis are summarized, and compared to the individual bounds, in Table \ref{Tab:Margin}. Finally, we give some information about the fit. The correlation matrix of the couplings $\{c_\gamma,  c_{W t}, c_{Wb} , c_g, c_Y, 
 \tilde{c}_\gamma,  \tilde{c}_{W t}, \tilde{c}_{Wb}, \tilde{c}_g, \tilde{c}_Y\}$
is given by
\begin{equation}
  \begin{bmatrix}[ccccc |  ccccc]
      1.00  &  0.17  &  0.32  & -0.35  &  0.27  &  0.00  &  0.00  &  0.00  &  0.00  &  0.00  \\
      0.17  &  1.00  &  0.57  & -0.04  &  0.02  &  0.03  &  0.03  &  0.02  &  0.02  &  0.02  \\
      0.32  &  0.57  &  1.00  & -0.14  &  0.11  &  0.02  &  0.01  & -0.04  & -0.03  & -0.03  \\
     -0.35  & -0.04  & -0.14  &  1.00  & -0.93  &  0.00  &  0.00  &  0.00  &  0.00  &  0.00  \\
      0.27  &  0.02  &  0.11  & -0.93  &  1.00  &  0.00  &  0.00  &  0.00  &  0.00  &  0.00  \\
  \hline     0.00  &  0.03  &  0.02  &  0.00  &  0.00  &  1.00  &  1.00  &  0.82  &  0.89  &  0.72 \\
      0.00  &  0.03  &  0.01  &  0.00  &  0.00  &  1.00  &  1.00  &  0.86  &  0.92  &  0.78 \\
      0.00  &  0.02  & -0.04  &  0.00  &  0.00  &  0.82  &  0.86  &  1.00  &  0.98  & 0.88 \\
      0.00  &  0.02  & -0.03  &  0.00  &  0.00  &  0.89  &  0.92  &  0.98  &  1.00  & 0.91 \\
      0.00  &  0.02  & -0.03  &  0.00  &  0.00  &  0.72  &  0.78  &  0.88  &  0.91  & 1.00  
  \end{bmatrix}\ .
\end{equation}
The off-diagonal entries connecting the real and imaginary couplings are small indicating that there is little correlation between them. This is not surprising as most observables are only sensitive to either the real or the imaginary couplings. The minimum $\chi^2$ of the multidimensional fit is $\chi^2 = 17$, with 45 experimental entries and 10 fit parameters, leading to a
$\chi^2$ per degree of freedom, $\chi^2/dof \sim 0.5$.

Table \ref{Tab:Margin} shows that the global fit allows for relative large values of the couplings $C_{\alpha}$, especially for the real and imaginary top Yukawa. One might wonder if the EFT approach is still valid in this regime, or if dimension-eight effects, coming for example from double insertions of dimension-six operators, start to become important. By turning on the partial dimension-eight corrections to collider observables and $b\rightarrow s \gamma$
given in Sections \ref{sect:collider} and \ref{sect:rareB}, we checked that the bounds obtained in the global fit with central matrix elements are not significantly affected.

\section{Minimal Flavor Violation  scenario}
\label{sect:MFV}
In this section we study the top-Higgs couplings in Eq. \eqref{eq:Leff} in the context  of Minimal Flavor Violation (MFV) \cite{D'Ambrosio:2002ex}.
In the absence of Yukawa couplings of the Higgs to fermions, the SM Lagrangian is invariant under   $U(3)^3$ transformations acting on the  family indices of the 
quark fields $q_L, u_R, d_R$.    In the SM,  the  flavor symmetry ${\cal G}_F =SU(3)_{q_L} \times SU(3)_{u_R} \times SU(3)_{d_R} $  is broken by the Yukawa couplings. MFV assumes that this also holds for possible BSM physics and that the only
spurions that break ${\cal G}_F$ are the Yukawa couplings.

The operators in Eq. \eqref{eq:Leff} all break chiral symmetry. In MFV,  their flavor structure is highly constrained and can be obtained by inserting the appropriate combinations of  Yukawa matrices $Y_u$   and  $Y_d$ that make the operators  formally invariant under ${\cal G}_F$. 
For example, to leading order  in Yukawa insertions one has   $\Gamma_{B,W,g}^{u},  Y_u' \propto  Y_u$  and $\Gamma_{W}^{d}  \propto  Y_d$. More invariants can be constructed by inserting an arbitrary number of products 
$Y_u Y_u^\dagger$ or $Y_d Y_d^\dagger$, followed by $Y_u$ or $Y_d$~\cite{D'Ambrosio:2002ex}.
We  restrict  ourselves to the case in which only the top Yukawa coupling is treated as $\mathcal O(1)$, while the Yukawas of the other quarks are considered small.
We then consider operators that contain at most one insertion of light quark Yukawas and an arbitrary number of insertions of $y_t$.
Under this assumption we can disregard insertions of $Y_d Y_d^\dagger$ and 
it is possible to show that an arbitrary polynomial of $(Y_u Y_u^\dagger)$  induces the same 
pattern of couplings  as a single insertion of $Y_u Y^\dagger_u$.

Within MFV extended by the additional assumptions described above, the 
chirality-flipping top couplings can only  have  a limited number of flavor structures.
Working in the weak basis in which $Y_d$ is diagonal, $Y_d = \lambda_d$, and $Y_u = V_{\textrm{CKM}}^{\dagger} \lambda_u$, with $\lambda_{d} = \textrm{diag}(y_d,y_s,y_b)$ and
$\lambda_u = \textrm{diag}(y_u,y_c,y_t)$, the only possible structures for up-type operators (like $C_g$, $C_\gamma$, $C_Y$, and $C_{Wt}$ 
related to $\Gamma_{B,W,g}^{u},  Y_u'$) 
are
\begin{equation}\label{MFVup}
Y_u = V_{\textrm{CKM}}^\dagger \lambda_u, \qquad  
P \left( Y_u Y^\dagger_u \right) Y_u \sim V_{\textrm{CKM}}^\dagger \left( \begin{array}{ccc} 
0 & 0 & 0 \\
0 & 0 & 0 \\
0 & 0 & y_t P(y_t^2) \\
\end{array} \right)~,   
\end{equation}
where $P(X)$ denotes a polynomial function of $X$. 
For down-type operators, like $C_{Wb}$  (related to $\Gamma_W^b$) or the $b$ quark dipole moments, we only have
\begin{equation}\label{MFVdown}
Y_d =  \lambda_d, \qquad   
P \left( Y_u Y^\dagger_u \right) Y_d \sim 
P(y_t^2) \cdot 
\left( \begin{array}{ccc} 
V_{td} V_{td}^*  y_d & V_{ts} V_{td}^*  y_s & V_{tb} V_{td}^*  y_b \\
V_{td} V_{ts}^*  y_d & V_{ts} V_{ts}^*  y_s & V_{tb} V_{ts}^*  y_b \\
V_{td} V_{tb}^*  y_d & V_{ts} V_{tb}^*  y_s & V_{tb} V_{tb}^*  y_b \\
\end{array} \right).
\end{equation}
Rotating the $u$ quark to the mass basis ($u_L \rightarrow V^\dagger_{\textrm{CKM}} u_L$), it becomes clear that the operators $C_g$, $C_\gamma$, $C_Y$, and $C_{Wt}$
correspond to the MFV structure $P\left( Y_u Y^\dagger_u \right) Y_u$. The situation is more complicated for the down-type operator $C_{Wb}$, which does not correspond to  
any of the structures in Eq. \eqref{MFVdown}, but would be generated by three  (or more) insertions of the down Yukawa, $\left( Y_d Y^\dagger_d \right) Y_d$, under the assumption that powers of 
$y_d$ and $y_s$ are small with respect to $y_b$ and can be neglected.  We will focus in the rest of this section on up-type operators and neglect $C_{Wb}$.

Even under the assumption that $y_t \gg y_{u,c}$, we cannot simply set $y_u$ and $y_c$ to zero and reduce the two structures in Eq. \eqref{MFVup} to a single one.
Instead, we have to consider the scaling of a given observable in powers of the light quark Yukawa couplings. 
For example, in the case of the nucleon EDM, the first structure in Eq. \eqref{MFVup} induces a $u$ quark EDM or CEDM proportional to $y_u$.
As shown in Sec. \ref{sec:RGEs}, the second structure also induces $u$ dipoles that are proportional to $y_u y_t^2  P(y_t^2)$. Thus the two structures in Eq. \eqref{MFVup} contribute to the nucleon EDM at the same order in light quark Yukawas, and are independent.

Thus, in  a generic MFV scenario in which arbitrary insertions of  $y_t$ are allowed, the operator basis of Eq. \eqref{eq:Leff} needs to be extended 
\begin{equation}
{\cal L}_{\rm eff}^{\rm BSM}   =
\sum_{\alpha \in \{ Y, g, \gamma, Wt\} }  \ C_\alpha  \, O_\alpha 
+ C^\prime_\alpha  \, O^\prime_\alpha + {\rm h.c.} \ ,
\qquad  \quad
\label{eq:Leff_mfv}
 \end{equation}
where
\begin{subequations}
\label{eq:operatorsMFV}
\bea
O^\prime_Y  &=&   -  v \, \bar{u}_L \lambda_u  u_R  \left( v h  + \frac{3}{2}  h^2 + \frac{1}{2}  \frac{h^3}{v}  \right) \ ,  
\\
O^\prime_\gamma &=&  - \frac{e Q_t}{2}  v  \, \bar{u}_L  \lambda_u  \sigma_{\mu \nu} \left( F^{\mu \nu} - t_W  Z^{\mu \nu} \right) u_R \, \left(1 + \frac{h}{v}\right)   \ ,
\qquad 
\\
O^\prime_g &=&  - \frac{g_s}{2}  v  \, \bar{u}_L  \lambda_u \sigma_{\mu \nu} G^{\mu \nu} u_R \, \left(1 + \frac{h}{v}\right)   \ ,
\\
O^\prime_{Wt} \!\!   &=& -g v \bigg[  \frac{1}{\sqrt{2}}  \bar{d}_L  V_{CKM}^\dagger \lambda_u \simu  u_R W_{\mu\nu}^- 
+ \!  \! \bar u_L \lambda_u \simu u_R \bigg(\frac{1}{2c_W} Z_{\mu\nu}+i g W_\mu^-W_\nu^+\bigg)\bigg]\bigg(1+\frac{h}{v}\bigg) \nn \\
\eea
\end{subequations}
Differently from the operators $\mathcal O_\alpha$, $\mathcal O_\alpha^\prime$
induce couplings of the $u$ and $c$ quarks that are proportional to $y_u$ and $y_c$, respectively. 
In this context, we wish to  address the  following questions: 
\begin{itemize}
\item  How do the $u$ and $c$ couplings (implied by linear MFV) affect the constraints on the top-Higgs couplings? 
\item   Do we have enough information to put stringent bounds on 
the top couplings once we include both flavor structures, $C_\alpha$ and $C_\alpha^\prime$ at the same time? 
\end{itemize}
We focus only on the imaginary parts of the couplings. For the real parts, the best constraints come from top physics, while the $u$ and $c$ couplings are poorly constrained.

\begin{table}
\center 
\begin{tabular}{||c|c|c||}
\hline \hline 
Coupling                  &  {\bf R-fit}   &  {\bf Central}   \\
\hline
$v^2 \tilde{c}^\prime_\gamma$ &   $[-0.6 , 0.6 ] \cdot 10^{-3}$  &    $[-4,4] \cdot 10^{-4}$	\\
$v^2 \tilde{c}^\prime_{Wt}$   &   $[-1.2 , 1.2 ] \cdot 10^{-3} $ & $ [-3.3, 3.3] \cdot 10^{-4} $	\\
$v^2 \tilde{c}^\prime_g$      &   $[-3.8,3.8] \cdot 10^{-2} $	    &       $[-0.8, 0.8] \cdot 10^{-5} $ \\
$v^2 \tilde{c}^\prime_Y$      &   $[-1.4,1.4] \cdot 10^{-2}$	&   $  [-1.3,  1.3] \cdot 10^{-2}$  \\
\hline \hline
\end{tabular}
\caption{Allowed regions for the couplings $\tilde{c}_\alpha '$ in the linear MFV scenario, under different 
treatments of the theoretical uncertainties (R-fit versus central).}\label{MFV1}
\end{table}

To address the first question we assume $C_\alpha = 0$, or equivalently, we work in linear MFV. This assumption is explicitly realized in perturbative models where additional insertions of the Yukawa couplings,
such as the structure $(Y_u Y_u^\dagger) Y_u$, are loop suppressed.
In Table \ref{MFV1} we list the 90 \% C.L.\ bound on the coefficients $\tilde{c}_\gamma'$, $\tilde{c}_{Wt}'$, $\tilde{c}_Y'$, and $\tilde{c}_g'$, obtained 
under the assumption of linear MFV, and treating the hadronic uncertainties with the R-fit  and central methods. 
For the R-fit analysis we see that the bound on ${\tilde c}_\gamma^\prime$ is a factor of  two   stronger  than the bound 
on ${\tilde c}_\gamma$ in Table~\ref{Tab:Individual}.
This can be understood from the tree-level contribution of the $u$ quark EDM to the neutron EDM which does not suffer from hadronic uncertainties because of the good control of the nucleon tensor charges.  Furthermore,  the charm EDM only provides small contributions  to the light quark (C)EDMs and the Weinberg operator, such that there is no room for cancellations.

The bounds on $\tilde c_g^\prime$, $\tilde{c}^\prime_{Wt}$, and $\tilde{c}^\prime_Y$ are identical to those on $\tilde c_g$, $\tilde{c}_{Wt}$, and $\tilde{c}_Y$ in Table~\ref{Tab:Individual}, because they are all dominated by the contribution of the top couplings to the electron EDM. 
The contributions of the $u$  (mainly through the $u$ (C)EDM) and $c$ quark (mainly through the Weinberg operator) to the neutron EDM 
can cancel  with the existing theoretical uncertainties.
A comparison with Table~\ref{Tab:Individual} reveals that  in the central case,  the bounds on $\tilde c_{\alpha}^{\,\prime}$ are always better 
than the bounds on $c_{\alpha}$, with the exception of the Yukawa coupling. This again illustrates the impact of hadronic uncertainties. 

\begin{figure}
\center
\includegraphics[width=7cm]{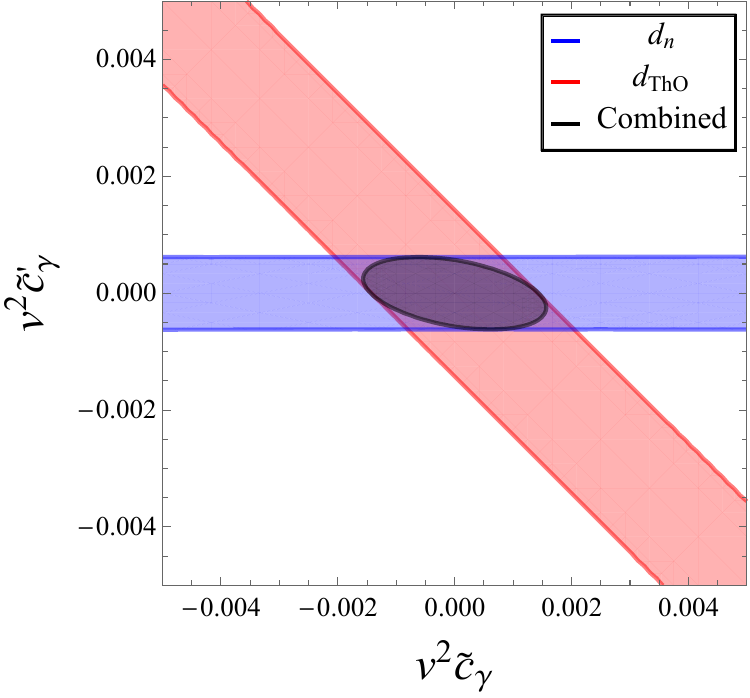}
\caption{90\% C.L.\  allowed regions in the   $v^2 \tilde{c}_{\gamma} $- $v^2  \tilde{c}_{\gamma}^\prime$ plane, 
using the R-fit method to treat theoretical uncertainties  in  the relevant   hadronic matrix elements.}\label{fig:MFV}
\end{figure}

To address the second question,  we  study the case in which both couplings, $C_\alpha$ and $C_\alpha^\prime$, are generated by BSM physics. 
We turn on one class of operators at a time.  The top couplings are  now proportional to   $\tilde{c}_\alpha + \tilde{c}_\alpha^\prime$,  while the $u$ and $c$ couplings are proportional to $\tilde{c}_\alpha^\prime$.  
Strictly speaking there is then no correlation between the  top  and light flavor couplings. 
The top anomalous couplings are, because of their contribution to the electron EDM, constrained at the same level as in the non-MFV case.   
On the other hand, the theoretical uncertainties are large enough that the contributions of the $u$ and $c$ quark to the 
neutron EDM can cancel, leading to no constraint on  $\tilde{c}_\alpha^\prime$ with the exception of $\tilde{c}_\gamma^\prime$.   For $\tilde{c}_\gamma$ and $\tilde{c}^\prime_\gamma$ we find that both 
couplings are very well constrained (illustrated in Figure~\ref{fig:MFV}) 
\begin{equation}
|v^2 \tilde{c}_\gamma|  < 1.5 \cdot 10^{-3} \qquad |v^2 \tilde{c}^\prime_\gamma|  < 0.6 \cdot 10^{-3}.
\end{equation}
For the other couplings,   $\tilde{c}_{Wt}$-$\tilde{c}_{Wt}^\prime$, $\tilde{c}_g$-$\tilde{c}_g^\prime$, and $\tilde c_Y $-$ \tilde{c}_Y^\prime$,  there exists a free direction, 
the direction in which $\tilde{c}_\alpha + \tilde{c}_\alpha^\prime = 0$. 
Because of the electron EDM limit, the coupling to the top quark remains bound at the same level as in the non-MFV analysis, despite the free direction in the $\tilde c_\alpha - \tilde c^\prime_\alpha$ plane. Finally,  using central values for the matrix elements would lead to bounds on both $\tilde{c}_\alpha$ and $\tilde{c}_\alpha^\prime$ for all couplings.

\section{Conclusions}
\label{sect:conclusion}

In this work  we have  discussed  probes of  chirality-flipping 
top-Higgs  couplings, including  both CP-conserving  and CP-violating  interactions. 
Working to leading order (dimension-six) in the framework of the SM-EFT, 
the chirality-flipping   interactions  involving top and Higgs fields are characterized by five complex couplings. 
We  have  studied direct and indirect constraints on these couplings, 
the latter arising from both high- and low-energy observables  
(for a synopsis see Tables~\ref{OverviewReal} and \ref{OverviewIm}). 
We have derived bounds on anomalous couplings under three scenarios: 1) we allow one operator at a time to 
be generated at the high scale  (still allowing for both CP-even and CP-odd couplings). 
The results of this analysis are summarized in Figs.~\ref{fig:Cgamma} and  \ref{fig:CWb} and in Table~\ref{Tab:Individual}.
2) We have performed a global analysis by allowing all chirality-flipping top couplings at the high scale, 
with results summarized in Figs.~\ref{marginalized1} and \ref{marginalized3} and  Table~\ref{Tab:Margin}.
3) Finally, we have studied the top couplings in the context of Minimal Flavor Violation, which enforces relations among the top and 
lighter flavor anomalous couplings.

The overarching message emerging from our single-operator analysis is that  indirect probes 
put stronger constraints on the couplings than direct probes. 
Our major result is that even after properly taking into account the  hadronic and nuclear  uncertainties, 
EDMs dominate the bounds on all  the CPV  top couplings.  In particular,  
bounds on the  top EDM  (weak EDM)  are improved by  three orders of magnitude over the previous literature, 
leading to $|d_t| <  5 \cdot 10^{-20}$ $e$ cm ($90\%$ C.L.). 
In the Minimal Flavor Violation framework, we find that  top CPV couplings are bound at the same level or stronger than in the non-MFV case. 

Also for the CP-even couplings we find that indirect probes are very powerful. In the single operator analysis, Higgs production and decay signal strengths, electroweak precision observables, and the $b \rightarrow s \gamma$
branching ratio provide better constraints on $c_g$, $c_Y$, $c_\gamma$, and $c_{Wb}$ than direct observables. The only exception is $c_{Wt}$, which is mainly constrained by the helicity fractions of $W$ bosons produced in top quark decays.

If BSM physics simultaneously  generates  several operators at the scale $\Lambda$, 
cancellations are possible. For example,  a relatively large top EDM ($\tilde{c}_\gamma$) can be compatible with 
the absence of a signal in the ThO experiment, if an electron EDM is also generated at the  scale $\Lambda$, 
with exactly the right size to cancel the renormalization-group contribution from  $\tilde{c}_\gamma$, at the level of a few permil.     
This implies a  very non-trivial conspiracy 
among the couplings of the underlying model and still sets a powerful constraint on any BSM dynamics.

Another possibility is that new physics generates all the couplings of Eq.~(\ref{eq:Leff}) at the 
matching scale $\Lambda$.  In this case we  have quantified  the effect of cancellations 
by performing a global analysis with  five  complex  couplings. 
For the real part of the couplings we find that most bounds are not significantly affected. The exception are the top chromo-magnetic dipole moment $c_g$ and 
the correction to the top Yukawa $c_Y$, whose contributions to Higgs production can conspire to partially cancel. 
In particular, large corrections to the top Yukawa, up to 50\%, are still allowed. 
Future measurements of the $t\bar t$ and $t\bar t h$ cross sections  at the LHC, especially in regions where the contribution of $c_g$ is enhanced   \cite{Aguilar-Saavedra:2014iga},
will help to further improve these bounds.
For the imaginary part of the couplings, fixing the hadronic and nuclear matrix elements to  their central values, 
we find that the marginalized bounds are one-to-two orders of magnitude weaker than in the single operator analysis.  
However, as illustrated in Fig.~\ref{edmvswdm}, this requires strong correlations among different couplings again pointing to non-trivial effects in the underlying model. 

The above conclusions are blurred  by a more  conservative  treatment of theoretical uncertainties, 
such as the R-fit method.  In this case  unconstrained directions remain in the ten-dimensional parameter space. In this light, an inclusion of CP-odd collider observables into our analysis would also be very interesting. In any case, the unconstrained directions  underscore both the importance of having complementary ``orthogonal'' probes and 
the  importance of improved  calculations  of  the hadronic and nuclear matrix elements needed 
to relate EDMs to CP-violating operators.

\vspace{0.5cm}

\noindent {\bf Acknowledgements} --  We thank Aneesh Manohar for correspondence on Refs.~\cite{Jenkins:2013zja,Jenkins:2013wua, Alonso:2013hga},  Nicolas Mileo for correspondence on Ref.~\cite{Mileo:2016mxg}, and we are grateful to Miko\l aj Misiak for pointing out Refs.\ \cite{Czakon:2015exa,Misiak:2015xwa} to us.
We also acknowledge stimulating discussions with Sacha Davidson and Joachim Brod. 
EM thanks Simone Alioli for help in the calculation of the single top cross section. 
VC and EM  acknowledge support by the US DOE Office of Nuclear Physics and by the LDRD program at Los Alamos National Laboratory.
WD and JdV  acknowledge  support by the Dutch Organization for Scientific Research (NWO) 
through a RUBICON  and VENI grant, respectively.

\bibliographystyle{h-physrev3} 
\bibliography{bibliography}

\begin{thebibliography}{100}

\bibitem{Abe:1995hr}
CDF, F.~Abe {\em et~al.},
\newblock Phys. Rev. Lett. {\bf 74}, 2626 (1995), hep-ex/9503002.

\bibitem{Abachi:1995iq}
D0, S.~Abachi {\em et~al.},
\newblock Phys. Rev. Lett. {\bf 74}, 2632 (1995), hep-ex/9503003.

\bibitem{Kaplan:1991dc}
D.~B. Kaplan,
\newblock Nucl. Phys. {\bf B365}, 259 (1991).

\bibitem{Agashe:2006wa}
K.~Agashe, G.~Perez, and A.~Soni,
\newblock Phys. Rev. {\bf D75}, 015002 (2007), hep-ph/0606293.

\bibitem{Carena:2008vj}
M.~Carena, G.~Nardini, M.~Quiros, and C.~E.~M. Wagner,
\newblock Nucl. Phys. {\bf B812}, 243 (2009), 0809.3760.

\bibitem{Atwood:1992vj}
D.~Atwood, A.~Aeppli, and A.~Soni,
\newblock Phys. Rev. Lett. {\bf 69}, 2754 (1992).

\bibitem{Choudhury:2012np}
D.~Choudhury and P.~Saha,
\newblock JHEP {\bf 08}, 144 (2012), 1201.4130.

\bibitem{Baumgart:2012ay}
M.~Baumgart and B.~Tweedie,
\newblock JHEP {\bf 03}, 117 (2013), 1212.4888.

\bibitem{Biswal:2012dr}
S.~S. Biswal, S.~D. Rindani, and P.~Sharma,
\newblock Phys. Rev. {\bf D88}, 074018 (2013), 1211.4075.

\bibitem{Bernreuther:2013aga}
W.~Bernreuther and Z.-G. Si,
\newblock Phys. Lett. {\bf B725}, 115 (2013), 1305.2066,
\newblock [Erratum: Phys. Lett.B744,413(2015)].

\bibitem{Hioki:2013hva}
Z.~Hioki and K.~Ohkuma,
\newblock Phys. Rev. {\bf D88}, 017503 (2013), 1306.5387.

\bibitem{Aguilar-Saavedra:2014iga}
J.~A. Aguilar-Saavedra, B.~Fuks, and M.~L. Mangano,
\newblock Phys. Rev. {\bf D91}, 094021 (2015), 1412.6654.

\bibitem{Bramante:2014gda}
J.~Bramante, A.~Delgado, and A.~Martin,
\newblock Phys. Rev. {\bf D89}, 093006 (2014), 1402.5985.

\bibitem{Englert:2014oea}
C.~Englert, D.~Goncalves, and M.~Spannowsky,
\newblock Phys. Rev. {\bf D89}, 074038 (2014), 1401.1502.

\bibitem{Rindani:2015vya}
S.~D. Rindani, P.~Sharma, and A.~W. Thomas,
\newblock JHEP {\bf 10}, 180 (2015), 1507.08385.

\bibitem{Gaitan:2015aia}
R.~Gaitan, E.~A. Garces, J.~H.~M. de~Oca, and R.~Martinez,
\newblock Phys. Rev. {\bf D92}, 094025 (2015), 1505.04168.

\bibitem{Bernreuther:2015yna}
W.~Bernreuther, D.~Heisler, and Z.-G. Si,
\newblock JHEP {\bf 12}, 026 (2015), 1508.05271.

\bibitem{CorderoCid:2007uc}
A.~Cordero-Cid, J.~M. Hernandez, G.~Tavares-Velasco, and J.~J. Toscano,
\newblock J. Phys. {\bf G35}, 025004 (2008), 0712.0154.

\bibitem{Fael:2013ira}
M.~Fael and T.~Gehrmann,
\newblock Phys. Rev. {\bf D88}, 033003 (2013), 1307.1349.

\bibitem{Bouzas:2012av}
A.~O. Bouzas and F.~Larios,
\newblock Phys. Rev. {\bf D87}, 074015 (2013), 1212.6575.

\bibitem{Bouzas:2013jha}
A.~O. Bouzas and F.~Larios,
\newblock Phys. Rev. {\bf D88}, 094007 (2013), 1308.5634.

\bibitem{Rontsch:2015una}
R.~R\"{o}ntsch and M.~Schulze,
\newblock JHEP {\bf 08}, 044 (2015), 1501.05939.

\bibitem{Grzadkowski:2008mf}
B.~Grzadkowski and M.~Misiak,
\newblock Phys. Rev. D {\bf 78}, 077501 (2008), 0802.1413.

\bibitem{Drobnak:2010ej}
J.~Drobnak, S.~Fajfer, and J.~F. Kamenik,
\newblock Phys. Rev. {\bf D82}, 114008 (2010), 1010.2402.

\bibitem{GonzalezSprinberg:2011kx}
G.~A. Gonzalez-Sprinberg, R.~Martinez, and J.~Vidal,
\newblock JHEP {\bf 07}, 094 (2011), 1105.5601,
\newblock [Erratum: JHEP05,117(2013)].

\bibitem{Drobnak:2011aa}
J.~Drobnak, S.~Fajfer, and J.~F. Kamenik,
\newblock Nucl. Phys. {\bf B855}, 82 (2012), 1109.2357.

\bibitem{Rindani:2011pk}
S.~D. Rindani and P.~Sharma,
\newblock JHEP {\bf 11}, 082 (2011), 1107.2597.

\bibitem{Rindani:2011gt}
S.~D. Rindani and P.~Sharma,
\newblock Phys. Lett. {\bf B712}, 413 (2012), 1108.4165.

\bibitem{Cao:2015doa}
Q.-H. Cao, B.~Yan, J.-H. Yu, and C.~Zhang,
\newblock (2015), 1504.03785.

\bibitem{Hioki:2015env}
Z.~Hioki and K.~Ohkuma,
\newblock Phys. Lett. {\bf B752}, 128 (2016), 1511.03437.

\bibitem{Aguilar:2015vsa}
R.~Romero~Aguilar, A.~O. Bouzas, and F.~Larios,
\newblock Phys. Rev. {\bf D92}, 114009 (2015), 1509.06431.

\bibitem{Schulze:2016qas}
M.~Schulze and Y.~Soreq,
\newblock (2016), 1603.08911.

\bibitem{Birman:2016jhg}
J.~L. Birman, F.~D\'eliot, M.~C.~N. Fiolhais, A.~Onofre, and C.~M. Pease,
\newblock (2016), 1605.02679.

\bibitem{BhupalDev:2007ftb}
P.~S. Bhupal~Dev, A.~Djouadi, R.~M. Godbole, M.~M. Muhlleitner, and S.~D.
  Rindani,
\newblock Phys. Rev. Lett. {\bf 100}, 051801 (2008), 0707.2878.

\bibitem{Brod:2013cka}
J.~Brod, U.~Haisch, and J.~Zupan,
\newblock JHEP {\bf 1311}, 180 (2013).

\bibitem{Dolan:2014upa}
M.~J. Dolan, P.~Harris, M.~Jankowiak, and M.~Spannowsky,
\newblock Phys. Rev. D {\bf 90}, 073008 (2014), 1406.3322.

\bibitem{Demartin:2014fia}
F.~Demartin, F.~Maltoni, K.~Mawatari, B.~Page, and M.~Zaro,
\newblock Eur. Phys. J. C {\bf 74}, 3065 (2014), 1407.5089.

\bibitem{Kobakhidze:2014gqa}
A.~Kobakhidze, L.~Wu, and J.~Yue,
\newblock JHEP {\bf 10}, 100 (2014), 1406.1961.

\bibitem{Khatibi:2014bsa}
S.~Khatibi and M.~M. Najafabadi,
\newblock Phys. Rev. D {\bf 90}, 074014 (2014), 1409.6553.

\bibitem{Demartin:2015uha}
F.~Demartin, F.~Maltoni, K.~Mawatari, and M.~Zaro,
\newblock Eur. Phys. J. C {\bf 75}, 267 (2015), 1504.00611.

\bibitem{Chen:2015rha}
Y.~Chen, D.~Stolarski, and R.~Vega-Morales,
\newblock Phys. Rev. D {\bf 92}, 053003 (2015), 1505.01168.

\bibitem{Buckley:2015vsa}
M.~R. Buckley and D.~Goncalves,
\newblock Phys. Rev. Lett. {\bf 116}, 091801 (2016), 1507.07926.

\bibitem{Mileo:2016mxg}
N.~Mileo, K.~Kiers, A.~Szynkman, D.~Crane, and E.~Gegner,
\newblock (2016), 1603.03632.

\bibitem{Kamenik:2011dk}
J.~F. Kamenik, M.~Papucci, and A.~Weiler,
\newblock Phys. Rev. D {\bf 85}, 071501 (2012), 1107.3143,
\newblock [Erratum: Phys. Rev.D88,no.3,039903(2013)].

\bibitem{Zhang:2012cd}
C.~Zhang, N.~Greiner, and S.~Willenbrock,
\newblock Phys. Rev. {\bf D86}, 014024 (2012), 1201.6670.

\bibitem{deBlas:2015aea}
J.~de~Blas, M.~Chala, and J.~Santiago,
\newblock JHEP {\bf 09}, 189 (2015), 1507.00757.

\bibitem{Buckley:2015nca}
A.~Buckley {\em et~al.},
\newblock Phys. Rev. {\bf D92}, 091501 (2015), 1506.08845.

\bibitem{Buckley:2015lku}
A.~Buckley {\em et~al.},
\newblock JHEP {\bf 04}, 015 (2016), 1512.03360.

\bibitem{Bylund:2016phk}
O.~B. Bylund, F.~Maltoni, I.~Tsinikos, E.~Vryonidou, and C.~Zhang,
\newblock (2016), 1601.08193.

\bibitem{Gorbahn:2014sha}
M.~Gorbahn and U.~Haisch,
\newblock JHEP {\bf 06}, 033 (2014), 1404.4873.

\bibitem{Chien:2015xha}
Y.~T. Chien, V.~Cirigliano, W.~Dekens, J.~de~Vries, and E.~Mereghetti,
\newblock JHEP {\bf 02}, 011 (2016), 1510.00725,
\newblock [JHEP02,011(2016)].

\bibitem{Cirigliano:2016njn}
V.~Cirigliano, W.~Dekens, J.~de~Vries, and E.~Mereghetti,
\newblock (2016), 1603.03049.

\bibitem{Buchmuller:1985jz}
W.~Buchm{\"u}ller and D.~Wyler,
\newblock Nucl. Phys. B {\bf 268}, 621 (1986).

\bibitem{Grzadkowski:2010es}
B.~Grzadkowski, M.~Iskrzynski, M.~Misiak, and J.~Rosiek,
\newblock JHEP {\bf 1010}, 085 (2010), 1008.4884.

\bibitem{Jenkins:2013zja}
E.~E. Jenkins, A.~V. Manohar, and M.~Trott,
\newblock JHEP {\bf 10}, 087 (2013), 1308.2627.

\bibitem{Jenkins:2013wua}
E.~E. Jenkins, A.~V. Manohar, and M.~Trott,
\newblock JHEP {\bf 01}, 035 (2014), 1310.4838.

\bibitem{Alonso:2013hga}
R.~Alonso, E.~E. Jenkins, A.~V. Manohar, and M.~Trott,
\newblock JHEP {\bf 04}, 159 (2014), 1312.2014.

\bibitem{Aad:2015yem}
ATLAS, G.~Aad {\em et~al.},
\newblock JHEP {\bf 04}, 023 (2016), 1510.03764.

\bibitem{Bernreuther:1992be}
W.~Bernreuther, O.~Nachtmann, P.~Overmann, and T.~Schroder,
\newblock Nucl. Phys. {\bf B388}, 53 (1992),
\newblock [Erratum: Nucl. Phys.B406,516(1993)].

\bibitem{Brandenburg:1992be}
A.~Brandenburg and J.~P. Ma,
\newblock Phys. Lett. {\bf B298}, 211 (1993).

\bibitem{Bernreuther:1993hq}
W.~Bernreuther and A.~Brandenburg,
\newblock Phys. Rev. {\bf D49}, 4481 (1994), hep-ph/9312210.

\bibitem{Choi:1997ie}
S.~Y. Choi, C.~S. Kim, and J.~Lee,
\newblock Phys. Lett. {\bf B415}, 67 (1997), hep-ph/9706379.

\bibitem{Sjolin:2003ah}
J.~Sjolin,
\newblock J. Phys. {\bf G29}, 543 (2003).

\bibitem{Antipin:2008zx}
O.~Antipin and G.~Valencia,
\newblock Phys. Rev. {\bf D79}, 013013 (2009), 0807.1295.

\bibitem{Gupta:2009wu}
S.~K. Gupta, A.~S. Mete, and G.~Valencia,
\newblock Phys. Rev. {\bf D80}, 034013 (2009), 0905.1074.

\bibitem{Gupta:2009eq}
S.~K. Gupta and G.~Valencia,
\newblock Phys. Rev. {\bf D81}, 034013 (2010), 0912.0707.

\bibitem{Hayreter:2015ryk}
A.~Hayreter and G.~Valencia,
\newblock Phys. Rev. {\bf D93}, 014020 (2016), 1511.01464.

\bibitem{Dekens:2013zca}
W.~Dekens and J.~de~Vries,
\newblock JHEP {\bf 1305}, 149 (2013), 1303.3156.

\bibitem{Hisano:2012cc}
J.~Hisano, K.~Tsumura, and M.~J. Yang,
\newblock Phys.Lett. {\bf B713}, 473 (2012), 1205.2212.

\bibitem{Degrassi:2005zd}
G.~Degrassi, E.~Franco, S.~Marchetti, and L.~Silvestrini,
\newblock JHEP {\bf 0511}, 044 (2005), hep-ph/0510137.

\bibitem{Kaplan:1988ku}
D.~B. Kaplan and A.~Manohar,
\newblock Nucl. Phys. B {\bf 310}, 527 (1988).

\bibitem{Grojean:2013kd}
C.~Grojean, E.~E. Jenkins, A.~V. Manohar, and M.~Trott,
\newblock JHEP {\bf 1304}, 016 (2013).

\bibitem{Bhattacharya:2015rsa}
T.~Bhattacharya, V.~Cirigliano, R.~Gupta, E.~Mereghetti, and B.~Yoon,
\newblock Phys. Rev. {\bf D92}, 114026 (2015), 1502.07325.

\bibitem{Elias-Miro:2013gya}
J.~Elias-Miró, J.~R. Espinosa, E.~Masso, and A.~Pomarol,
\newblock JHEP {\bf 08}, 033 (2013), 1302.5661.

\bibitem{Elias-Miro:2013mua}
J.~Elias-Miro, J.~R. Espinosa, E.~Masso, and A.~Pomarol,
\newblock JHEP {\bf 11}, 066 (2013), 1308.1879.

\bibitem{Aebischer:2015fzz}
J.~Aebischer, A.~Crivellin, M.~Fael, and C.~Greub,
\newblock (2015), 1512.02830.

\bibitem{Weinberg:1989dx}
S.~Weinberg,
\newblock Phys. Rev. Lett. {\bf 63}, 2333 (1989).

\bibitem{Wilczek:1976ry}
F.~Wilczek and A.~Zee,
\newblock Phys. Rev. D {\bf 15}, 2660 (1977).

\bibitem{BraatenPRL}
E.~Braaten, C.-S. Li, and T.-C. Yuan,
\newblock Phys. Rev. Lett. {\bf 64}, 1709 (1990).

\bibitem{Agashe:2014kda}
Particle Data Group, K.~A. Olive {\em et~al.},
\newblock Chin. Phys. {\bf C38}, 090001 (2014).

\bibitem{Barr:1990vd}
S.~M. Barr and A.~Zee,
\newblock Phys. Rev. Lett. {\bf 65}, 21 (1990).

\bibitem{Gunion:1990iv}
J.~Gunion and D.~Wyler,
\newblock Phys.Lett. {\bf B248}, 170 (1990).

\bibitem{Abe:2013qla}
T.~Abe, J.~Hisano, T.~Kitahara, and K.~Tobioka,
\newblock JHEP {\bf 1401}, 106 (2014), 1311.4704.

\bibitem{Jung:2013hka}
M.~Jung and A.~Pich,
\newblock JHEP {\bf 04}, 076 (2014), 1308.6283.

\bibitem{Dekens:2014jka}
W.~Dekens {\em et~al.},
\newblock JHEP {\bf 07}, 069 (2014), 1404.6082.

\bibitem{Dicus:1989va}
D.~A. Dicus,
\newblock Phys.Rev. {\bf D41}, 999 (1990).

\bibitem{Boyd:1990bx}
G.~Boyd, A.~K. Gupta, S.~P. Trivedi, and M.~B. Wise,
\newblock Phys.Lett. {\bf B241}, 584 (1990).

\bibitem{Peskin:1990zt}
M.~E. Peskin and T.~Takeuchi,
\newblock Phys. Rev. Lett. {\bf 65}, 964 (1990).

\bibitem{Peskin:1991sw}
M.~E. Peskin and T.~Takeuchi,
\newblock Phys. Rev. {\bf D46}, 381 (1992).

\bibitem{Barbieri:2004qk}
R.~Barbieri, A.~Pomarol, R.~Rattazzi, and A.~Strumia,
\newblock Nucl. Phys. {\bf B703}, 127 (2004), hep-ph/0405040.

\bibitem{Berthier:2015oma}
L.~Berthier and M.~Trott,
\newblock JHEP {\bf 05}, 024 (2015), 1502.02570.

\bibitem{Berthier:2015gja}
L.~Berthier and M.~Trott,
\newblock JHEP {\bf 02}, 069 (2016), 1508.05060.

\bibitem{Contino:2016jqw}
R.~Contino, A.~Falkowski, F.~Goertz, C.~Grojean, and F.~Riva,
\newblock (2016), 1604.06444.

\bibitem{Aaltonen:2013wca}
CDF, D0, T.~A. Aaltonen {\em et~al.},
\newblock Phys. Rev. {\bf D89}, 072001 (2014), 1309.7570.

\bibitem{Aad:2014kva}
ATLAS, G.~Aad {\em et~al.},
\newblock Eur. Phys. J. {\bf C74}, 3109 (2014), 1406.5375.

\bibitem{Chatrchyan:2013faa}
CMS, S.~Chatrchyan {\em et~al.},
\newblock JHEP {\bf 02}, 024 (2014), 1312.7582,
\newblock [Erratum: JHEP02,102(2014)].

\bibitem{Aad:2014fwa}
ATLAS, G.~Aad {\em et~al.},
\newblock Phys. Rev. {\bf D90}, 112006 (2014), 1406.7844.

\bibitem{Chatrchyan:2012ep}
CMS, S.~Chatrchyan {\em et~al.},
\newblock JHEP {\bf 12}, 035 (2012), 1209.4533.

\bibitem{ATLAS-CONF-2014-007}
CERN Report No. ATLAS-CONF-2014-007, 2014 (unpublished).

\bibitem{Khachatryan:2014iya}
CMS, V.~Khachatryan {\em et~al.},
\newblock JHEP {\bf 06}, 090 (2014), 1403.7366.

\bibitem{ATLAS-CONF-2015-079}
CERN Report No. ATLAS-CONF-2015-079, 2015 (unpublished).

\bibitem{CMS-PAS-TOP-16-003}
CERN Report No. CMS-PAS-TOP-16-003, 2016 (unpublished).

\bibitem{Atwood:1994vm}
D.~Atwood, A.~Kagan, and T.~G. Rizzo,
\newblock Phys. Rev. {\bf D52}, 6264 (1995), hep-ph/9407408.

\bibitem{Haberl:1995ek}
P.~Haberl, O.~Nachtmann, and A.~Wilch,
\newblock Phys. Rev. {\bf D53}, 4875 (1996), hep-ph/9505409.

\bibitem{Czakon:2011xx}
M.~Czakon and A.~Mitov,
\newblock Comput. Phys. Commun. {\bf 185}, 2930 (2014), 1112.5675.

\bibitem{Czakon:2013goa}
M.~Czakon, P.~Fiedler, and A.~Mitov,
\newblock Phys. Rev. Lett. {\bf 110}, 252004 (2013), 1303.6254.

\bibitem{Botje:2011sn}
M.~Botje {\em et~al.},
\newblock (2011), 1101.0538.

\bibitem{Lai:2010vv}
H.-L. Lai {\em et~al.},
\newblock Phys. Rev. {\bf D82}, 074024 (2010), 1007.2241.

\bibitem{Martin:2009iq}
A.~D. Martin, W.~J. Stirling, R.~S. Thorne, and G.~Watt,
\newblock Eur. Phys. J. {\bf C63}, 189 (2009), 0901.0002.

\bibitem{Ball:2012cx}
R.~D. Ball {\em et~al.},
\newblock Nucl. Phys. {\bf B867}, 244 (2013), 1207.1303.

\bibitem{Franzosi:2015osa}
D.~Buarque~Franzosi and C.~Zhang,
\newblock Phys. Rev. {\bf D91}, 114010 (2015), 1503.08841.

\bibitem{Beenakker:2001rj}
W.~Beenakker {\em et~al.},
\newblock Phys. Rev. Lett. {\bf 87}, 201805 (2001), hep-ph/0107081.

\bibitem{Beenakker:2002nc}
W.~Beenakker {\em et~al.},
\newblock Nucl. Phys. {\bf B653}, 151 (2003), hep-ph/0211352.

\bibitem{Reina:2001sf}
L.~Reina and S.~Dawson,
\newblock Phys. Rev. Lett. {\bf 87}, 201804 (2001), hep-ph/0107101.

\bibitem{Dawson:2002tg}
S.~Dawson, L.~H. Orr, L.~Reina, and D.~Wackeroth,
\newblock Phys. Rev. {\bf D67}, 071503 (2003), hep-ph/0211438.

\bibitem{Frederix:2011zi}
R.~Frederix {\em et~al.},
\newblock Phys. Lett. {\bf B701}, 427 (2011), 1104.5613.

\bibitem{Degrande:2012gr}
C.~Degrande, J.~M. Gerard, C.~Grojean, F.~Maltoni, and G.~Servant,
\newblock JHEP {\bf 07}, 036 (2012), 1205.1065,
\newblock [Erratum: JHEP03,032(2013)].

\bibitem{Hayreter:2013kba}
A.~Hayreter and G.~Valencia,
\newblock Phys. Rev. {\bf D88}, 034033 (2013), 1304.6976.

\bibitem{Abazov:2009ii}
D0, V.~M. Abazov {\em et~al.},
\newblock Phys. Rev. Lett. {\bf 103}, 092001 (2009), 0903.0850.

\bibitem{Aaltonen:2010jr}
CDF, T.~Aaltonen {\em et~al.},
\newblock Phys. Rev. {\bf D82}, 112005 (2010), 1004.1181.

\bibitem{Aad:2015eto}
ATLAS, G.~Aad {\em et~al.},
\newblock JHEP {\bf 01}, 064 (2016), 1510.03752.

\bibitem{Chatrchyan:2014tua}
CMS, S.~Chatrchyan {\em et~al.},
\newblock Phys. Rev. Lett. {\bf 112}, 231802 (2014), 1401.2942.

\bibitem{CDF:2014uma}
CDF, D0, T.~A. Aaltonen {\em et~al.},
\newblock Phys. Rev. Lett. {\bf 112}, 231803 (2014), 1402.5126.

\bibitem{Bordes:1994ki}
G.~Bordes and B.~van Eijk,
\newblock Nucl. Phys. {\bf B435}, 23 (1995).

\bibitem{Stelzer:1997ns}
T.~Stelzer, Z.~Sullivan, and S.~Willenbrock,
\newblock Phys. Rev. {\bf D56}, 5919 (1997), hep-ph/9705398.

\bibitem{Harris:2002md}
B.~W. Harris, E.~Laenen, L.~Phaf, Z.~Sullivan, and S.~Weinzierl,
\newblock Phys. Rev. {\bf D66}, 054024 (2002), hep-ph/0207055.

\bibitem{Campbell:2009ss}
J.~M. Campbell, R.~Frederix, F.~Maltoni, and F.~Tramontano,
\newblock Phys. Rev. Lett. {\bf 102}, 182003 (2009), 0903.0005.

\bibitem{Zhang:2010dr}
C.~Zhang and S.~Willenbrock,
\newblock Phys. Rev. {\bf D83}, 034006 (2011), 1008.3869.

\bibitem{Zhang:2016omx}
C.~Zhang,
\newblock Phys. Rev. Lett. {\bf 116}, 162002 (2016), 1601.06163.

\bibitem{Aaltonen:2012rz}
CDF, D0, T.~Aaltonen {\em et~al.},
\newblock Phys. Rev. {\bf D85}, 071106 (2012), 1202.5272.

\bibitem{Aad:2012ky}
ATLAS, G.~Aad {\em et~al.},
\newblock JHEP {\bf 06}, 088 (2012), 1205.2484.

\bibitem{Chatrchyan:2013jna}
CMS, S.~Chatrchyan {\em et~al.},
\newblock JHEP {\bf 10}, 167 (2013), 1308.3879.

\bibitem{Khachatryan:2014vma}
CMS, V.~Khachatryan {\em et~al.},
\newblock JHEP {\bf 01}, 053 (2015), 1410.1154.

\bibitem{Boudreau:2013yna}
J.~Boudreau, C.~Escobar, J.~Mueller, K.~Sapp, and J.~Su,
\newblock (2013), 1304.5639.

\bibitem{Czarnecki:2010gb}
A.~Czarnecki, J.~G. Korner, and J.~H. Piclum,
\newblock Phys. Rev. {\bf D81}, 111503 (2010), 1005.2625.

\bibitem{Khachatryan:2014jba}
CMS, V.~Khachatryan {\em et~al.},
\newblock Eur. Phys. J. {\bf C75}, 212 (2015), 1412.8662.

\bibitem{Aad:2015gba}
ATLAS, G.~Aad {\em et~al.},
\newblock Eur. Phys. J. {\bf C76}, 6 (2016), 1507.04548.

\bibitem{Heinemeyer:2013tqa}
LHC Higgs Cross Section Working Group, S.~Heinemeyer {\em et~al.},
\newblock (2013), 1307.1347.

\bibitem{Harlander:2002wh}
R.~V. Harlander and W.~B. Kilgore,
\newblock Phys. Rev. Lett. {\bf 88}, 201801 (2002), hep-ph/0201206.

\bibitem{Anastasiou:2002yz}
C.~Anastasiou and K.~Melnikov,
\newblock Nucl. Phys. {\bf B646}, 220 (2002), hep-ph/0207004.

\bibitem{Ravindran:2003um}
V.~Ravindran, J.~Smith, and W.~L. van Neerven,
\newblock Nucl. Phys. {\bf B665}, 325 (2003), hep-ph/0302135.

\bibitem{Anastasiou:2002wq}
C.~Anastasiou and K.~Melnikov,
\newblock Phys. Rev. {\bf D67}, 037501 (2003), hep-ph/0208115.

\bibitem{Harlander:2002vv}
R.~V. Harlander and W.~B. Kilgore,
\newblock JHEP {\bf 10}, 017 (2002), hep-ph/0208096.

\bibitem{Spira:1997dg}
M.~Spira,
\newblock Fortsch. Phys. {\bf 46}, 203 (1998), hep-ph/9705337.

\bibitem{Grinstein:1991cd}
B.~Grinstein and M.~B. Wise,
\newblock Phys. Lett. {\bf B265}, 326 (1991).

\bibitem{Greiner:2011tt}
N.~Greiner, S.~Willenbrock, and C.~Zhang,
\newblock Phys. Lett. {\bf B704}, 218 (2011), 1104.3122.

\bibitem{Drobnak:2011wj}
J.~Drobnak, S.~Fajfer, and J.~F. Kamenik,
\newblock Phys. Lett. B {\bf 701}, 234 (2011), 1102.4347.

\bibitem{Mertens:2011ts}
P.~Mertens and C.~Smith,
\newblock JHEP {\bf 08}, 069 (2011), 1103.5992.

\bibitem{Bertolini:2014sua}
S.~Bertolini, A.~Maiezza, and F.~Nesti,
\newblock Phys. Rev. {\bf D89}, 095028 (2014), 1403.7112.

\bibitem{Lenz:2010gu}
A.~Lenz {\em et~al.},
\newblock Phys. Rev. D {\bf 83}, 036004 (2011), 1008.1593.

\bibitem{Altmannshofer:2012az}
W.~Altmannshofer and D.~M. Straub,
\newblock JHEP {\bf 08}, 121 (2012), 1206.0273.

\bibitem{Altmannshofer:2011gn}
W.~Altmannshofer, P.~Paradisi, and D.~M. Straub,
\newblock JHEP {\bf 04}, 008 (2012), 1111.1257.

\bibitem{Misiak:2006zs}
M.~Misiak {\em et~al.},
\newblock Phys. Rev. Lett. {\bf 98}, 022002 (2007), hep-ph/0609232.

\bibitem{Lunghi:2006hc}
E.~Lunghi and J.~Matias,
\newblock JHEP {\bf 04}, 058 (2007), hep-ph/0612166.

\bibitem{Benzke:2010tq}
M.~Benzke, S.~J. Lee, M.~Neubert, and G.~Paz,
\newblock Phys. Rev. Lett. {\bf 106}, 141801 (2011), 1012.3167.

\bibitem{Kagan:1998ym}
A.~L. Kagan and M.~Neubert,
\newblock Eur. Phys. J. {\bf C7}, 5 (1999), hep-ph/9805303.

\bibitem{Misiak:2015xwa}
M.~Misiak {\em et~al.},
\newblock Phys. Rev. Lett. {\bf 114}, 221801 (2015), 1503.01789.

\bibitem{Czakon:2015exa}
M.~Czakon {\em et~al.},
\newblock JHEP {\bf 04}, 168 (2015), 1503.01791.

\bibitem{Buras:1998raa}
A.~J. Buras,
\newblock {Weak Hamiltonian, CP violation and rare decays},
\newblock in {\em {Probing the standard model of particle interactions.
  Proceedings, Summer School in Theoretical Physics, NATO Advanced Study
  Institute, 68th session, Les Houches, France, July 28-September 5, 1997. Pt.
  1, 2}}, pp. 281--539, 1998, hep-ph/9806471.

\bibitem{Charles:2004jd}
CKMfitter Group, J.~Charles {\em et~al.},
\newblock Eur. Phys. J. {\bf C41}, 1 (2005), hep-ph/0406184.

\bibitem{Amhis:2014hma}
Heavy Flavor Averaging Group (HFAG), Y.~Amhis {\em et~al.},
\newblock (2014), 1412.7515.

\bibitem{Baron:2013eja}
ACME Collaboration, J.~Baron {\em et~al.},
\newblock Science {\bf 343}, 269 (2014), 1310.7534.

\bibitem{Baker:2006ts}
C.~A. Baker {\em et~al.},
\newblock Phys. Rev. Lett. {\bf 97}, 131801 (2006), hep-ex/0602020.

\bibitem{Afach:2015sja}
J.~Pendlebury {\em et~al.},
\newblock Phys. Rev. {\bf D92}, 092003 (2015), 1509.04411.

\bibitem{Griffith:2009zz}
W.~C. Griffith {\em et~al.},
\newblock Phys. Rev. Lett. {\bf 102}, 101601 (2009).

\bibitem{Graner:2016ses}
B.~Graner, Y.~Chen, E.~G. Lindahl, and B.~R. Heckel,
\newblock (2016), 1601.04339.

\bibitem{Kumar:2013qya}
K.~Kumar, Z.-T. Lu, and M.~J. Ramsey-Musolf,
\newblock {Working Group Report: Nucleons, Nuclei, and Atoms},
\newblock in {\em {Community Summer Study 2013: Snowmass on the Mississippi
  (CSS2013) Minneapolis, MN, USA, July 29-August 6, 2013}}, 2013, 1312.5416.

\bibitem{Chupp:2014gka}
T.~Chupp and M.~Ramsey-Musolf,
\newblock Phys. Rev. {\bf C91}, 035502 (2015), 1407.1064.

\bibitem{PhysRevLett.86.22}
M.~A. Rosenberry and T.~E. Chupp,
\newblock Phys. Rev. Lett. {\bf 86}, 22 (2001).

\bibitem{Parker:2015yka}
R.~Parker {\em et~al.},
\newblock Phys. Rev. Lett. {\bf 114}, 233002 (2015), 1504.07477.

\bibitem{Baluni:1978rf}
V.~Baluni,
\newblock Phys. Rev. D {\bf 19}, 2227 (1979).

\bibitem{Peccei:1977hh}
R.~D. Peccei and H.~R. Quinn,
\newblock Phys. Rev. Lett. {\bf 38}, 1440 (1977).

\bibitem{Pospelov_review}
M.~Pospelov and A.~Ritz,
\newblock Annals Phys. {\bf 318}, 119 (2005), hep-ph/0504231.

\bibitem{Guo:2015tla}
F.~K. Guo {\em et~al.},
\newblock Phys. Rev. Lett. {\bf 115}, 062001 (2015), 1502.02295.

\bibitem{Bhattacharya:2015esa}
T.~Bhattacharya, V.~Cirigliano, R.~Gupta, H.-W. Lin, and B.~Yoon,
\newblock Phys. Rev. Lett. {\bf 115}, 212002 (2015), 1506.04196.

\bibitem{Shindler:2015aqa}
A.~Shindler, T.~Luu, and J.~de~Vries,
\newblock Phys. Rev. {\bf D92}, 094518 (2015), 1507.02343.

\bibitem{Alexandrou:2015spa}
C.~Alexandrou {\em et~al.},
\newblock Phys. Rev. {\bf D93}, 074503 (2016), 1510.05823.

\bibitem{Shintani:2015vsx}
E.~Shintani, T.~Blum, T.~Izubuchi, and A.~Soni,
\newblock Phys. Rev. {\bf D93}, 094503 (2016), 1512.00566.

\bibitem{deVries:2012ab}
J.~de~Vries, E.~Mereghetti, R.~G.~E. Timmermans, and U.~van Kolck,
\newblock Annals Phys. {\bf 338}, 50 (2013), 1212.0990.

\bibitem{Bsaisou:2014oka}
J.~Bsaisou, U.-G. Mei{\ss}ner, A.~Nogga, and A.~Wirzba,
\newblock Annals Phys. {\bf 359}, 317 (2015), 1412.5471.

\bibitem{Skripnikov}
L.~V. Skripnikov, A.~N. Petrov, and A.~V. Titov,
\newblock The Journal of Chemical Physics {\bf 139},  (2013).

\bibitem{Fleig:2014uaa}
T.~Fleig and M.~K. Nayak,
\newblock J. Molec. Spectrosc. {\bf 300}, 16 (2014), 1401.2284.

\bibitem{Bhattacharya:2015wna}
PNDME, T.~Bhattacharya {\em et~al.},
\newblock Phys. Rev. {\bf D92}, 094511 (2015), 1506.06411.

\bibitem{Pospelov_qCEDM}
M.~Pospelov and A.~Ritz,
\newblock Phys. Rev. D {\bf 63}, 073015 (2001), hep-ph/0010037.

\bibitem{Pospelov_deuteron}
O.~Lebedev, K.~A. Olive, M.~Pospelov, and A.~Ritz,
\newblock Phys. Rev. D {\bf 70}, 016003 (2004), hep-ph/0402023.

\bibitem{Hisano1}
J.~Hisano, J.~Y. Lee, N.~Nagata, and Y.~Shimizu,
\newblock Phys. Rev. D {\bf 85}, 114044 (2012), 1204.2653.

\bibitem{Pospelov_Weinberg}
D.~A. Demir, M.~Pospelov, and A.~Ritz,
\newblock Phys. Rev. D {\bf 67}, 015007 (2003), hep-ph/0208257.

\bibitem{Hisano2}
K.~Fuyuto, J.~Hisano, and N.~Nagata,
\newblock Phys. Rev. D {\bf 87}, 054018 (2013), 1211.5228.

\bibitem{Pospelov_piN}
M.~Pospelov,
\newblock Phys. Lett. B {\bf 530}, 123 (2002), hep-ph/0109044.

\bibitem{Dmitriev:2003sc}
V.~F. Dmitriev and R.~A. Sen'kov,
\newblock Phys. Rev. Lett. {\bf 91}, 212303 (2003), nucl-th/0306050.

\bibitem{deJesus:2005nb}
J.~H. de~Jesus and J.~Engel,
\newblock Phys. Rev. {\bf C72}, 045503 (2005), nucl-th/0507031.

\bibitem{Ban:2010ea}
S.~Ban, J.~Dobaczewski, J.~Engel, and A.~Shukla,
\newblock Phys. Rev. {\bf C82}, 015501 (2010), 1003.2598.

\bibitem{Dzuba:2009kn}
V.~A. Dzuba, V.~V. Flambaum, and S.~G. Porsev,
\newblock Phys. Rev. A {\bf 80}, 032120 (2009), 0906.5437.

\bibitem{Engel:2013lsa}
J.~Engel, M.~J. Ramsey-Musolf, and U.~van Kolck,
\newblock Prog. Part. Nucl. Phys. {\bf 71}, 21 (2013), 1303.2371.

\bibitem{Eversmann:2015jnk}
JEDI, D.~Eversmann {\em et~al.},
\newblock Phys. Rev. Lett. {\bf 115}, 094801 (2015), 1504.00635.

\bibitem{deVries2011b}
J.~de~Vries {\em et~al.},
\newblock Phys. Rev. C {\bf 84}, 065501 (2011), 1109.3604.

\bibitem{Bsaisou:2014zwa}
J.~Bsaisou {\em et~al.},
\newblock JHEP {\bf 03}, 104 (2015), 1411.5804,
\newblock [Erratum: JHEP05,083(2015)].

\bibitem{Stetcu:2008vt}
I.~Stetcu, C.-P. Liu, J.~L. Friar, A.~C. Hayes, and P.~Navratil,
\newblock Phys.Lett. {\bf B665}, 168 (2008), 0804.3815.

\bibitem{Song:2012yh}
Y.-H. Song, R.~Lazauskas, and V.~Gudkov,
\newblock Phys.Rev. {\bf C87}, 015501 (2013), 1211.3762.

\bibitem{Yamanaka:2015qfa}
N.~Yamanaka and E.~Hiyama,
\newblock Phys. Rev. {\bf C91}, 054005 (2015), 1503.04446.

\bibitem{Yamanaka:2016itb}
N.~Yamanaka, T.~Yamada, E.~Hiyama, and Y.~Funaki,
\newblock (2016), 1603.03136.

\bibitem{Hanneke:2008tm}
D.~Hanneke, S.~Fogwell, and G.~Gabrielse,
\newblock Phys. Rev. Lett. {\bf 100}, 120801 (2008), 0801.1134.

\bibitem{Bennett:2006fi}
Muon g-2, G.~W. Bennett {\em et~al.},
\newblock Phys. Rev. {\bf D73}, 072003 (2006), hep-ex/0602035.

\bibitem{Aoyama:2012wj}
T.~Aoyama, M.~Hayakawa, T.~Kinoshita, and M.~Nio,
\newblock Phys. Rev. Lett. {\bf 109}, 111807 (2012), 1205.5368.

\bibitem{Bouchendira:2010es}
R.~Bouchendira, P.~Clade, S.~Guellati-Khelifa, F.~Nez, and F.~Biraben,
\newblock Phys. Rev. Lett. {\bf 106}, 080801 (2011), 1012.3627.

\bibitem{Bona:2007qt}
SuperB, M.~Bona {\em et~al.},
\newblock (2007), 0709.0451.

\bibitem{Nishida:2011dh}
S.~Nishida,
\newblock {Experimental Prospects for $B \to X_{s/d} \gamma$ and $B -> X_s
  \ell^+\ell^-$},
\newblock in {\em {CKM unitarity triangle. Proceedings, 6th International
  Workshop, CKM 2010, Warwick, UK, September 6-10, 2010}}, 2011, 1102.1045.

\bibitem{CMS:2013xfa}
CMS,
\newblock {Projected Performance of an Upgraded CMS Detector at the LHC and
  HL-LHC: Contribution to the Snowmass Process},
\newblock in {\em {Community Summer Study 2013: Snowmass on the Mississippi
  (CSS2013) Minneapolis, MN, USA, July 29-August 6, 2013}}, 2013, 1307.7135.

\bibitem{ATL-PHYS-PUB-2014-016}
CERN Report No. ATL-PHYS-PUB-2014-016, 2014 (unpublished).

\bibitem{D'Ambrosio:2002ex}
G.~D'Ambrosio, G.~Giudice, G.~Isidori, and A.~Strumia,
\newblock Nucl.Phys. {\bf B645}, 155 (2002), hep-ph/0207036.

\end{thebibliography}

\end{document}